\documentclass[preprint,showpacs,preprintnumbers,amsmath,amssymb,aps,nofootinbib,floatfix]{revtex4}

\newcommand{\tmpred}{red }
\newcommand{\tmpblue}{blue }
\newcommand{\tmpgreen}{green }
\newcommand{\tmpcyan}{cyan }
\newcommand{\tmpmagenta}{magenta }
\newcommand{\tmpblack}{black }
\newcommand{\tmpand}{and }

\usepackage{graphicx}
\usepackage{dcolumn}
\usepackage{bm}

\newcommand{\BE}{\begin{equation}}
\def\EE{\end{equation}}
\def\BEA{\begin{eqnarray}}
\def\EEA{\end{eqnarray}}
\def\EL{\nonumber\\}

\newcommand{\ie}{{\em i.e.\ }}
\newcommand{\eg}{{\em e.\,g.\ }}

\newcommand{\etal}{{\it et al.}}
\def\chpt{\raise0.4ex\hbox{$\chi$}PT}
\def\schpt{S\raise0.4ex\hbox{$\chi$}PT}
\def\figref#1{Fig.~\ref{fig:#1}}
\def\Figref#1{Figure~\ref{fig:#1}}
\def\figrefs#1#2{Figs.~\ref{fig:#1} and \ref{fig:#2}}
\def\Figrefs#1#2{Figures~\ref{fig:#1} and \ref{fig:#2}}

\def\Figrefthree#1#2#3{Figures~\ref{fig:#1}, \ref{fig:#2}, and \ref{fig:#3}}

\def\figrefto#1#2{Figs.~\ref{fig:#1} -- \ref{fig:#2}}

\def\secref#1{Sec.~\ref{sec:#1}}
\def\secrefs#1#2{Secs.~\ref{sec:#1} and \ref{sec:#2}}
\def\Secref#1{Section~\ref{sec:#1}}
\def\tabref#1{Table~\ref{tab:#1}}
\def\tabrefs#1#2{Tables~\ref{tab:#1} and \ref{tab:#2}}

\def\leftvec{\raise1.5ex\hbox{$\leftarrow$}\kern-.85em}
\def\half{{\scriptstyle \raise.2ex\hbox{${1\over2}$}}}
\def\threehalves{{\scriptstyle \raise.15ex\hbox{${3\over2}$}}}
\def\third{{\scriptstyle \raise.15ex\hbox{${1\over3}$}}}
\def\third{{\scriptstyle \raise.15ex\hbox{${1\over3}$}}}
\def\twothirds{{\scriptstyle \raise.15ex\hbox{${2\over3}$}}}
\def\fourth{{\scriptstyle \raise.15ex\hbox{${1\over4}$}}}
\def\gtwid{\raise.3ex\hbox{$>$\kern-.75em\lower1ex\hbox{$\sim$}}}
\def\ltwid{\raise.3ex\hbox{$<$\kern-.75em\lower1ex\hbox{$\sim$}}}

\def\cM{{\cal M}}

\def\cO{{\cal O}}

\def\Eqs#1#2{Equations~(\ref{eq:#1}) and (\ref{eq:#2})}
\def\eq#1{Eq.~(\ref{eq:#1})}
\def\eqs#1#2{Eqs.~(\ref{eq:#1}) and (\ref{eq:#2})}
\def\eqsthree#1#2#3{Eqs.~(\ref{eq:#1}), (\ref{eq:#2}) and (\ref{eq:#3})}
\def\eqsfour#1#2#3#4{Eqs.~(\ref{eq:#1}), (\ref{eq:#2}), (\ref{eq:#3}) and (\ref{eq:#4})}

\def\prd#1{Phys.\ Rev.\ {\bf D#1}}

\def\berlin{Nucl.\ Phys.\ {\bf B} (Proc.\ Suppl.) {\bf 106-107} (2002)}
\def\boston{Nucl.\ Phys.\ {\bf B} (Proc.\ Suppl.) {\bf 119} (2003)}
\def\tsukubathree{Nucl.\ Phys.\ {\bf B} (Proc.\ Suppl.) {\bf 129-130} (2004)}
\def\fermilabtwo{presented at the International Symposium,
{\it Lattice 2004}, Fermilab, June 21--26, 2004, to be published
in Nucl.\ Phys.\ {\bf B} (Proc.\ Suppl.)}

\def\MeV{{\rm Me\!V}}
\def\GeV{{\rm Ge\!V}}
\def\ie{{\it i.e.},\ }
\def\eg{{\it e.g.},\ }
\def\et{{\it et al.}}

\def\vs{{\it vs.}\ }
\def\Dslash{{D \!\!\!\!/}}

\newcommand{\msbar}{\text{$\overline{\text{MS}}$}}

\begin{document}

\title{Light pseudoscalar decay constants, quark masses, and low energy constants
from three-flavor lattice QCD}
\author{\phantom{MILC Collaboration}}
\affiliation{{\rm MILC Collaboration}}

\author{C.~Aubin}
\affiliation{Department of Physics, Washington University, St.~Louis, MO 63130, USA}

\author{C.~Bernard} 
\affiliation{Department of Physics, Washington University, St.~Louis, MO 63130, USA}

\author{C.~DeTar} 
\affiliation{Physics Department, University of Utah, Salt Lake City, UT 84112, USA}

\author{Steven Gottlieb} 
\affiliation{Department of Physics, Indiana University, Bloomington, IN 47405, USA}

\author{E.B.~Gregory} 
\affiliation{Department of Physics, University of Arizona, Tucson, AZ 85721, USA}

\author{U.M.~Heller} 
\affiliation{American Physical Society, One Research Road, Box 9000, Ridge, NY 11961-9000}

\author{J.E.~Hetrick}
\affiliation{Physics Department, University of the Pacific, Stockton, CA 95211, USA}

\author{J.~Osborn} 
\affiliation{Physics Department, University of Utah, Salt Lake City, UT 84112, USA}

\author{R.~Sugar}
\affiliation{Department of Physics, University of California, Santa Barbara, CA 93106, USA}

\author{D.~Toussaint} 
\affiliation{Department of Physics, University of Arizona, Tucson, AZ 85721, USA}

\date{\today}

\begin{abstract}
As part of our program of lattice simulations of three flavor QCD with improved
staggered quarks, we have calculated pseudoscalar meson masses and decay
constants for a range of valence quark masses and sea quark masses on lattices
with lattice spacings of about 0.125 fm and 0.09 fm. We fit the lattice data to
forms computed with ``staggered chiral perturbation theory.''  
Our results provide a sensitive test of the lattice simulations, and especially
of the chiral behavior, including the effects of chiral logarithms.
We find: $f_\pi  =   129.5 \pm 0.9\pm 3.5 \; \MeV$,
$ f_K  =   156.6 \pm 1.0\pm 3.6 \; \MeV $, and
$f_K/f_\pi   =  1.210(4)(13)$,
where the errors are statistical and systematic. 
Following a recent paper by Marciano, our value of
$f_K/f_\pi$ implies $|V_{us}|=0.2219(26)$.  
Further, we obtain
$m_u/m_d = 0.43(0)(1)(8)$,
where the errors are from statistics, simulation systematics, 
and electromagnetic effects, respectively.  
The partially quenched data can also be used to determine several of the
constants of the low energy chiral effective Lagrangian: in particular we find
$2L_8 - L_5= -0.2(1)(2) \; \times 10^{-3}$ at chiral scale $m_\eta$,
where the errors are statistical and systematic. 
This provides an alternative (though not independent)
way of estimating $m_u$; the value of $2L_8 - L_5$ is far outside the range
that would allow the up quark to be massless. 
Results for $m_s^{\msbar}$, $\hat m^\msbar$, and $m_s/\hat m$ can be obtained from
the same lattice data and chiral fits, and have been presented previously in 
joint work with the HPQCD and UKQCD collaborations. Using the perturbative mass renormalization
reported in that work, we obtain $ m_u^\msbar =  1.7(0)(1)(2)(2)\;\MeV$ and 
$m_d^\msbar =  3.9(0)(1)(4)(2)\;\MeV$ at scale $2\;\GeV$, with errors
from statistics, simulation, perturbation theory,
and electromagnetic effects, respectively.
\end{abstract}

\maketitle

\section{INTRODUCTION}
\label{sec:intro}

Using lattice QCD techniques, the masses and 
decay constants of light pseudoscalar mesons can be determined
with high precision at fixed quark mass and lattice spacing.
Assuming that the chiral and continuum extrapolations are under control, one can
therefore calculate from first principles a number of physically important quantities, including

\begin{itemize}

\item Pion and kaon leptonic decay constants, $f_\pi$ and $f_K$, and their ratio. 

\item Low energy (``Gasser-Leutwyler'' \cite{GASSER_LEUTWYLER}) constants $L_i$, in particular
$L_5$, $L_4$, and the combinations $2L_8-L_5$ and $2L_6-L_4$.

\item Quark mass ratios, such as $m_s/\hat m$, where $\hat m$ is the average
of the $u$ and $d$ quark masses, and $m_u/m_d$. 

\item Absolute quark mass values, if the mass renormalization constant is known
perturbatively or nonperturbatively.

\end{itemize}

The comparison of $f_\pi$ and $f_K$ with experiment provides a sensitive
test of lattice methods and algorithms.  A precise determination of
$f_K$, or $f_K/f_\pi$ may in fact be turned around to determine the magnitude
of the CKM element $V_{us}$, as emphasized recently by Marciano \cite{Marciano:2004uf}.
The quark masses are fundamental parameters of the Standard Model, and hence are 
phenomenologically   and intrinsically interesting. Of special importance here is the up quark mass:
if $m_u$ or $m_u/m_d$ can be bounded away from zero with small enough errors, it can
rule out $m_u=0$ as a solution to the strong CP problem \cite{STRONG_CP,Kaplan:1986ru}.\footnote
{We note that Creutz \protect{\cite{CREUTZ}} has argued that the statement $m_u=0$ is
not physically meaningful and therefore cannot be a resolution of  
the CP problem.  Since we find a non-zero
value for $m_u$ here, we are not forced to face this issue directly in the current work.}
Finally, the Gasser-Leutwyler parameters give a concise summary
of  the properties of low energy QCD.
In particular the combination $2L_8-L_5$ provides an alternative (although not independent)
handle on the up  quark mass \cite{COHEN,Nelson:tb}.

Extracting these important quantities is predicated on being able to control
the chiral and continuum extrapolations.  The improved staggered (Kogut-Susskind, KS) 
quarks \cite{IMP_ACTION,SYM_GAUGE}
used here have the advantage in this respect of allowing us to simulate at quite small
quark mass: Our lowest $m_\pi/m_\rho$ value is $\approx 0.3$, a
pion mass of roughly $250\, \MeV$. 
On the other hand, these extrapolations are complicated by the fact that
a single  staggered quark field describes four species of quarks. We call this
degree of freedom ``taste'' to distinguish it from physical flavor. We simulate the latter
by introducing distinct staggered fields for each nondegenerate quark flavor; while we 
handle the former by taking the fourth root of the staggered quark determinant.

The fact that taste symmetry is violated at finite lattice spacing leads to both
practical and theoretical complications. 
The improvement of the fermion action \cite{IMP_ACTION} reduces the
splittings among pseudoscalar mesons of various tastes to
$\cO(\alpha_S^2 a^2)$; yet the splittings are still numerically large,
especially on our coarser lattices. This practical problem
makes it impossible to fit our data with continuum chiral perturbation theory (\chpt)
expressions (see Refs.~\cite{MILC_SPECTRUM} and \cite{Aubin:2003ne}, as well as discussion
in \secref{discussion}).  Instead, we must use ``staggered chiral perturbation theory'' (\schpt)
\cite{LEE_SHARPE,CB_FSB,CA_CB1,CA_CB2}, which includes discretization effects within the
chiral expansion.   Using \schpt, we can take the chiral and continuum
limits at the same time, and arrive at physical results with rather small systematic
errors.  

Theoretically, it is not obvious that, in the presence of taste-violations, the fourth  
root procedure commutes with the limit of lattice spacing $a\to 0$.  
Assuming that perturbation theory for
the standard KS theory without the fourth root correctly reproduces a continuum  four-taste
theory, then the fourth root trick is correct in perturbation theory \cite{CB_MG_PQCHPT}, 
since it just multiplies each virtual quark loop by $1/4$.  However, nonperturbatively,
the fourth root version is almost certainly 
not ultra-local at finite lattice spacing, and the possibility
remains that it violates locality (and therefore universality) in the continuum limit.
We believe that existing checks \cite{BIG_PRL,GOTTLIEB_LAT03,MILC_SPECTRUM2,MASON_ALPHA}
of the formalism against experimental
results already make this possibility unlikely.  The current work adds more evidence
that the method gives results that agree well with experiment and have the proper
chiral behavior, up to controlled taste-violating effects that vanish in the
continuum limit.  However the question is not yet settled.  We discuss this
further in \secref{fourth-root} and briefly refer to other recent work that addresses
the issue.

This violation of taste symmetry arises because the full axial symmetry (at $m_q=0$)
is broken to a single U(1) subgroup on the lattice.   This means that only one
of the pseudoscalars, which we call the ``Goldstone meson'', has
its mass and decay constant protected from renormalization.
A study of pion masses and decay constants by the JLQCD collaboration \cite{JLQCD_FPI}
explored the masses and decay constants of all of the pseudoscalars in a
quenched calculation.   We concentrate almost exclusively here on Goldstone 
mesons, thus avoiding the necessity for renormalization.  

We have generated a large ``partially quenched'' data set of Goldstone meson
masses and decay constants
using three flavors of  improved KS sea
quarks.  These quantities have been computed with a wide range of sea quark
masses (with $m_u=m_d\not=m_s$), and on lattices with 
lattice spacings of about 0.125 fm and 0.09 fm.
We have 8 or 9
different valence quark masses available for each set of sea quark masses and
lattice spacing.  This data may be fit to chiral-logarithm forms from \schpt, which at present
have been computed for Goldstone mesons only \cite{CA_CB1,CA_CB2}.  However, since the masses
for mesons of other tastes enter into the one-loop chiral logarithms of the Goldstone
mesons,  some control over those masses is also needed.  We have computed most non-Goldstone
``full QCD'' (valence masses equal to sea masses) pion masses on most of our lattices.  We can
fit that data to the tree-level (LO) \schpt\ form, and use the results for splitting and slopes
as input to the NLO terms for the Goldstone mesons.  There is, of course, a NNLO error in this
procedure, which we estimate in \secref{NNLO}.

The outline of the rest of this paper is as follows:
\Secref{methodology} explains the methodology used to compute raw lattice
results (at fixed $a$ and fixed quark mass). In \secref{simulations}, we describe
the details of our simulations.  We present a first look at the raw data in \secref{first-results}.
Taste violations are discussed in \secref{taste-violations},
followed in \secref{schpt} by a detailed description of our \schpt\ fitting forms.  Relevant
results from weak coupling perturbation  theory are collected in \secref{perturbation-theory}.
At the current level of precision, electromagnetic and isospin-violating effects cannot
be ignored, and we discuss the necessary corrections and the attendant systematic errors in
\secref{EM}. \Secref{fits} then presents the \schpt\ fits, including a description of fit
ranges (in quark mass), an inventory of all fit parameters, the resulting fits, and
a discussion of various issues relevant to the extraction of physical results.  The discussion
includes details of the continuum extrapolation, the evidence for chiral logarithms,
an estimate of the systematic errors associated with using a (slightly) mass-dependent
renormalization scheme, a critical look at the applicability and convergence
of the chiral perturbation theory on our data set, 
bounds on residual finite
volume effects, 
and some comments relevant to the fourth-root trick.
In \secref{results}, we present our final results, tabulate the systematic errors,
and discuss prospects for improving the current determinations.

In collaboration with
the HPQCD and UKQCD groups, we have previously reported results for
$m_s^{\msbar}$, the average $u$-$d$ quark mass $\hat m^\msbar$,  and $m_s/\hat m$ \cite{strange-mass}.
The data sets and chiral fits described in detail here are the same ones that were used
in Ref.~\cite{strange-mass}.

\section{METHODOLOGY}
\label{sec:methodology}

For the axial current corresponding to the unbroken (except by quark mass) axial
symmetry, the decay constant $f_{PS}$ can be found from the matrix element
of $\bar \psi \gamma_5 \psi$ between the vacuum and the pseudoscalar meson.
In terms of the one component staggered fermion field  $\bar \psi \gamma_5 \psi$ corresponds
to the operator
\begin{equation} {\cal O}_P (t) = \bar\chi^a(\vec x,t) (-1)^{\vec x + t} \chi^a(\vec x,t) \ .
\end{equation}
Here $a$ is a summed color index.
The relevant matrix element can be obtained from a pseudoscalar propagator
using ${\cal O}_P$ as both the source and sink operator: 
\begin{equation} P_{PP}(t) = \frac{1}{V_s}\sum_{\vec y} \langle {\cal O}_P(\vec x,0) {\cal O}_P(\vec y,t) \rangle
= C_{PP} e^{-m_{PS} t}
+ {\rm excited\ state\ contributions}\ \  , \end{equation}
where $m_{PS}$ is the mass of the pseudoscalar, and $V_s$ is the spatial volume.

The decay constant is obtained from $C_{PP}$ by \cite{JLQCD_FPI,TOOLKIT}
\begin{equation}\label{fpi_eq}
 f_{PS} = (m_x+m_y) \sqrt{\frac{V_s}{4} } \sqrt{\frac{C_{PP}}{m_{PS}^3}} \ ,
 \end{equation}
where $m_x$ and $m_y$ are the two valence quark masses in the pseudoscalar meson.
Throughout this paper we use the convention where the experimental value of $f_\pi$ is 
approximately $131\;\MeV$.
Note that in computing this meson propagator we
must take care to normalize  the lattice Dirac matrix as $M=am+\Dslash$.
The four in the denominator arises from the number of tastes natural to the Kogut-Susskind
formulation. (See unnumbered equations between Eqs. 7.2 and 7.3 in Ref.~\cite{TOOLKIT}.)\footnote{We 
thank C.\ Davies, G.\ P.\ Lepage, J.\ Shigemitsu and M.\ Wingate
for help in getting this normalization correct.}

However, the point operator ${\cal O}_P$
has large overlap with excited states.   For calculating masses it is customary to use an extended
source operator that suppresses these overlaps, together with a point sink.  In our case, this
extended operator is a ``Coulomb wall,'' \ie we fix to the lattice Coulomb gauge and sum over
all lattice points on a timeslice:
\begin{equation} {\cal O}_W (t) = \sum_{\vec x, \vec y} \bar\chi(\vec x,t)(-1)^{\vec x + t} \chi(\vec y,t)\ \ \ .\end{equation}
We can calculate propagators with any source or sink operator we wish. Ignoring excited
state contributions, we have for example
\begin{equation}  \langle {\cal O}_P(\vec x,0) {\cal O}_W(t) \rangle  = C_{PW}  e^{-m_\pi t} \ \ \ .\end{equation}
We will use the shorthand ``PP'' for point-source point-sink propagators, 
``WP'' for Coulomb-wall-source point-sink propagators, ``PW'' for
point-source Coulomb-wall-sink propagators, and ``WW'' for
Coulomb-wall source and sink propagators.
In previous calculations of pseudoscalar decay constants the relation
$ C_{PP} = C_{WP}^2 /C_{WW}$ has often been used to get the point-point
amplitude.  However, the wall-wall propagator has large statistical fluctuations
and severe problems with excited states, as was discussed in Ref.~\cite{JLQCD_FPI}.
To be able to use the PP operator to get $C_{PP}$ directly, rather
than indirectly by way of the ratio formula, one needs much better
statistics.  We do this by replacing
the point source with a ``random-wall'' source, which
simulates many point sources.  We set the source on each site 
of a time slice to a three component
complex unit vector with a random direction in color space,
and use this as the source for a conjugate gradient inversion 
to compute the quark propagator,
whose magnitude is squared to produce the Goldstone pion propagator.
Thus, contributions to a meson propagator where 
the quark and antiquark originate
on different spatial sites will average to zero and, after dividing by the spatial
lattice volume, this source can be used instead of ${\cal O}_P$.

Figures~\ref{masses_fig} and \ref{amps_fig} show masses and amplitudes from
pion propagators with random-wall and Coulomb-wall sources and sinks.
In Fig.~\ref{masses_fig}, we can see that extraction of masses from the ``WW''
propagators is almost hopeless.  Including an excited state helps, but statistical
errors become very large.  In Fig.~\ref{amps_fig}, the WW amplitudes are also
slower to plateau, though not as bad as the masses.  As a consistency check, note
that the WP and PW amplitudes are equal, and the masses extracted from the
diagonal PP and WW propagators approach their value from above (since excited
states must contribute to these propagators with the same sign as the ground state).
As an additional illustration of the difficulties with using the
Coulomb-wall---Coulomb-wall propagator, Fig.~\ref{prop_ratio_fig} plots the
ratio of the point-point pion propagator (using the random wall source) to the alternative
$P_{PW} P_{WP} /P_{WW}$ (with a different mass than in Figs.~\ref{masses_fig} and 
\ref{amps_fig}).  While this ratio is approaching one, it is clear that we would
either need very large minimum time in the fit or a careful removal of excited
states to use the ``WW'' propagators.

Given the problems with the WW propagators, we have opted to use only the
Coulomb-wall---point-sink and random-wall---point-sink propagators.
We performed a simultaneous fit to these two propagators, with an amplitude
for each propagator and a common mass.  In these fits the WP propagator
dominates the determination of the mass; while the amplitude of the PP
propagator is required for computing the decay constant.
Since the combination $C_{PP}/m_\pi^3$ is needed for determining $f_\pi$
and the mass and amplitude in a fit to a meson propagator are strongly
correlated, we used this combination as one of our fitting parameters.
That is, we fit the point---point and wall---point meson correlators to
\begin{eqnarray} \label{FITFORM_EQ}
     P_{PP} &=& m_\pi^3 \, A_{PP} e^{-m_\pi t} \EL
     P_{WP} &=& m_\pi^3 \, A_{WP} e^{-m_\pi t} \end{eqnarray}
so that $A_{PP}$ is the desired combination $C_{PP}/m_\pi^3$.
Since the correlation between $m_\pi$ and the propagator amplitude is
positive, the statistical error on the quantity $C_{PP}/m_\pi^3$ is
somewhat smaller than a naive combination of the errors on $C_{PP}$
and $m_\pi$.

\section{SIMULATIONS}
\label{sec:simulations}

These calculations were made on lattices generated with a one loop
Symanzik and tadpole improved gauge action\cite{SYM_GAUGE,LEPAGE_MACKENZIE} and an
order $a^2$ tadpole improved Kogut-Susskind quark action \cite{IMP_ACTION}.
Parameters of most of the lattices, as well as the light hadron spectrum,
are in Ref.~\cite{MILC_SPECTRUM,MILC_SPECTRUM2}.  The determination of the static quark
potential, used here to set the lattice spacing, is presented in
Ref.~\cite{MILC_SPECTRUM,MILC_SPECTRUM2,MILC_POTENTIAL}.   
In addition to the runs tabulated in
Ref.~\cite{MILC_SPECTRUM}, we now have a 
partially completed run with $a\hat m'=0.005$ and $am'_s=0.05$.  
(Here and below, the primes on masses indicate that they
are the dynamical quark masses used in the simulations, not the physical
masses $\hat m$ and $m_s$.)
In addition,
we have results from two runs at a finer lattice spacing, $a \approx 0.09$ fm,
with quark masses of $a\hat m',am'_s = 0.0124,0.031$ and $0.0062,0.031$.  These
runs, with $\hat m'=0.4 m'_s$ and $0.2 m'_s$, are analogous to the coarse lattice
runs with $a\hat m',am'_s = 0.02,0.05$ and $0.01,0.05$ respectively.
All of these lattices have a spatial size of about 2.5 fm with the exception
of the $a\hat m',am'_s = 0.005,0.05$ run, where the spatial size is about 3.0 fm.
Table~\ref{RUNTABLE} lists the parameters of the runs used here.

We note here that the values for $am'_s$ were
approximately tuned from the vector to pseudoscalar meson mass
ratio in initial runs with fairly heavy quarks.  Our best determinations of
the physical strange quark mass at these lattice spacings turned out to be
lower by 8 to 22\% (coarse) and 6 to 12\% (fine) than the nominal values $m'_s$,
where the range depends on whether or not taste-violating terms (as determined
by \schpt\ fits) are set to zero before demanding that $m_\pi$ and $m_K$ take
their physical values on a given lattice.

\begin{table}[t]
\begin{center}
\setlength{\tabcolsep}{1.5mm}
\begin{tabular}{|l|l|l|l|l|l|l|l|}
\hline
$a\hat m'$ / $am'_s$  & \hspace{-1.0mm}$10/g^2$ &dims. & lats. & $am_\pi$ & $am_K$ & $af_\pi$ & $af_K$  \\
\hline
0.03  / 0.05   & 6.81 &$20^3\times64$ & 262 & 0.37787(18) & 0.43613(19) &       0.11452(31) & 0.12082(31) \\
0.02  / 0.05   & 6.79 &$20^3\times64$ & 485 & 0.31125(16) & 0.40984(21) &       0.10703(18) & 0.11700(21) \\
0.01  / 0.05   & 6.76 &$20^3\times64$ & 608 & 0.22447(17) & 0.38331(24) &       0.09805(14) & 0.11281(17) \\
0.007  / 0.05   & 6.76 &$20^3\times64$ & 447 & 0.18891(20) & 0.37284(27) &      0.09364(20) & 0.11010(28) \\
0.005  / 0.05   & 6.76 &$24^3\times64$ & 137 & 0.15971(20) & 0.36530(29) &      0.09054(33) & 0.10697(40) \\
\hline
0.0124  / 0.031   & 7.11 &$28^3\times96$ & 531 & 0.20635(18) & 0.27217(21) &    0.07218(16) & 0.07855(17) \\
0.0062  / 0.031   & 7.09 &$28^3\times96$ & 583 & 0.14789(18) & 0.25318(19) &    0.06575(13) & 0.07514(17) \\
\hline
\end{tabular}
\caption{Parameters of the simulations in units of the lattice
spacing.  The first four columns are the dynamical quark masses $a\hat m'/am'_s$, the
gauge coupling $10/g^2$, the lattice dimensions, and the number of 
configurations used in these calculations.
The remaining four columns are the ``diagonal'' pseudoscalar masses
and amplitudes, with valence quark masses equal to the sea quark
masses.
The masses shown here come from a separate spectrum calculation, using more
source time slices than were
used in the partially quenched calculations, and using more lattices at $a\hat m'=0.03$.
Equation \protect{\ref{eq:R1_FIT_EQ}} can be used to express these masses
and decay constants in units of $r_1$.
\label{RUNTABLE}
}
\end{center}
\end{table}

Pseudoscalar propagators were calculated on lattices separated by six units
of simulation time, using two source time slices per lattice.  For the
coarse lattices, nine valence quark masses were used, ranging from $0.1 m'_s$
to $m'_s$; while for the fine lattices eight masses ranging from $0.14 m'_s$
to $m'_s$ were used. 
In all but one of the runs, the source slices were taken at different points in
successive lattices, which leads to smaller autocorrelations than using the
same source time slices on all lattices.
The effects of the remaining correlations among the sample lattices were estimated
in two ways.   First, jackknife error estimates for the masses and decay constants
were made eliminating one lattice at a time, and again eliminating four successive 
lattices.  Secondly, an integrated autocorrelation time was estimated by
summing the autocorrelations of the single elimination jackknife results
over separations from one to five samples (six to thirty simulation time units)
$\tau_{int} = \sum_1^5 2 C_i$, where $C_i$ is the normalized autocorrelation of
jackknife results omitting lattices separated by $6i$ time units.  The error
estimate including the effects of autocorrelations is a factor of $\sqrt{1+\tau_{int}}$
larger than the error from the single elimination jackknife fit.  Table \ref{AUTOCOR_TABLE}
summarizes the results of these tests.
The numbers in Table  \ref{AUTOCOR_TABLE} vary a lot, consistent with the well known
difficulties in measuring autocorrelations on all but the longest runs.  Since
we actually expect the autocorrelations to be smooth functions of the quark mass,
we account for them by increasing all the elements of the covariance matrix
by an approximate average of these factors squared, $(1.10)^2$, which is equivalent 
to increasing error estimates by a factor of 1.10.

\begin{table}[t]
\begin{center}
\setlength{\tabcolsep}{1.5mm}
\begin{tabular}{|l|l|l|l|l|l|l|}
\hline
$a\hat m'$ / $am'_s$  & \hspace{-1.0mm}$10/g^2$ & $\frac{\Delta m^2(4)}{\Delta m^2(1)}$ &
$\frac{\Delta f(4)}{\Delta f(1)}$ & $\tau_{int,m}$ & $\tau_{int,f}$ \\
\hline
0.03  / 0.05   & 6.81 & 1.10 & 1.16 & 0.25 & 0.15  \\
0.02  / 0.05   & 6.79 & 1.07 & 1.00 & 0.01 & -0.09  \\
0.01  / 0.05   & 6.76 & 1.28 & 1.12 & 0.30 & 0.27  \\
0.007  / 0.05   & 6.76 & 1.05 & 0.90 & -0.02 & -0.03  \\
0.005  / 0.05   & 6.76 & 1.06 & 1.20 & -0.04 & -0.04  \\
\hline
0.00124  / 0.031   & 7.11 & 1.10 & 1.13 & 0.25 & 0.15  \\
0.00062  / 0.031   & 7.09 & 1.10 & 0.95 & 0.22 & -0.01  \\
\hline
\end{tabular}
\caption{Estimates of the effects of autocorrelations.
$\Delta m^2(4)/\Delta m^2(1) $ is the ratio of error estimates for the
squared pion mass between jackknife estimates with a block size of four
and a block size of one.  $\Delta f^2(4)/\Delta f^2(1) $ is the same thing
for the decay constant.
$\tau_{int,m}$ and $\tau_{int,f}$ are the integrated autocorrelation times
for the squared pion mass and decay constant.
All of these numbers are averaged over the valence quark masses.
\label{AUTOCOR_TABLE}
}
\end{center}
\end{table}

Propagators were fit to Eq.~\ref{FITFORM_EQ} using a minimum time distance of
$20a$ for the coarse lattices and $30a$ for the fine lattices.  At these distances,
the contamination from excited states is at most comparable to the statistical errors.
For example, Fig.~\ref{dmin-fine} shows results for pion masses and amplitudes as a function
of minimum fitting distance for one of the fine runs. Since our other systematic errors
are significantly larger than statistical errors (see \secref{results}), we can neglect 
the systematic effect due to excited states.

For each run, the propagator fitting produced a pion mass and decay constant
for each combination of valence quark masses. 
We call the two valence quarks in a particular meson $x$ and $y$;
there are 
45 different combinations of $m_x$,$m_y$ for the coarse
lattices and 36 for the fine, although, as described in \secref{mass-subsets},
the largest valence quark masses were not used in all of the fits.   
All of the masses and decay amplitudes from a single run are correlated.
For each run with $N$ samples,
a covariance matrix describing the fluctuations of all of these
numbers was made by doing a single elimination jackknife fit, omitting one
lattice at a time, and rescaling the covariance matrix of the jackknife fits
by $(N-1)^2$.
A single elimination jackknife, rather than one where larger blocks were
omitted, was used because getting a reliable covariance matrix requires
a number of samples large compared to the dimension of the matrix.
Then, to account for autocorrelations, this covariance matrix was rescaled by
the factor estimated above.
Finally,  to allow simultaneous fitting of the meson decay constants and
masses from all of the runs as a function of valence and sea quark
masses, the covariance
matrices from the individual runs were combined into a large block-diagonal
covariance matrix.  (Runs with different sea quark masses or gauge couplings
are independent, so correlations between different runs can be set to zero.)

Fitting the pseudoscalar propagators produces masses and decay constants
in units of the lattice spacing $a$, and to convert to physical units
we must estimate $a$ from a calculation of some dimensional quantity
whose value is known.   This amounts to saying that we are calculating
ratios of these quantities to some other quantity calculated from these
simulations.  We express our results in units of a length obtained
from the static quark potential, $r_1$, where
$r_1^2\, F(r_1) = 1.0$ \cite{SOMMER,MILC_POTENTIAL}.  This has the advantage
that $r_1$ can be accurately determined in units of the lattice spacing.
But $r_1$ is not a directly measurable quantity, and its physical value must
in turn be obtained from some other quantities.
We have calculated the static quark potential in all of these runs, and fit
it to determine $r_1/a$.  To smooth out statistical fluctuations in these
values, we then computed a ``smoothed $r_1$'' by fitting the $r_1/a$ values
to a smooth function.   A simple form, which gives a good fit
over the range of quark masses and gauge coupling used here, is \cite{MILC_SPECTRUM2}
\begin{equation}\label{eq:R1_FIT_EQ}
 \log(r_1/a) = C_{00} + C_{10} (10/g^2 - 7) + C_{01} am_{tot} +
 C_{20} (10/g^2 - 7)^2\ ,\end{equation}
where $m_{tot} = 2 \hat m'+ m'_{s}$.
The results of the fit are
\begin{eqnarray}\label{eq:R1_RESULTS}
C_{00}= 1.258(3) && C_{10}=0.937(9) \nonumber \\
C_{01}= -0.83(3) && C_{20}=-0.27(2)
\end{eqnarray}

When we need an absolute lattice scale, we start with the
scale from $\Upsilon$ $2S$-$1S$ or $1P$-$1S$ splittings,
determined by the HPQCD group \cite{BIG_PRL,HPQCD_PRIVATE}.
This gives a scale $a^{-1}=1.588(19)$ GeV on the coarse $0.01/0.05$ lattices,
and $a^{-1}=2.271(28)$ GeV on the fine $0.0062/0.031$ lattices.
For light quark masses $\ltwid m_s/2$, the mass dependence of these quantities
and of $r_1$ appears to be slight, and we neglect it.  With our smoothed values
of $r_1/a$, we then get $r_1= 0.324(4)$ fm on the coarse lattices and
$r_1= 0.320(4)$ fm on the fine lattices.

To extrapolate $r_1$ to the continuum, we first
assume that the dominant discretization errors go like $\alpha_S a^2$.  Using
$\alpha_V(q^*)$ \cite{Davies:2002mv} (with scale $q^*=\pi/a$) for $\alpha_S$
gives a ratio $(\alpha_S a^2)_{\rm fine}/(\alpha_S a^2)_{\rm coarse}=0.427$.
Extrapolating away the discretization errors linearly then results in
$r_1=0.317(7)$ fm in the continuum. 
However, taste-violating effects,
while formally $\cO(\alpha^2_S a^2)$ and hence subleading, are known to
be at least as important as the leading errors in some case.  Therefore, one should
check if the result changes when the errors are assumed to go like $\alpha^2_S a^2$.
Taking $\alpha_S=\alpha_V(3.33/a)$ gives a ratio
$(\alpha^2_S a^2)_{\rm fine}/(\alpha^2_S a^2)_{\rm coarse}=0.375$; while
a direct lattice measurement of the taste-splittings gives a ratio of $0.35$.
Extrapolating linearly to the continuum then implies
$r_1=0.318(7)$ fm or $r_1=0.319(6)$ fm, respectively, in agreement with the
previous result.  For our final result, we use an ``average'' ratio of 0.4 and
add the effect of varying this ratio in quadrature with the statistical error.
We obtain $r_1=0.317(7)$ fm.
A systematic error of $0.03$ fm in $r_1$ from
our choice of fitting methods is omitted since it is common to all
our runs and cancels out in the final results here. 
Using our current value $r_0/r_1= 1.472(7)$,
the result for $r_1$ implies $r_0$ is about 7\% smaller than the standard phenomenological
choice $r_0=0.5\;$fm, although the difference is within the expected 
range of error of the phenomenological
estimates \cite{SOMMER}.

\section{First Look at Results}
\label{sec:first-results}

Figures~\ref{mpisq_coarse_fig} and \ref{fpi_coarse_fig} present
pseudoscalar masses and decay constants in units of $r_1$
as functions of the valence
quark masses for several different light quark masses.  All of these
points are from the lattices with $a \approx 0.125$ fm.
Figure~\ref{mpisq_coarse_fig} also contains pion masses where
the sea quark mass varies along with the valence quark masses.

Figures~\ref{mpisq_extrap_fig} and \ref{fpi_extrap_fig} show
the effect of changing the lattice spacing.  For lattice spacings
$a \approx 0.125$ fm and $a \approx 0.09$ fm we show results with
$\hat m'=0.4m'_s$ and $\hat m'=0.2m'_s$, again in units of $r_1$.  The horizontal axis
is again the sum of the valence quark masses in the meson.
These figures also show a crude extrapolation to $a=0$, made by
taking a linear extrapolation in $\alpha_S a^2 $ using pairs of points
with the same $\hat m'/m'_s$.   In Fig.~\ref{mpisq_extrap_fig}
one pair of extrapolated points has diagonal lines showing the
data points that were extrapolated to produce this point.
In hindsight, $m'_s$ used in the $a\approx 0.09$ fm
runs was smaller than that used in the $a \approx 0.125$ fm runs, as
indicated by the fact that the finer lattice points fall slightly to
the left of the corresponding coarse lattice points.

\section{Taste symmetry violations}
\label{sec:taste-violations}

As mentioned above, we use the term ``taste'' to denote the different staggered-fermion species
resulting from doubling.
At finite lattice spacing, taste
symmetry is violated.  Although the improved staggered action
reduces the taste violating effects to $\cO(\alpha_S^2 a^2)$ from $\cO(\alpha_S a^2)$ 
with unimproved staggered fermions, the violations are still quite significant
numerically.  

Figure~\ref{fig:splittings} shows the splittings between pions of various tastes
on our coarse lattices. There are 16 such pions, $\pi_B$, where $B={5,\mu5,\mu\nu,\mu,I}$ 
($\nu >\mu$) labels taste matrices in the taste Clifford algebra generated by Euclidean 
gamma matrices $\xi_\mu$. The $\pi_5$ is the Goldstone (pseudoscalar taste) pion, whose mass is required to
vanish in the chiral limit by the exact (non-singlet) lattice axial symmetry.
All the pions in \figref{splittings} are flavor-{\it charged},
\ie $\pi^+$ mesons.  Thus there are no contributions
from disconnected graphs, even for the taste-singlet $\pi^+_I$.  The
approximate ``accidental'' $SO(4)$ identified by Lee and Sharpe \cite{LEE_SHARPE} is
clearly a good symmetry:  there is near degeneracy between $\pi^+_{05}$ and 
$\pi^+_{i5}$, between $\pi^+_{0i}$ and $\pi^+_{ij}$, and between $\pi^+_{0}$ and
$\pi^+_{i}$. When we assume such degeneracy, we can think of the index $B$ as running
over the multiplets $5,A,T,V,I$ with degeneracies 1, 4, 6, 4, 1, respectively.

The fit in \figref{splittings} is to the tree-level chiral form given in
Refs.~\cite{LEE_SHARPE,CA_CB1}: 
\begin{equation}\label{eq:tree-masses}
        m^2_{\pi^+_B} = 2\mu_{\rm tree} \hat m  + a^2\Delta_B \ .
\end{equation}
The slope, $\mu_{\rm tree}$, is the same for all tastes, but there are constant
splittings for each non-Goldstone multiplet ($\Delta_5=0$).
Although the fit is poor (chiral logs, including taste-violations, are needed), 
it does give the pion squared masses
within a few per cent: The biggest deviation, 7\%,  is for the Goldstone pion at the 
lowest mass; most other deviations are $\sim\!2\%$. 

\tabref{splittings} shows the values of $a^2\Delta_B$ coming from the fit on the coarse
lattices.  On the fine lattices, we have measured non-Goldstone
pion masses only on the set with quark masses $0.0124$, $0.031$. So we directly
compare the splittings with those of the corresponding coarse lattice (masses $0.02$, $0.05$).
The fine-lattice splittings are smaller by a common factor of $0.35$,
within errors.
This is consistent with the expectation that taste violations go like
$\cO(\alpha^2_S a^2)$. Indeed, if we take $\alpha_S=\alpha_V(q^*)$ \cite{Davies:2002mv} and choose
$q^*=\pi/a$ because taste violations occur at the scale of the cutoff, we find
\begin{equation}\label{eq:alpha2-a2}
\frac{(\alpha^2_V(q^*\!\!=\!\pi/a)\; a^2)_{\rm fine}}{(\alpha^2_V(q^*\!\!=\!\pi/a)\; a^2)_{\rm coarse}} 
= 0.372 \ .
\end{equation}

\begin{table}[t]
\begin{center}
\setlength{\tabcolsep}{1.5mm}
\def\arraystretch{.8}
\begin{tabular}{|c|c|c|}
\noalign{\vspace{0.5cm}}
\hline
taste ($B$)  & $r_1^2 (a^2 \Delta_B)_{\rm coarse}$ & $\frac{(a^2 \Delta_B)_{\rm fine}}{(a^2 \Delta_B)_{\rm coarse}}$   \\
\hline
  A & 0.205(2)    &  0.344(23)        \\
  T & 0.327(4)    &  0.353(18)        \\
  V & 0.439(5)    &  0.347(22)        \\
  I & 0.537(15)   &  0.384(33)        \\
\hline
\end{tabular}
\caption{Mass-squared splittings in units of $r_1$ for the coarse lattices, and
the ratio of fine to coarse splittings. Results from  tastes that are degenerate under
the accidental $SO(4)$ have been combined.
\label{tab:splittings}
}
\end{center}
\end{table}

The ratio of taste-violating terms between fine and coarse lattices is an input to the chiral
fits for Goldstone pions discussed below.  
The measured splitting ratio of $0.35$ is used as a central value.  The
error can be estimated by varying $q^*$ in \eq{alpha2-a2}: $q^*=\pi/(2a)$ gives a ratio of
$0.324$; while $2\pi/a$ gives $0.398$. 
We take $0.3$--$0.4$  as an appropriate range for our analysis
of systematics.

We warn the reader here that the notation in \eq{tree-masses} can be slightly misleading.
We have shown explicitly the $a^2$ factor in the taste-violating splitting, $a^2\Delta_B$,
but this does not mean that $\Delta_B$ itself is independent of lattice spacing, or even that it
approaches a non-zero constant in the continuum limit. Indeed,
the argument above implies that $\Delta_B$ is a slowly varying function of $a$ that
goes like $\alpha_V^2(\pi/a)$ for small $a$.  A similar comment applies
to the other taste-violating parameters introduced in \secref{NLO}:  the $a^2$ dependence is always
shown explicitly, but dependence on $a$ through the coupling is hidden.

In physical units, the splittings on the coarse lattices are quite large.  The largest
is for the taste-singlet pion:
$a^2\Delta_I\approx(450\;\MeV)^2$; while the smallest, for the taste axial-vector pion, is
$a^2\Delta_A\approx(280\;\MeV)^2$.
Given the size of these splittings, which are discretization errors, 
it is not surprising that the lattice data is not well fit by
continuum chiral perturbation theory (\chpt) forms.
Figure \ref{fig:fpi-vs-m-no-taste-viols} shows such an attempted fit for the Goldstone
$f_\pi$ to the standard NLO partially quenched
continuum form \cite{SHARPE_SHORESH} plus analytic NNLO terms.  More details about this
fit will be explained below, when we discuss the corresponding fits that take into
account taste violations.  For the moment, we simply remark that the minuscule confidence
level (${\rm CL}\approx 10^{-250}$; $\chi^2/{\rm d.o.f.}= 8.77$ with 204 degrees of freedom) shows how hard it is to ignore lattice artifacts at the level of chiral
logarithms.

\section{Staggered chiral perturbation theory} 
\label{sec:schpt}

Lee and Sharpe \cite{LEE_SHARPE} found the chiral Lagrangian
that describes a single staggered field.
Their Lagrangian includes the effects of
taste violations at $\cO(a^2)$ as well as the standard violations of
chiral symmetry from mass terms at $\cO(m_q)$, where $m_q$ is
a generic quark mass. They introduced a power counting that considers
$m_q$ and $a^2$ to be of the same
order, which is appropriate here: In Fig.~\ref{fig:splittings}
the splittings are comparable to the squared meson masses. Tree level
(LO) is thus  $\cO(m_q,a^2)$; chiral logs appear at one-loop (NLO) and are
$\cO(m_q^2,m_qa^2,a^4).\;$\footnote{Throughout this paper, we define the order of
a contribution to be the order of the corresponding term in the chiral Lagrangian.
This is the simplest way to keep the power counting consistent between
decay constants and meson masses, although it does
lead to the unnatural statement that the tree-level $f_\pi$ is ``$\cO(m_q)$''
since it comes from the kinetic energy term in the chiral Lagrangian.
What matters ultimately is only the relative size of contributions:
the first correction to the tree-level
value of $m^2_\pi$ or $f_\pi$ is smaller by one power of $m_q$.}

The Lee-Sharpe Lagrangian is not directly appropriate to the calculations
here because it has only one flavor (one staggered field).
Aubin and Bernard \cite{CB_FSB,CA_CB1,CA_CB2} 
have generalized Ref.~\cite{LEE_SHARPE}
to $n$ staggered flavors and shown how to accommodate the
$\root 4 \of {\rm Det}$ trick in loop calculations.  This is what is meant
by ``staggered chiral perturbation theory,'' \schpt.

Continuum chiral perturbation theory can be thought of as an expansion in the dimensionless
quantity 
\begin{equation}\label{eq:chiqdef}
\chi_q \equiv \frac{2\mu m_q}{8\pi^2 f_\pi^2} \ .
\end{equation}
where ${2\mu m_q}$ is the tree-level mass of a $q\bar q$ meson.
For physical kaons, we expect the relevant quark mass parameter to be 
$\chi_{ud,s}\equiv (\chi_{ud}+\chi_s)/2\approx 0.18$ (where $\chi_{ud}$  is the average
value for the $u$ and $d$ quarks); this is reasonable
given the experimental result $f_K/f_\pi\approx 1.22$. 

Staggered chiral perturbation theory is
a joint expansion in $\chi_q$ and $\chi_{a^2}$, which measures the size of the $\cO(a^2)$ taste violations: 
\begin{equation}\label{eq:chia2def}
\chi_{a^2} \equiv \frac{a^2\overline{\Delta}}{8\pi^2 f_\pi^2} \ ,
\end{equation}
where $a^2\overline{\Delta}$ is a ``typical'' taste-violating term. Taking for $a^2\overline{\Delta}$
the average meson splitting (see \eq{delta-av} below),  we have
$a^2\overline{\Delta}\approx (350\; \MeV)^2$ 
and $\chi_{a^2}\approx 0.09$ on the coarse lattices; $a^2\overline{\Delta}\approx (200\; \MeV)^2$
and $\chi_{a^2}\approx 0.03$ on the fine lattices.  If one instead uses the larger of the $\cO(a^2)$ 
taste-violating hairpin parameters \cite{CA_CB1,CA_CB2}, $a^2\delta'_A$,
to estimate $\overline{\Delta}$ and $\chi_{a^2}$, one gets slightly smaller values.

\subsection{NLO forms}
\label{sec:NLO}

One-loop chiral logs and analytic terms have
been calculated in \schpt\ for Goldstone meson masses \cite{CA_CB1} and decay
constants \cite{CA_CB2}.  
Partially quenched results are included, so all forms needed to
fit the numerical data are available.

References \cite{CA_CB1,CA_CB2} express the chiral logarithms in terms
of
\begin{eqnarray}\label{eq:chiral-log1}
        \ell( m^2)& \equiv & m^2 \left(\ln \frac{m^2}{\Lambda_\chi^2}
         + \delta_1(mL)\right) 
        \\*
        \label{eq:chiral-log2}
         \tilde \ell(m^2)& \equiv & -\left(\ln
        \frac{m^2}{\Lambda_\chi^2} + 1\right) + \delta_3(mL) \ ,
\end{eqnarray}
where $\Lambda_\chi$ is the chiral scale, and $L$ is the spatial dimension.
The finite volume correction terms $\delta_1$ and $\delta_3$ are 
\cite{CB_FSB} 
\begin{eqnarray}\label{eq:delta1}
        \delta_1(mL) & = & 4
                \sum_{\vec r\ne 0}
                \frac{K_1(|\vec r|mL)}{mL|\vec r|} \ , \\*
        \label{eq:delta3}
        \delta_3(mL) & =& 2 \sum_{\vec r\ne 0}
                K_0(|\vec r|mL)\ ,
\end{eqnarray}
where $K_0$ and $K_1$ are Bessel functions of imaginary argument, and $\vec  r$,
which labels the various
periodic images, is a three-dimensional vector with integer components.
We have assumed here that corrections due to the finite time extent 
are negligible; this is true for our lattices, for which the
time dimension is between $2.7$ and $3.4$ times greater than the spatial dimension.
The function $\ell(m^2)$ in \eq{chiral-log1} arises from tadpole diagrams with a single meson
propagator; $\tilde \ell(m^2)$ in \eq{chiral-log2} comes from double-poles, which are present only in the
partially quenched (and quenched) cases, not in the full QCD limit.  In practice, we compute
the sum in \eq{delta1} or \eq{delta3} with cutoff  $|\vec r|\le N$, where $N$ is an integer, and
increment $N$ by 1 until the sum changes by a fractional amount $\le \epsilon$.  To be
conservative, we take $\epsilon=10^{-9}$ for central-value fits.  However, a much
weaker criterion, $\epsilon=0.001$, is adequate to reduce the error in the sum
well below our statistical errors, and we often use the weaker criterion for alternative
fits in the systematic error estimates.

In the generic case relevant to our data ($m_u=m_d\equiv \hat m\not=m_s$ and no degeneracies between
valence and sea quarks), the NLO \schpt\ expressions  for a meson $P$ composed of valence
quarks $x$ and $y$ are \cite{CA_CB1,CA_CB2}
\begin{eqnarray}\label{eq:m-NLO}
        \frac{(m^{\rm NLO}_{P^+_5})^2}
        {\left( m_x+m_y \right)}&=& \mu \Biggl\{1 +
        \frac{1}{16\pi^2f^2}\Bigg(
        \frac{2}{3} \sum_{j} R^{[3,2]}_{j}(\{\cM^{[3]}_{XY_I}\})\;
         \ell(m^2_{j}) \nonumber \\*
        &&\hspace{-0.6truein} -2a^2\delta'_V \sum_{j} R^{[4,2]}_{j}(\{\cM^{[4]}_{XY_V}\})\, \ell(m^2_{j})
        -2a^2\delta'_A \sum_{j} R^{[4,2]}_{j}(\{\cM^{[4]}_{XY_A}\})\, \ell(m^2_{j})+a^2(L''+L') \Bigg)
         \nonumber \\*
        && \hspace{-0.2truein} 
        +\frac{16\mu_{\rm tree}}{f^2}\left(2L_8-L_5\right)
        \left(m_x+m_y\right) 
        +\frac{32\mu_{\rm tree}}{f^2}\left(2L_6-L_4\right)
         \left(2\hat m+m_s \right) \Biggr\} \\*
\label{eq:f-NLO}
        f^{\rm NLO}_{P^+_5} & = & f\Biggl\{ 1
        + \frac{1}{16\pi^2 f^2}
         \Biggl[-\frac{1}{32}\sum_{Q,B} \ell\left(m^2_{Q_B}\right)
        + \frac{1}{6}\Biggl(R^{[2,2]}_{X_I}(\{\cM^{[2]}_{X_I}\})
        \tilde\ell(m^2_{X_I})
         \nonumber \\* &&
        +R^{[2,2]}_{Y_I}(\{\cM^{[2]}_{Y_I}\})\tilde\ell(m^2_{Y_I})
        +\sum_{j}
        D^{[2,2]}_{j,X_I}(\{\cM^{[2]}_{X_I}\})\ell(m^2_{j}) \nonumber \\* &&
        +\sum_{j}D^{[2,2]}_{j,Y_I}(\{\cM^{[2]}_{Y_I}\})\ell(m^2_{j})
        -2\sum_{j}R^{[3,2]}_{j}(\{\cM^{[3]}_{XY_I}\})\ell(m^2_{j})\Biggr)
         \nonumber \\* &&
        +\frac{1}{2}a^2 \delta'_V\Biggl(  R^{[3,2]}_{X_V}(\{\cM^{[3]}_{X_V}\})
        \tilde\ell(m^2_{X_V})
        +   R^{[3,2]}_{Y_V}(\{\cM^{[3]}_{Y_V}\})\tilde\ell(m^2_{Y_V})
          \nonumber \\* &&
        + \sum_{j}  D^{[3,2]}_{j,X_V}(\{\cM^{[3]}_{X_V}\})\ell(m^2_{j})
        +\sum_{j}D^{[3,2]}_{j,Y_V}(\{\cM^{[3]}_{Y_V}\})\ell(m^2_{j})
          \nonumber \\* &&
        +2\sum_{j}R^{[4,2]}_{j}
        (\{\cM^{[4]}_{XY_V}\})\ell(m^2_{j})
         \Biggr)        + \Bigl( V \to A \Bigr) + a^2(L''-L')\Biggr]
          \nonumber \\* &&
        + \frac{8\mu_{\rm tree}}{f^2}L_5\left( m_x + m_y \right)
        + \frac{16\mu_{\rm tree}}{f^2}L_4\left( 2\hat m + m_s\right)
        \Biggr\} \ .
\end{eqnarray}
Here $\mu$ and $f$ are the continuum chiral parameters, $\delta'_V$ and $\delta'_A$ are LO 
taste-violating parameters (hairpins), 
$L_i$ are the NLO Gasser-Leutwyler \cite{GASSER_LEUTWYLER}
coefficients, and 
$L'$ and $L''$ are linear
combinations  of the
taste-violating NLO coefficients. The reason for using the tree-level $\mu_{\rm tree}$ parameter
from \eq{tree-masses} in the $L_i$ terms will be explained in \secref{NNLO}.
$X_\Xi$ and $Y_\Xi$ are flavor-neutral mesons of taste $\Xi$ made of $x,\bar x$ and $y,\bar y$
quarks, respectively, and $U_\Xi$, $D_\Xi$, and $S_\Xi$ are corresponding flavor-neutral mesons made from $u$, $d$, and $s$ sea quarks, respectively. The index
$Q$ runs over the 6 mesons made from one valence and one sea quark, and
$B$ runs over the 16 meson tastes. 
The residues $R^{[n,k]}_j$ and $D^{[n,k]}_{j,i}$ in \eqs{m-NLO}{f-NLO} are defined
in Refs.~\cite{CA_CB1,CA_CB2}.  For completeness, we quote them here:
\begin{equation}\label{eq:residues}
        R_j^{[n,k]}\left(\left\{\cM\right\}\!;\!\left\{\mu\right\}\right)
         \equiv  \frac{\prod_{a=1}^k (\mu^2_a- m^2_j)}
        {\prod_{\ell=1}^{'n} (m^2_\ell - m^2_j)}\ .
\end{equation}
\begin{equation}\label{eq:residues2}
        D_{j, i}^{[n,k]}\left(\left\{\cM\right\}\!;
        \!\left\{\mu\right\}\right) \equiv -\frac{d}{d m^2_{i}}
        R_{j}^{[n,k]}\left(\left\{\cM\right\}\!;
        \!\left\{\mu\right\}\right)\ .
\end{equation}
Each of these residues is a function of two sets of masses, the ``denominator'' set
$\{\cM\}=\{m_1,m_2,\ldots, m_n\}$ and the ``numerator'' set $\{\mu\}=\{\mu_1,\mu_2,\ldots,\mu_k\}$.
The indices $j$ and $i$, $1\le j,i \le n$, refer to particular denominator masses; the
prime on the product in the denominator of \eq{residues} means that $\ell=j$ is omitted.

In \eqs{m-NLO}{f-NLO}, 
the denominator mass-set arguments are shown explicitly; they are  
\begin{eqnarray}\label{eq:denom-mass-sets}
        \{\cM^{[2]}_{X_I}\}& \equiv & \{ m_{X_I}, m_{\eta_I} \}\ , \nonumber \\*
        \{\cM^{[2]}_{Y_I}\}& \equiv & \{ m_{Y_I}, m_{\eta_I} \}\ , \nonumber \\*
        \{\cM^{[3]}_{XY_I}\}& \equiv & \{ m_{X_I}, m_{Y_I}, m_{\eta_I} \}\ , \nonumber \\*
        \{\cM^{[3]}_{X_V}\}& \equiv & \{ m_{X_V}, m_{\eta_V}, m_{\eta'_V} \}\ , \\
        \{\cM^{[3]}_{Y_V}\}& \equiv & \{ m_{Y_V}, m_{\eta_V}, m_{\eta'_V} \}\ , \nonumber \\*
        \{\cM^{[4]}_{XY_V}\}& \equiv & \{ m_{X_V}, m_{Y_V}, m_{\eta_V}, m_{\eta'_V} \}\ . \nonumber
\end{eqnarray}
The index $j$ in \eqs{m-NLO}{f-NLO} is summed over the denominator masses.
Sets for axial-vector taste ($A$) are found from the corresponding vector taste ($V$) sets by taking $V\to A$
in \eq{denom-mass-sets}.
The masses  $m_{\eta_I}$, $m_{\eta_V}$, $m_{\eta'_V}$ are given by \cite{CA_CB1}
\begin{eqnarray}\label{eq:eigenvalues}
	m_{\eta_I}^2 & = & \frac{m_{U_I}^2}{3}+
        \frac{2m_{S_I}^2}{3} \ , \nonumber \\*
        m_{\eta_V}^2 & = & \frac{1}{2}\left( m_{U_V}^2 + m_{S_V}^2 +
        \frac{3}{4}a^2\delta'_V - Z
        \right)\ , \nonumber \\*
        m_{\eta'_V}^2 & = &  \frac{1}{2}\left( m_{U_V}^2 + m_{S_V}^2 +
        \frac{3}{4}a^2\delta'_V +  Z \right) \ ; \\*
         Z & \equiv &\sqrt{\left(m_{S_V}^2-m_{U_V}^2\right)^2
         - \frac{a^2\delta'_V}{2}
        \left(m_{S_V}^2-m_{U_V}^2\right) +\frac{9(a^2\delta'_V)^2}{16}
         } \ . \nonumber
\end{eqnarray}
The numerator mass-set arguments of the residues
in \eqs{m-NLO}{f-NLO} are not shown explicitly because they are 
always \begin{equation}\label{eq:num_mass_sets}
        \{ \mu^{[2]}_\Xi \} \equiv \{ m_{U_\Xi},m_{S_\Xi}   \} \ ,
\end{equation}
where the taste label $\Xi$  is taken  equal to the taste of the denominator set.

Degeneracies among the various masses in
\eqs{m-NLO}{f-NLO} occur quite often in our data set.  In particular, ``partially quenched
pions'' have $m_x=m_y$  and hence $m_{X_B}=m_{Y_B}$ for each taste $B$.  Similarly ``partially quenched kaons''
have $m_y=m_s$ and hence $m_{Y_B}=m_{S_B}$.  Going to full QCD introduces additional
degeneracies $m_X=m_Y=m_U$ (for pions) or $m_X=m_U$ (for kaons). Further, the 
accidental degeneracy $m_{Y_I}=m_{\eta_I}$ appears in our data when $am_y=0.04$,
$a\hat m'=0.02$, $am'_s=0.05$ (coarse) or $am_y=0.0248$,
$a\hat m'=0.0124$, $am'_s=0.031$ (fine).  Formulas for many of these degenerate cases
appear in Refs.~\cite{CA_CB1,CA_CB2}. For the other cases, one can carefully take
appropriate limits in \eqs{m-NLO}{f-NLO}, or, more conveniently,
return to the original integrands in Refs.~\cite{CA_CB1,CA_CB2} 
and take the limits before performing the momentum integrations.  

Because \eqs{m-NLO}{f-NLO} are quite complicated, it is useful to write down
a simple result that shows more clearly how taste violations
change the continuum chiral behavior. The pion decay constant in full QCD with
two (degenerate) flavors ($m_x=m_y=m_u=m_d\equiv \hat m$,
with the strange quark integrated out) is particularly simple.
In that case, the result corresponding to \eq{f-NLO}  is
\begin{eqnarray}\label{eq:fpi-NF2}
        f^{\rm 1-loop}_{\pi^+_5} && = f\Biggl\{ 1 +
        \frac{1}{16\pi^2 f^2}
         \Biggl[-2\left(\frac{1}{16}\sum_{B}
          \ell(m^2_{\pi_B})\right)
         \nonumber \\* 
&& \hspace{0.6truein}-4\Biggl( 
        \ell(m^2_{\eta'_V})- \ell(m^2_{\pi_V}) \Biggr)
         - 4\Biggl( 
        \ell(m^2_{\eta'_A})- \ell(m^2_{\pi_A}) 
        \Biggr) + a^2(L''-L') \Biggr]\nonumber\\*
   &&\hspace{0.6truein} + \frac{8\mu_{\rm tree}}{f^2}L_5 (2\hat m) +
\frac{16\mu}{f^2}L_4 \left( 2\hat m\right)
    \Biggr\}  \ ,
\end{eqnarray}
with $B$ running as usual over the 16 possible tastes, and
\begin{eqnarray}\label{eq:eigenvalues-NF2}
        m_{\pi_B}^2  &\equiv& m_{U_B}^2 = m_{D_B}^2\ ,   \nonumber \\*
        m_{\eta'_V}^2 & = &  m_{\pi_V}^2 + \frac{1}{2}a^2\delta'_V\ ,  \nonumber \\*
        m_{\eta'_A}^2 & = &  m_{\pi_A}^2 + \frac{1}{2}a^2\delta'_A\  . 
\end{eqnarray}
In \eq{fpi-NF2}, the term multiplied by $-2$ gives the average of all
tastes and becomes the standard $SU(2)_L\times SU(2)_R$ chiral logarithm
in the continuum limit, when all tastes are degenerate. The terms multiplied
by $-4$ clearly vanish in the continuum limit 
because $m_{\pi_V}=m_{\eta'_V}$ and $m_{\pi_A}=m_{\eta'_A}$ 
when $a^2\delta'_V=0=a^2\delta'_A$.

Since $\hat m'$ is significantly less than $m'_s$ for many of our runs, 
\eq{fpi-NF2} is often not a bad approximation to the chiral behavior of our
(full QCD) data. It will be useful in the discussion of finite volume effects in
\secref{FINITE-VOLUME}.

We note here that Refs.~\cite{LEE_SHARPE,CA_CB1,CA_CB2} explicitly include in the
chiral Lagrangian the effects of
terms in the $\cO(a^2)$ staggered-quark Symanzik action that violate the taste symmetries. There are
also ``generic'' $\cO(a^2)$ terms in the Symanzik action that have the same 
symmetries as the continuum QCD action and are not included explicitly.  An example is
$a^2 \bar \psi D^2 D_\mu (\gamma_\mu \otimes I)\psi$, where $\gamma_\mu$ and $I$ act on spin and taste
indices, respectively.  The effect of such terms on the chiral Lagrangian is to produce
$\cO(a^2)$ variation in physical parameters such as $f$, $\mu$,
and, at higher order in $m_q$, $L_i$.  
We build the possibility of such generic variation in physical parameters into the chiral fits
below.\footnote{Before comparing its values for different 
lattice spacings, the parameter $\mu$
must be renormalized by the inverse of the mass renormalization constant. 
See \protect{\secref{perturbation-theory}}.}
Since our staggered action is $a^2$ tadpole improved, we expect such generic
variation to be of size $\alpha_S a^2 \Lambda_{\rm QCD}\approx 2\%$.   When we extrapolate the physical
parameters to the continuum, we will need to know how $\alpha_S a^2$ changes from the coarse
to fine lattice. As in the case of taste violations, such discretization errors occur at the scale of
the cutoff. Therefore, we use $\alpha_S=\alpha_V(q^*=\pi/a)$ for central values, and allow $q^*$
to vary between $\pi/(2a)$ and $2\pi/a$ for the error estimate.  We have
\begin{equation}\label{eq:alpha-a2}
\frac{(\alpha_V(q^*\!\!=\!\pi/a)\; a^2)_{\rm fine}}{(\alpha_V(q^*\!\!=\!\pi/a)\; a^2)_{\rm coarse}} 
= 0.427 \ ,
\end{equation}
and a range for this ratio of 0.398 to 0.441.

As they stand, \eqs{m-NLO}{f-NLO} are slightly inconvenient because the
renormalization of the $\cO(m_qa^2)$ analytic NLO parameters $L'$ and $L''$ under a change in
the chiral scale $\Lambda_\chi$ is complicated and involves the physical $L_i$ parameters.  
This is due to the fact that
the meson masses multiplying the logarithms include $\cO(a^2)$ splittings. It is more
natural, therefore, to redefine the $L'$ and $L''$ by
associating particular $\cO(a^2)$ terms  with the $L_i$.
We make the replacements
\begin{eqnarray}
        \frac{16\mu_{\rm tree}}{f^2}(2L_8-L_5) (m_x+m_y) 
	& \to & \frac{16}{f^2}(2L_8-L_5)\Big[\mu_{\rm tree}(m_x+m_y) + a^2\Delta_I\Big] \ , \nonumber \\*
        \frac{32\mu_{\rm tree}}{f^2}\left(2L_6-L_4\right) \left(2\hat m+m_s \right) 
	& \to & \frac{32}{f^2}(2L_6-L_4)\Big[\mu_{\rm tree}(2\hat m+m_s) + \threehalves a^2\Delta_I\Big] \ , \nonumber \\*
         \frac{8\mu_{\rm tree}}{f^2}L_5\left( m_x + m_y \right)
	& \to & \frac{8}{f^2}L_5\Big[\mu_{\rm tree}(m_x+m_y) + a^2\Delta_{av}\Big] \ , \nonumber \\*
         \frac{16\mu_{\rm tree}}{f^2}L_4\left( 2\hat m + m_s\right)
	& \to & \frac{16}{f^2}L_4\Big[\mu_{\rm tree}(2\hat m+m_s) + \threehalves a^2\Delta_{av}\Big] \ , 
\label{eq:replacements}
\end{eqnarray}
where $a^2\Delta_I$ is given by \eq{tree-masses}, and
\begin{equation}\label{eq:delta-av}
	 a^2\Delta_{av} \equiv \frac{a^2}{16}( \Delta_5+4\Delta_A+6\Delta_T+4\Delta_V+\Delta_I)
\end{equation}
is the average splitting. After these redefinitions, a change in $\Lambda_\chi$ 
renormalizes $L'$ according to:
\begin{equation}\label{eq:Lambdap-scale}
L'(\tilde \Lambda_\chi) = L'(\Lambda_\chi) + 2(\delta'_A+\delta'_V)\;\ln(\tilde \Lambda_\chi^2/\Lambda_\chi^2) \ ;
\end{equation}
while $L''$ is independent of scale. From this we would expect that $L'$ is comparable in size
to $\delta'_A+\delta'_V$, an expectation that is borne out by the fits. 
The $L_i$ renormalize by:
\begin{equation}\label{eq:Li-scale}
L_i(\tilde \Lambda_\chi) = L_i(\Lambda_\chi) + \frac{C_i}{256\pi^2} \ln(\tilde 
\Lambda_\chi^2/\Lambda_\chi^2)\ ,
\end{equation}
with
\begin{eqnarray}\label{eq:Ci}
C_4=-1\;; &\qquad& C_5= -3\;;\cr 
2C_6-C_4=-2/9\;; &\qquad& 2C_8-C_5= 4/3\;.
\end{eqnarray}

\subsection{NNLO Terms }
\label{sec:NNLO}

As we will see below, the high statistical precision of our data requires us to go beyond
the NLO formulas, even for subsets of the data that include only the lighter valence quark masses.
We include explicitly all NNLO physical analytic parameters, \ie
all analytic terms of $\cO(m_q^3)$.  
There are five such terms for $m^2_{P^+}$ and an additional five for $f_{P^+}$ \cite{Aubin:2003ne}.
Expressed in terms of $\chi_q$ defined in \eq{chiqdef}, they are given by:
\begin{eqnarray}\label{eq:m-NNLO}
\frac{(m^{\rm NNLO}_{P^+_5})^2}
        {\left( m_x+m_y \right)}&=&
	\mu\Big( 1+ {\rm NLO} + \beta^{(m)}_1 
               \left(\chi_x + \chi_y\right)^2 
+ \beta^{(m)}_2 \left(2\chi_{ud} + \chi_s\right)^2 
+ \beta^{(m)}_3 \left(\chi_x + \chi_y\right)
            \left(2\chi_{ud} + \chi_s\right) \nonumber\\
&&\qquad\quad +\; \beta^{(m)}_4 \left(\chi_x - \chi_y\right)^2 
+ \beta^{(m)}_5 \left(2\chi^2_{ud} + \chi^2_s\right) \Big)\ , \\
\label{eq:f-NNLO}
	f^{\rm NNLO}_{P^+_5}& = & f\Big( 1+ {\rm NLO} + \beta^{(f)}_1 
               \left(\chi_x + \chi_y\right)^2 
+ \beta^{(f)}_2 \left(2\chi_{ud} + \chi_s\right)^2 
+ \beta^{(f)}_3 \left(\chi_x + \chi_y\right)
            \left(2\chi_{ud} + \chi_s\right) \nonumber\\
&&\qquad\quad +\; \beta^{(f)}_4 \left(\chi_x - \chi_y\right)^2 
+ \beta^{(f)}_5 \left(2\chi^2_{ud} + \chi^2_s\right) \Big)\ ,
\end{eqnarray}
where ``NLO'' denotes the lower order contributions, (the corrections
to the leading ``1'' in \eqs{m-NLO}{f-NLO}, with the substitutions in \eq{replacements}).
The interchange symmetries among valence quarks 
$x \leftrightarrow y$ and  sea quarks $u \leftrightarrow d \leftrightarrow s \leftrightarrow u$  
restrict the form of the NNLO corrections.  
These terms were obtained independently
in Ref.~\cite{Sharpe:2003vy}.

Possible analytic taste-violating terms at NNLO of $\cO(m_q^2 a^2)$ are included
implicitly by allowing the $L_i$ ($\cO(m_q^2)$ terms) to vary with lattice spacing.\footnote{
It is not hard to show that allowing the $L_i$ to vary with $a$ generates
all possible $\cO(m_q^2 a^2)$ contributions to masses and decay constants of Goldstone pions at rest.  
However, it is at this order that Lorentz-violating terms can affect the Goldstone pions
\cite{CA_CB1}, so one
would expect to find slight differences at fixed lattice spacing
between masses and decay constants calculated here and those for
pions of non-zero 3-momentum.}
However, such variation can be caused either by taste-violating terms in the Symanzik Lagrangian
or by terms with the same symmetries as the continuum operators but with explicit factors
of $a^2$ (\ie by generic discretization errors on the $L_i$).
As explained above, the generic discretization errors are expected to be $\approx\! 2\%$; while the 
new taste-violating terms could change the apparent value of  $L_i$  between coarse and 
fine sets by order $\chi^{\rm coarse}_{a^2}-\chi^{\rm fine}_{a^2}\approx 6\%$.  For our preferred fits, we use 
Bayesian priors \cite{BAYSE} to restrict  the differences in the $L_i$ on coarse and fine sets
to be at most $7.5\%$ (for a 3-$\sigma$ variation); while in alternative fits used to assess
the systematic error, we relax or tighten this restriction. (See \secref{params}.)
Note that when we extrapolate the $L_i$ to the continuum, we 
have no {\it a priori} way to distinguish variation as
$\alpha_S a^2$ (generic discretization errors) from variation as $\alpha_S^2 a^2$ (taste violations).  
Therefore, we consider both types of variation (\ie \eqs{alpha-a2}{alpha2-a2}, as 
well as the ranges of these ratios) and include the difference in the systematic error.
In practice, these alternative fits and these assumptions 
in how the $L_i$ are extrapolated to the
continuum contribute only a small fraction of the total systematic error. (See \tabref{Li-syst}.)  

At NNLO, $\cO(m_q a^4)$ analytic terms may also affect $m^2_{P^+_5}$ and $f_{P^+_5}$.  
We neglect such terms, which would make contributions similar to that of $L'$ and $L''$ in \eqs{m-NLO}{f-NLO},
but multiplied by $a^4$ instead of $a^2$.  Because $\chi_{a^2}$ is only 0.09 in the worst case, we would
expect the error thereby induced to be at most $\chi^2_{a^2} \sim 1\%$.  In fact, since we
consider a generous range for how taste-violating $a^2$ terms may vary as we go from coarse to fine 
lattices (see discussion immediately following \eq{alpha2-a2}), the
effects of  $\cO(m_q a^4)$ analytic terms should already be
included in our systematic error estimates.

The final possible NNLO terms in the Lagrangian are $\cO(a^6)$.  However, it is easy to
see that such terms do not contribute to $m^2_{P^+_5}$ and $f_{P^+_5}$. The Goldstone theorem
requires that $m^2_{P^+_5}$ be proportional to at least a single power of quark mass ($m_x+m_y$ in this
case), and terms in the Lagrangian must have at least two derivatives to make analytic
contributions to $f_{P^+_5}$ through Noether's theorem or wave-function renormalization. 

In addition to analytic terms at NNLO, there are, of course, NNLO chiral logarithms (from 2-loop graphs,
as well as 1-loop graphs that involve NLO parameters).
These non-analytic terms have not been calculated in \schpt, but in any case are not 
expected to be important here:
Wherever the quark masses or splittings  are large enough for the analytic NNLO terms to be significant, 
the NNLO logarithms should be slowly varying and well approximated by analytic terms.  
As discussed in \secref{params}, the NNLO terms 
make a difference primarily in the  interpolation around $m_s$,
not in the extrapolation to $\hat m$.  The systematic errors inherent in our treatment of
the NNLO terms are estimated by varying the masses we fit to 
and the Bayesian priors governing these terms and their changes with $a$, as well as by
adding still higher (NNNLO) terms.

There are also NNLO effects induced by the ambiguity in the parameters one puts into
NLO expressions.  In particular, we have at present expressed the  
``chiral coupling,'' $1/(16\pi^2f^2)$ in \eqs{m-NLO}{f-NLO}, 
in terms of bare (tree-level) parameter $f$.  Replacing $f$ with the experimental value
of $f_\pi$, say, would generate a difference at NNLO.  As we discuss below,
the difference between $f_\pi$ and $f$ is significant: $\approx\! 13\%$.  
If we had the full NNLO expression,
including 2-loop effects, then the ambiguity would be resolved up to terms of NNNLO.  But in 
the present case there is no {\it a priori}\/ way to decide this issue.

We argue, however, that putting a
{\it physical}\/ parameter in the chiral coupling
($f\to f_\pi$) is likely to result in a better
convergent \chpt.
  This is similar to  
the argument for using a physical, rather than bare,  coupling in weak-coupling 
perturbation theory \cite{LEPAGE_MACKENZIE}.
In practice, we consider three versions of the fits:
\begin{itemize}
\item[]{(1)} Fix coupling as 
$1/(16\pi^2f_\pi^2)$.

\item[]{(2)} Leave coupling as 
$1/(16\pi^2f^2)$.

\item[]{(3)} Write coupling as 
$\omega/(16\pi^2f_\pi^2)$ and treat $\omega$ as an additional fit parameter:
either allow it to vary freely or force it to vary
around 1 using Bayesian priors.
\end{itemize}
Good fits are possible with all three choices. 
Both because of the argument above, and because it guarantees that 
the NLO chiral logarithms for very light quarks have the expected weight 
relative to the tree level terms,
we take choice (1) for our central values.  
Choice (3), with its extra parameter, results in the highest confidence levels of the three.
When $\omega$ is allowed to vary freely, its value decreases as higher quark masses
are included in the fits, reaching $\omega\approx 0.6$ by set {\it II}\/ (see \secref{mass-subsets}).
This is similar to replacing $f\to f_K$, perhaps not surprising for fits that must cover a range of valence
masses up to a large fraction of $m_s$.  But for the quantities computed here,
all of which are sensitive to
the chiral behavior at low quark mass, 
we do not include fits with $\omega$ free since we expect $1/(16\pi^2f_\pi^2)$,
not $1/(16\pi^2f_K^2)$,
to be  the correct weight for the logarithms in the low mass regime.
We still allow fits with range $\omega=1.0\pm0.1$ because 10\% is roughly the difference
between the physical $f^2_\pi$ and its value in the chiral limit.
As discussed in \secref{results}, the main effects of including
fits with arbitrary $\omega$ would be to increase the systematic error in $f_\pi$ 
by about $1\;\MeV$
(with a corresponding effect on $f_K/f_\pi$) and to
double the simulation error on $m_u/m_d$ 
(which error, however, is small compared to unknown electromagnetic effects).

Good fits with choice (2) require $f\,\gtwid\, f_K$ (equivalent to $\omega\approx 0.6$ in choice (3))
and quite large NNLO terms. In addition, the $\cO(a^2)$ NLO
parameter $a^2L''$ becomes unreasonably large ($\sim (630\,\MeV)^2$ on the coarse lattices). For these
reasons we exclude choice (2) fits from the systematic error estimates;
including them would increase systematic errors by amounts comparable to those of arbitrary $\omega$ fits.

Similar considerations apply to the parameter $\mu$. In the analytic terms involving the $L_i$ in \eqs{m-NLO}{f-NLO}
(and hence in \eqs{m-NNLO}{f-NNLO}),
we argue that it is best to put in $\mu_{\rm tree}$ from the linear (tree-level) fits,
\eq{tree-masses}.  The $L_i$ are then multiplied by actual squared meson masses (within the 
errors of \eq{tree-masses}).  This corresponds to how such terms are interpreted in
continuum \chpt\ analysis (see, \eg Ref.~\cite{GASSER_LEUTWYLER}).  An alternative, 
{\it a posteriori}, choice would be
to use the chiral limit of $m^2_{\pi}/(2\hat m)$ coming from the full NNLO fit.  
This would replace $\mu_{\rm tree}$
in \eqs{m-NLO}{f-NLO} by a number 5\% smaller. 
Since \eq{tree-masses} gives a maximum of 7\% errors, we choose that larger value as the systematic effect.  
It would however 
be unreasonable to replace $\mu_{\rm tree}$ in \eqs{m-NLO}{f-NLO} or  (\eqs{m-NNLO}{f-NNLO}) 
by the fit parameter $\mu$ itself. That is because the 
effective value of $\mu$ is corrected by terms involving $m_s$ that do not go
away in the chiral limit for the light quarks.  Indeed, one might expect corrections of at least
$\chi_{ud,s}+\chi^2_{ud,s}\approx 20\%$.  In practice
the fit parameter $\mu$ from our preferred NNLO fit on the intermediate
valence mass set (subset {\it II}\/, \secref{mass-subsets}) is 29\% less than $\mu_{tree}$, which means
that $\mu(m_x+m_y)$ is significantly less than our measured value of $m^2_{P^+_5}\cong \mu_{\rm tree}(m_x+m_y)$.
This difference improves to 13\% in the fit to the lightest masses. 

As discussed in the introduction,
all meson masses appearing in the NLO chiral logarithms
in \eqs{m-NLO}{f-NLO}
are similarly evaluated using the previously determined values of the taste splittings, $a^2\Delta_B$,
and $\mu_{\rm tree}$ from the fit of our ``full QCD'' data for all meson tastes to \eq{tree-masses}.
In our results for masses and decay constants, 
the NNLO error introduced by this procedure is 
negligible. That is  because the effect of the small errors in masses in
the chiral logarithms on our extrapolated values is almost completely canceled by the effect
of variations of the analytic parameters in the fit.
We can check this by replacing $\mu_{\rm tree}$ in the fit by the ($5\%$ different) chiral limit
value; the effect is about $0.2\%$ on 
quark masses and less than $0.1\%$ on the decay constants,
in both cases much less than the total systematic error.
For the $L_i$, changing the value of $\mu$ in the chiral logarithms does not
completely cancel the effect of changing its value in the analytic terms, but there is
some cancellation.   Therefore the $7\%$ systematic effect in the $L_i$ discussed 
in the previous paragraph
remains a conservative estimate of the error.

\subsection{NNNLO terms}\label{sec:NNNLO}

We sometimes add some NNNLO terms of the following form:
\begin{eqnarray}\label{eq:m-NNNLO}
\frac{(m^{\rm NNNLO}_{P^+_5})^2}
        {\left( m_x+m_y \right)}&=&
	\mu\Big( 1+ {\rm NLO} + {\rm NNLO} +\rho^{(m)}
               \left(\chi_x + \chi_y\right)^3 
\Big)\ , \\
\label{eq:f-NNNLO}
	f^{\rm NNLO}_{P^+_5}& = & f\Big( 1+ {\rm NLO} +  {\rm NNLO}+ 
	\rho^{(f)} \left(\chi_x + \chi_y\right)^3 
\Big)\ ,
\end{eqnarray}
where NLO and NNLO represent the contributions from \eqsfour{m-NLO}{f-NLO}{m-NNLO}{f-NNLO}.
Since there are, of course, many additional NNNLO terms, it is nonsystematic to include only one each
for $m_{P^+_5}$ and $f_{P^+_5}$.  However, we pick these terms involving
valence masses because there is a steeper dependence on the valence masses than on the sea quark masses.
For lower quark masses, where we expect \chpt\ to work well, we fit to \eqs{m-NNNLO}{f-NNNLO}
only to estimate systematic errors due to the truncation of \chpt.  
When the fits include valence masses equal to or greater than $m'_s$,
we also use \eqs{m-NNNLO}{f-NNNLO} in order to improve the interpolation around the 
strange quark mass.  In
the former case, we find that the
values of $|\rho^{(f)}|$ and $|\rho^{(m)}|$ coming from the fits are typically less than 0.1; in both
cases they are always less than 0.2 (including when we fit to \eqs{m-NNNLOP}{f-NNNLOP} --- see \tabref{fit-params}).

Another form, used only for interpolations around the strange quark mass, adds on the square of the NLO
term as a mock-up of the effect of 2-loop chiral logarithms:
\begin{eqnarray}\label{eq:m-NNNLOP}
\frac{(m^{\rm NNNLO'}_{P^+_5})^2}
        {\left( m_x+m_y \right)}&=&
	\mu\Big( 1+ {\rm NLO} + {\rm NNLO} 
+\sigma^{(m)} \left({\rm NLO}\right)^2 
+\rho^{(m)} \left(\chi_x + \chi_y\right)^3 
\Big)\ , \\
\label{eq:f-NNNLOP}
	f^{\rm NNNLO'}_{P^+_5}& = & f\Big( 1+ {\rm NLO} +  {\rm NNLO} 
	+\sigma^{(f)} \left({\rm NLO}\right)^2 + \rho^{(f)} \left(\chi_x + \chi_y\right)^3 
\Big)\ ,
\end{eqnarray}
where again NLO and NNLO represent the contributions from \eqsfour{m-NLO}{f-NLO}{m-NNLO}{f-NNLO}.
The absolute values of the new coefficients $\sigma^{(m)}$ and $\sigma^{(f)}$ in the fits
are never greater than 0.14.

\section{Perturbation theory}\label{sec:perturbation-theory}
Because the axial current we use to compute decay constants is partially conserved, there is
no need for current renormalization.   Mass renormalization is however needed
to find continuum ($\msbar$) quark masses, as discussed in Ref.~\cite{strange-mass}.
Let $Z_m$ be the mass renormalization factor that 
connects the  bare lattice mass $(am)_0$ and the $\msbar$ mass at scale $\Lambda$:\footnote{$\Lambda$ was called $\mu$ in
\protect{\cite{strange-mass}}, a notation we avoid here for obvious reasons.}
\begin{equation}\label{eq:mass-renorm}
 m^{\msbar}(\Lambda) = Z_m({a\Lambda }) \frac{(am)_0}{a u_{0P}} \ .
\end{equation}
Here, unlike in Ref.~\cite{strange-mass}, we have shown explicitly the plaquette tadpole improvement factor
$u_{0P}$, necessary because the MILC improved staggered action
defines the lattice quark mass in a somewhat unconventional manner.

The renormalization factor $Z_m$ 
enters the analysis in another way.
As mentioned above, we need to renormalize the parameter $\mu$ if we wish to compare its
values at  different lattice spacings.  More precisely, we need the ratio
\begin{equation}\label{eq:Rm}
R_m \equiv \frac{Z_m(a^{\rm coarse}\Lambda )}{Z_m(a^{\rm fine}\Lambda )}\; \frac{u_{0P}^{\rm fine} }
{u_{0P}^{\rm coarse} }\ ,
\end{equation}
$R_m$ is in principle independent of $\Lambda$, although when $Z_m$ is evaluated at any given order
in perturbation theory, there is residual $\Lambda$ dependence from neglected higher order terms.
For definiteness, we take $\Lambda=2\,\GeV$.
$Z_m$ is given by \cite{strange-mass}
\begin{equation}
Z_m (a\Lambda)=\left(1+\alpha_V(q^*)\,Z_m^{(2)}(a\Lambda) +\cO(\alpha^2)\right)\ ,
\label{eq:Zm} 
\end{equation}
where $\alpha_V$ is determined from  small Wilson loops using third order perturbation theory \cite{Davies:2002mv,new_alpha},
the optimal scale $q^*$ is estimated using a 
second order BLM method~\cite{Hornbostel:2002af}, and $Z_m^{(2)}$  is \cite{Hein:2001kw,Becher:2002if,Q:thesis}
\begin{equation}
\label{eq:Zm2}
Z_m^{(2)}(a\Lambda) = \left(b-\frac4{3\pi}-\frac2\pi\ln(a\Lambda)\right),
\end{equation}
with $b\approx 0.5432$, correct to 0.1\%.
We have neglected the (tiny) $\cO(a)$ mass dependence of $b$, and hence of $Z_m(a{\Lambda})$.
From Ref.~\cite{strange-mass}, 
$q^*=2.335/a$ and $\alpha_V(q^*)= 0.252(5)$ on the coarse lattices;
$q^*=1.80/a$ and $\alpha_V(q^*)=0.247(4)$ on the fine.

To evaluate $R_m$, we use scale and plaquette values from the coarse $0.01/0.05$ 
and the fine $0.0062/0.031$ lattices, and neglect the
the small variation among the coarse or fine sets.  As mentioned
previously, the $\Upsilon$ splittings give \cite{BIG_PRL,HPQCD_PRIVATE}
$(a^{\rm coarse})^{-1}=1.588\,\GeV$ and $(a^{\rm fine})^{-1}=2.271\,\GeV$. 
The tadpole improvement factor are $u_{0P}^{\rm coarse}=0.8677$ and
$u_{0P}^{\rm fine}=0.8782$.  From \eqsthree{Rm}{Zm}{Zm2}, we find $R_m\approx0.958$.  
Then,
\begin{equation}\label{eq:mu-equality}
\mu^{\rm coarse} = R_m \mu^{\rm fine}
\end{equation}
with the above value of $R_m$ defines what we mean by ``equality'' of the parameter
$\mu$ on coarse and fine lattices.  Of course, \eq{mu-equality} may be violated
by generic $\cO(a^2)$ scaling violations ($\sim\!2\%$), as well as by
perturbative errors.
{\it A priori}, one expects a two-loop correction to $R_m$ of order
$\alpha_V^2$.  This is $\approx\! 6\%$.  In practice, fits have a confidence level
that is higher than those of our preferred fits
if we take $R_m \approx 0.87$ to $0.89$, \ie a 7 to 9\% difference
from $R_m=0.958$.  Although it is not possible to separate the perturbative errors from
the discretization errors in this difference, here and in Ref.~\cite{strange-mass}
we take
the larger value, $9\%$, as the conservative estimate of perturbative errors.  
This is $\approx\!1.5\alpha_V^2$.  For quantities that do not directly involve
perturbation theory, such as the decay constants and
the ratio $m_u/m_d$, we do not quote perturbative errors, {\it per se}\/. 
But $R_m$ still enters the chiral fits, so we include fits with $R_m = 0.87$
among the alternatives.

Another rough estimate of $R_m$ comes from $\mu_{\rm tree}$, \eq{tree-masses}.  Without
the proliferation of parameters at NLO and NNLO, the tree-level form
makes possible well-controlled fits on coarse and fine lattices separately. 
We get $R_m\approx 0.977$.  But note that \eq{tree-masses} can have up to $\sim\!7\%$
errors in describing the data, and there are also discretization errors in this
estimate.

\section{Electromagnetic and isospin-violating effects}\label{sec:EM}

Given the precision we are aiming at here, it is necessary to take into
account electromagnetic (EM)
and isospin-violating effects, at least in an approximate way.  Our simulation
is in isospin-symmetric  QCD, with $m_u$ set equal to $m_d$, and the electromagnetic
coupling, $e^2=4\pi\alpha_{EM}$, set to 0.  This means that
when we compare meson masses to experiment to determine
the physical quark masses $\hat m$ and $m_s$, we must first adjust the experimental
numbers to what they would be in a world without EM effects or isospin
violation.  This is particularly important for the pion, since the
difference between $m_{\pi^+}^2$ and $m_{\pi^0}^2$ is almost 7\%. 
Because the adjustment is only approximate, there are some residual systematic
errors on the quark masses, as discussed in Ref.~\cite{strange-mass}.

The decay constants, as well as the low energy constants $L_i$, 
are by definition pure QCD quantities,
so we do not have to take EM effects
directly into account in our determination.\footnote{However, the EM corrections
must be explicitly evaluated when 
the decay constants are compared to experiment \protect{\cite{Marciano:1993sh,PDG}}.}
Nevertheless,  there are indirect 
EM effects on $f_\pi$ and $f_K$,
which come in through the quark masses when
we extrapolate to the physical point. Isospin violations are irrelevant
for the $L_i$, which are defined to be mass independent. But for the
decay constants, there are both  direct and indirect 
isospin-violating effects, which we estimate
below.  The end result is that both the (indirect) EM and isospin-violating
errors on decay constants are very small, as long as we are careful to extrapolate to the
appropriate values of the quark mass in each case.  However the EM error on $m_u/m_d$ is 
large unless we are willing to assume that the EM effects on meson masses are accurately known.

Electromagnetism can be included in \chpt\ in a systematic way.
Dashen's theorem \cite{Dashen:eg} summarizes the EM effects on meson
masses at lowest nontrivial order in $e^2$ 
and the quark masses. It states that 
$m^2_{\pi^+}$ and $m^2_{K^+}$ receive equal $\cO(e^2)$ 
contributions in the chiral limit; 
while the $\pi^0$ and $K^0$ masses are unaffected. However,
there can be large and different EM contributions  
to $m^2_{\pi^+}$ and $m^2_{K^+}$ of
order $e^2\chi_{ud,s}$ \cite{Donoghue:hj,Urech:1994hd,Bijnens:1996kk,Gao:1996sa}.
Following Ref.~\cite{Nelson-thesis}, we let $\Delta_E$ parameterize 
violations of Dashen's theorem:
\begin{equation}\label{eq:Dashen}
(m^2_{K^+}-m^2_{K^0})_{\rm EM}= 
(1 + \Delta_E)(m^2_{\pi^+}-m^2_{\pi^0})_{\rm EM}\ .
\end{equation}
Then Refs.~\cite{Donoghue:hj,Urech:1994hd,Bijnens:1996kk,Gao:1996sa} 
suggest $\Delta_E\approx 1$.  
Most of these corrections are probably to the charged meson masses.
Indeed, the violation of Dashen's theorem for the $\pi^0$ is
$\cO(e^2\chi_{ud})$ \cite{Urech:1994hd} and therefore small. 
The EM contribution to $m^2_{K^0}$, on the other hand, 
is in principle the same order
as the violations of Dashen's theorem for the charged masses,
$e^2\chi_{ud,s}$ \cite{Urech:1994hd}.
Nevertheless, a large $N_c$, extended NJL model calculation \cite{Bijnens:1996kk} finds a tiny EM
correction to the $K^0$ mass at this order.   To be conservative, though, 
we allow for EM contributions to $m^2_{K^0}$ of
order of half the violations of Dashen's theorem, with unknown sign:
\begin{equation}\label{eq:K0-EM}
(m_{K^0}^2)_{\rm EM}\sim \pm(\Delta_E/2)(m^2_{\pi^+}-m^2_{\pi^0})_{\rm EM} \ . 
\end{equation}

The effects of isospin violation in the pion masses are quite small.
When $m_u\not=m_d$, $m^2_{\pi^0}$ gets a contribution of order $(\chi_u-\chi_d)^2$.
The isospin-violating splitting $(m_{\pi^+}-m_{\pi^0})_{\rm QCD}$ is estimated
as $0.17(3)\,\MeV$ in Ref.~\cite{GASSER_LEUTWYLER}, and as $0.32(20)\,\MeV$ in Ref.~\cite{Amoros:2001cp}.
In the kaon system, on the other hand, the effects of isospin violation are clearly important,
as is obvious from the fact that the experimental $K^+$--$K^0$ splitting is of opposite sign to that
in \eq{Dashen} for any $\Delta_E>-1$.  In our calculation, we can reduce the isospin violating effects in the 
kaon masses to the same order as in the pion system by focusing on the isospin averaged
quantity $(m_{K^0}^2 + m_{K^+}^2)/2$.  
We then neglect the remaining isospin violations in the meson masses.    
We have checked, using the estimates for $(m_{\pi^+}-m_{\pi^0})_{\rm QCD}$
above, that the indirect effect of such isospin violations on decay constants is extremely small:
$\ltwid 0.03\%$. 
These isospin violations were also neglected in the computation of quark masses in
Ref.~\cite{strange-mass}. We note, however, that including isospin violations could
have some small effect there, in particular on the result for the ratio $m_s/\hat m$.  
If $(m_{\pi^+}-m_{\pi^0})_{\rm QCD}$ is at the upper end of the range in \cite{Amoros:2001cp},
the central value for $m_s/\hat m$ in Ref.~\cite{strange-mass} could be changed from 27.4 to as low as 27.2.

Based on the above discussion, we may determine the
physical values of $\hat m$ and $m_s$ by
extrapolating the lattice squared meson masses to $m^2_{\hat \pi}$ and $m^2_{\hat K}$, given by
\begin{eqnarray}\label{eq:masses-EM}
m^2_{\hat \pi}& \equiv &m^2_{\pi^0}\nonumber \\
m^2_{\hat K}& \equiv & \half\left(m_{K^0}^2 + m_{K^+}^2 -(1+\Delta_E) ( m_{\pi^+}^2 - m_{\pi^0}^2)
\right) \ ,
\end{eqnarray}
where experimental
values are to be used on the right hand side.
Allowing for EM corrections to the $K^0$ mass, \eq{K0-EM},
replaces $\Delta_E\approx 1$ in \eq{masses-EM} with an effective value in the
range 0--2, which is in any case a conservative range for $\Delta_E$ that includes
the Dashen theorem result. Here and in
Ref.~\cite{strange-mass},  we take $\Delta_E=1$
for the central value, and use $0\le\Delta_E\le 2$
to estimate systematic errors in $\hat m$, $m_s$, and their ratio.

We can also estimate $m_u$ (or equivalently the ratios $m_u/m_d$ or $m_u/\hat m$) from our simulation.
Given $m_s$, we find $m_u$ 
by extrapolating in the light valence mass 
to the point where the $K^+$ has the mass $(m_{K^+})_{\rm QCD}$,
where ``QCD'' indicates that EM effects have been removed.  We take
\begin{equation}\label{eq:mK+-EM}
(m^2_{K^+})_{\rm QCD} \equiv  m_{K^+}^2 -(1+\delta_E) ( m_{\pi^+}^2 - m_{\pi^0}^2) \ ,
\end{equation}
with $\delta_E=1$ our central value, corresponding to $\Delta_E=1$ and vanishing EM correction
to the $K^0$ mass.  If we attribute the uncertainty in the 
effective value of $\Delta_E=1$ to the
uncertainty in \eq{K0-EM}, then we get a range $0.5\le\delta_E\le 1.5$. This produces only a small uncertainty
in $m_u/m_d$ because the variations in  $(m^2_{K^+})_{\rm QCD}$ and 
$m^2_{\hat K}$ are equal.  A more conservative assumption is that $\Delta_E$ arises primarily from
EM contributions to the $K^+$ mass.  This implies $\delta_E=\Delta_E$ and thus
$0\le\delta_E\le 2$, which we take as the range for estimating EM systematic errors in $m_u/m_d$.
Under this assumption those errors are quite large, $\sim\! 20\%$.
On the other hand, if we were for example to take $\Delta_E=0.84\pm0.25$ from Ref.~\cite{Bijnens:1996kk},
this error would be reduced to  $\sim\! 5\%$. 

There is an additional error on $m_u/m_d$ because we keep the light sea quarks with fixed masses $m_u=m_d=\hat m$
as we extrapolate in the light valence mass to $(m^2_{K^+})_{\rm QCD}$. 
The effect produces a fractional error in $m_u/m_d$ of  $\cO((m_u-m_d)^2)$, \ie of NNLO. This is
because terms of $\cO(m_u-m_d)$
cancel when expanding $m_u$ and $m_d$ around $\hat m \equiv (m_u+m_d)/2$.  We estimate the size of this effect using the
NNLO analytic terms;
from  \eq{m-NNLO} the only relevant coefficient is $\beta_5^{(m)}$.  This term gives a fractional error
$\beta_5^{(m)}(\chi_u^2 +\chi_d^2 - 2\chi^2_{ud})$. 
Using $\chi_{ud}\approx m_\pi^2/(8 \pi^2 f_\pi^2)\approx 0.014$,
our result $m_u/m_d\approx 0.43$, and 
$\beta_5^{(m)}\approx 1.86$ in the continuum limit 
from Fit C (see \secref{fit-results} and \tabref{fit-params}),
we find a negligible error $\approx\! 0.01\%$.

Our simulation directly determines decay constants of the charged mesons, $\pi^+$ and $K^+$, 
in the absence of electromagnetism and with $m_u=m_d$.   Since $m_{\hat \pi}$ is (approximately)
the $\pi^+$ mass in this limit, we must simply extrapolate $f_{\pi^+}$ to the point where the
mesons have the masses in \eq{masses-EM}, \ie to our
physical values of $\hat m$ and $m_s$. The situation with the kaons is rather different.  
It is $f_K^+$ that is measured experimentally,  not some isospin averaged decay constant
of $K^+$ and $K^0$.  We therefore should extract $f_K^+$ by extrapolating the light valence
quark to the physical value of $m_u$, not $\hat m$.   Despite the large uncertainty in 
 $m_u/m_d$ from EM effects, the indirect error induced in  $f_K^+$ through $m_u$ is
tiny, $\sim\!0.07\%$.  This is due to the fact that the decay constant changes slowly with mass:
It varies only by 20\% all the way from the $K$ to the $\pi$.

In principle there are also direct isospin-violating errors in the decay constants.
For $f_{K^+}$, there is an effect we can estimate from the coefficient $\beta_5^{(f)}$, \eq{f-NNLO}, similar to the one 
in $m_u/m_d$ 
from $\beta_5^{(m)}$. Since $\beta_5^{(f)}$ in our fits is $\approx\! -0.1$, \ie much smaller than
$\beta_5^{(m)}$, this effect is completely negligible.  For $f_{\pi^+}$, errors can also arise from
the coefficient $\beta_4^{(f)}$ because we assume $m_u=m_d$. But this coefficient is $\approx\! -0.06$,
again leading to a negligible effect.

\section{\schpt\ Fits}
\label{sec:fits}

We fit the partially quenched data for 
$m^2_{P^+_5}/(m_x + m_y)$ and $f_{P^+_5}$ together in all fits;
this helps to constrain the common $\cO(a^2)$ chiral parameters.  
Similarly, both coarse and fine data
are fit together, helping to constrain the overall lattice spacing dependence. 
Correlations between and among masses and decay constants 
within each sea-quark set are included, with the covariance matrix
computed as described in \secref{simulations}.

\subsection{Data subsets and fit ranges} 
\label{sec:mass-subsets}

Our lattice data is
very precise (0.1\% to 0.7\% on $m^2_{P^+_5}/(m_x+m_y)$, and 0.1\% to 0.4\% on $f_{P^+_5}$); while 
the \chpt\ expansion
parameter for the kaon,  $\chi_{ud,s}$,
is $\approx \! 0.18$.  Since $\chi_{ud,s}^2\approx0.03$, we cannot
expect NLO \chpt, which is missing corrections to $m^2_{P^+_5}/(m_x+m_y)$ or $f_{P^+_5}$ of order $\chi_{q}^2$, 
to work well for meson masses that are even an appreciable fraction of $m_K$.  NNLO \chpt, however,
may allow us to fit up to fairly near  $m_K$, because the missing
corrections at the kaon, $\chi_{ud,s}^3\approx0.006)$, are comparable to (but somewhat larger
than) the statistical accuracy of our data.  Of course, this is only a rough guide.
There are at least two sources of complications:  
(1) It is an idealization to imagine that the chiral expansion is governed by a single
mass parameter. Many different mesons contribute to chiral loops. 
Although we can restrict the valence masses in the fit, 
the $s$ sea quark mass in the simulations is fixed at $m'_s$.
Thus there will always be some contributions from fairly heavy mesons.
(2) Taste violations produce additional contributions to meson masses,
or, effectively, add another expansion parameter, $\chi_{a^2}$, \eq{chia2def}.

In practice 
we consider three different subsets of our complete (coarse and fine) partially quenched 
data set.  Compared to the strange sea quark mass in the simulations, $m'_s$, 
we can tolerate somewhat heavier valence masses on the fine lattices, since on those lattices
$m'_s$ exceeds $m_s$ by a smaller amount and contributions to meson masses
from taste splittings are smaller. The sets are:
\begin{itemize}
\item[$\bullet$]{} Subset {\it I}\/. \hfill\break
$m_x+m_y \le 0.40m'_s$ (coarse), and $m_x+m_y \le 0.54m'_s$ (fine). \hfill\break
47 valence mass combinations; 94 data points.

\item[$\bullet$]{} Subset {\it II}\/.\hfill\break
$m_x+m_y \le 0.70m'_s$ (coarse), and $m_x+m_y \le 0.80m'_s$ (fine).\hfill\break
120 valence mass combinations; 240 data points.

\item[$\bullet$]{} Subset {\it III}\/.\hfill\break
$m_x+m_y \le 1.10m'_s$ (coarse), and $m_x+m_y \le 1.14m'_s$ (fine).\hfill\break
208 valence mass combinations; 416 data points.
\end{itemize}
There are always twice as many data points as mass combinations because we are fitting 
$m^2_{P^+_5}/(m_x + m_y)$ and $f_{P^+_5}$ simultaneously.

On the fine lattices, $m'_s$ is about 10\% larger than $m_s$, so
we expect errors of order $(0.54\times1.1\times0.18)^2\approx\!1.1\%$ at NLO in subset {\it I}\/,
without even considering the effects of mixed valence-sea mesons or taste violations. Thus we do
not anticipate that the NLO form, \eqs{m-NLO}{f-NLO}, can fit the data, even on subset {\it I}\/.
Good fits should require at least the NNLO forms, \eqs{m-NNLO}{f-NNLO}, on all sets. 
Indeed, fitting with \eqs{m-NLO}{f-NLO} on subset {\it I}\/ gives 
minuscule confidence levels (${\rm CL}< 10^{-58}$;  $\chi^2/{\rm d.o.f.}= 6.39$ with 74 degrees of
freedom), and adding in those NNLO terms that involve sea quark masses (because 
$m'_s$ is not small) still results in ${\rm CL}< 10^{-13}$ 
($\chi^2/{\rm d.o.f.}= 3.00$ 
with 62 degrees of freedom).

We note that it is not practical to use valence masses below those in subset {\it I}\/ because
we rapidly run out of data, and in any case we cease to reduce significantly 
the masses of mesons made of a valence quark and a strange sea quark.

\subsection{Inventory of parameters and alternative fits} 
\label{sec:params}

Since there are a large number of fit parameters, we provide an inventory
before discussing the final fits in more detail.  Our standard NNLO fits on 
subsets  {\it I}\/ and  {\it II}\/ have the following number and types of parameters:

\begin{itemize}
\item[(a)]{{\bf LO}.}\hspace{0.05truein} 2 unconstrained parameters: $\mu$ [\eq{m-NLO}] and $f$ [\eq{f-NLO}].

\item[(b)]{{\bf NLO (physical)}.}\hspace{0.05truein}
4 unconstrained parameters: $2L_8\!-\!L_5$,  $2L_6\!-\!L_4$, 
$L_5$, $L_4$ [\eqsthree{m-NLO}{f-NLO}{replacements}].

\item[(c)]{{\bf NLO (taste-violating)}.}\hspace{0.05truein}
4 unconstrained parameters: $a^2\delta'_V$, $a^2\delta'_A$,
$a^2L'$, $a^2L''$ [\eqs{m-NLO}{f-NLO}]. 

\item[(d)]{{\bf NNLO (physical)}.}\hspace{0.05truein}
10 parameters:  $\beta^{(m)}_1,\dots, \beta^{(m)}_5$ [\eq{m-NNLO}], and  
$\beta^{(f)}_1,\dots, \beta^{(f)}_5$ [\eq{f-NNLO}]. For preferred fits, 
these are constrained with Bayesian
priors to have standard deviation of 1 around 0.  But alternative fits used for systematic
errors estimates leave these parameters unconstrained.  The difference in CL or
final results is small.

\item[(e)]{{\bf Scale}.}\hspace{0.05truein}
4 tightly constrained parameters that determine relative scale of different lattices:  
$C_{00}$ $C_{10}$, $C_{01}$, $C_{20}$ [\eq{R1_FIT_EQ}].
These are allowed to vary by 1 standard deviation
around values in \eq{R1_RESULTS}.  The (small) variation 
in these parameters makes very little difference
in CL or central values, but including the variation in the fit allows
us to incorporate the statistical errors in relative scale determination into the statistical
errors of our results.

\item[(f)]{{\bf Lattice spacing dependence}.}\hspace{0.05truein} 
16 parameters (usually tightly constrained) that control the fractional difference in the
physical fit parameters [(a), (b), and (d) above] between coarse and fine lattices.
In our preferred fits we allow the LO  parameters [(a)] to vary by 2\% (at the 
1 standard deviation level), 
and consider alternatives of ``0\%'' (\ie no variation in LO parameters 
between coarse and fine lattices parameters), 1\%, and 
4\% in estimating systematics.
For the NLO parameters [(b)],  the central choice is 2.5\%; alternatives
are 0\%, 1\%, 4\%, and 6\%.  For NNLO parameters [(d)], the central choice is 2.5\%; alternatives
are 0\%, 1\%, 4\%.   We also consider complete removal of the constraints for various small
subsets of the NLO and NNLO parameters.
\end{itemize}
In our standard NNLO fit there are thus a total of 40 parameters, of which 20 are generally tightly constrained.

We remind the reader that the NLO taste-violating parameters [(c) above] are forced to change by
a fixed ratio (in a given fit) in going from the coarse to fine lattices.
The point is that these parameters start at $\cO(\alpha_S^2 a^2)$, so we know how they change with
$a$, up to corrections that are higher order in $\alpha_S$ and/or $a$. A range for the ratio is
considered in assessing the systematic error (see discussion following \eq{alpha2-a2}).

The priors restricting the parameters governing lattice spacing dependence [(f) above] require further
explanation.  We note first that it is not possible, at least with the current data set, to remove
these restrictions on all the physical parameters,
allowing them to be arbitrarily different on coarse
and fine lattices. If we do that, the fit becomes unstable because there are
directions in parameter space in which the fit function is almost flat.  Some of these directions
can easily be seen in \eqs{m-NLO}{f-NLO}.  For example,  $a^2L'+a^2L''$ can grow large, compensated by
a decrease in $\mu$ and corresponding increases in 
$2L_8\!-\!L_5$, $2L_6\!-\!L_4$, $a^2\delta'_A$, $a^2\delta'_V$,
and the $\beta^{(m)}_i$ in \eq{m-NNLO}.  Only the first (continuum) log term in \eq{m-NLO}
is uncompensated, but we already know that the good fits allow a fairly large range in its coefficient,
the chiral coupling. (See discussion of the parameter $\omega$ in \secref{NNLO}.)  A similar mode
involves $a^2L''-a^2L'$ growing and $f$ decreasing in \eq{f-NLO}.  Even when such modes eventually
converge, the resulting fit is completely unphysical, with $20\%$--$100\%$ 
variation of results with lattice spacing and enormous NLO chiral corrections. 
Once the physical parameters are required to change by only a small amount with $a$, however, they cannot compensate for changes in taste-violating parameters like $a^2L'$ and $a^2L''$, and the runaway modes are damped.

Since the taste-violating NLO parameters ($a^2L'$ and $a^2L''$) are included explicitly,
the LO parameters $f$ and $\mu$ should have only generic ($\sim\!2\%$) changes with $a$,
which is our preferred choice for priors for their variations.  Of course, as mentioned 
following \eq{mu-equality}, differences in $\mu$ between coarse and fine lattices can also be due to perturbative
errors in the ratio $R_m$.  When we use the one-loop  value for $R_m$, 0.958, the fits prefer an $\approx 8\%$
difference in the renormalized $\mu$ value between coarse and fine, which is one indication of the size of
the perturbative error.

For most of the NLO physical parameters [(b)], the preferred 2.5\% prior for lattice-spacing dependence
is not restrictive, with the fits finding a change that is significantly smaller ($\ltwid 0.5\%$).
The exception is $2L_8\!-\!L_5$, for which the 2.5\% prior results in almost a 6\% difference (a $2.3\sigma$ effect),
suggesting that the corresponding NNLO taste-violating term has a sizable coefficient.  If we
instead remove any restriction on the $a$-dependence of $2L_8\!-\!L_5$ (while keeping the constraints on
the other physical parameters), $2L_8\!-\!L_5$ varies by $\approx\!  20 \%$ from coarse to
fine, which is sizable, $\sim\! 3(\chi^{\rm coarse}_{a^2}-\chi^{\rm fine}_{a^2})$. Perhaps generic and
taste-violating effects are both contributing significantly in the same direction.
Fortunately, the continuum extrapolated value of $2L_8\!-\!L_5$ changes by only 5\% when
we remove the restriction on its $a$-dependence. 
In any case, the systematic errors on $2L_8-L_5$ (as well as on the other $L_i$) are dominated by
the larger changes caused by varying the mass range and/or the details of the
chiral fits.  Thus, 
even if we were to take the full 20\% variation from coarse
to fine as the ``discretization error'' on this parameter
the final errors quoted 
in \secref{results} would hardly change at all. 

Our  preferred 2.5\% prior
for the NNLO physical parameters [(d)] is again generally not restrictive, with $\beta_2^{(m)}$
the only exception.   We note that allowing the NNLO terms to vary with $a$ is not 
systematic,\footnote{We thank Laurent Lellouch for  pointing this out.}  
because it effectively introduces some, but not all, NNNLO terms.
Therefore we consider an alternative in which NNLO physical parameters do not
vary with $a$ (``0\% priors''), but all other features of  the preferred fit are unchanged.  This fit
has lower confidence level (${\rm CL}=0.36$) than the preferred fit (${\rm CL}=0.65$), but gives
physical results that are very similar.

The alternative choices of priors for parameters (f)  also give good fits,  
except in the cases of 1\% (${\rm CL}\approx 5\times10^{-3}$) and 
``0\%'' priors\footnote{In this case there are 24 free parameters
in the fit.}  
(${\rm CL}\approx 10^{-4}$).
However we keep all the choices in the systematic
error analysis.  The 0\% priors case, where all physical parameters are fixed as a function of
$a$, gives results that are in no way extreme among all alternative fits considered in this work.

The preferred NNLO fit with parameters (a)--(f) above results in good confidence levels 
on data subsets {\it I}\/ and {\it II}\/.  This is all we need to extract the low energy
constants $L_i$, and it is acceptable for determining $f_\pi$ and $\hat m$.
But to determine $f_K$ and $m_s$ without an extrapolation to heavier valence masses, 
we need to fit to the data in subset {\it III}\/. We can then interpolate to the physical strange mass.
However the preferred NNLO fit, or variants thereof, does not give
good confidence level when applied to subset {\it III}\/.  Based on the discussion in \secref{mass-subsets},
this is not surprising, especially considering that the simulation mass $m'_s$ is larger than
the physical strange quark mass.  The problem, though, is that it is not possible to move
beyond NNLO in anything approaching a systematic way without introducing an unwieldy number of new parameters.  
If instead we fit subset {\it III}\/ to our rather {\it ad hoc} NNNLO forms in \secref{NNNLO}, we can
obtain acceptable fits.  But the high
quark masses involved, as well as the nonsystematic treatment of the higher order terms,
may introduce significant systematic errors in the low energy constants and
therefore in the extrapolation of $\hat m$ to the physical point, which is, of course, still needed for finding $m_s$ 
and $f_K$.  Thus such fits are
not acceptable for finding decay constants and quark masses.

Our solution to the above dilemma is to use the results for all LO and NLO parameters from the fits to subset
{\it II}\/ as inputs to the fits on subset {\it III}\/.  We then use the form
\eqs{m-NNNLOP}{f-NNNLOP} on subset {\it III}\/, but with the LO and NLO parameters, and their change with
lattice spacing, constrained to vary by
at most their statistical errors around their previously determined values.  The NNLO and NNNLO parameters
are left unconstrained.  Their variation with lattice spacing, which implicitly introduces
higher order mixed terms in  $\chi_q$ and $\chi_{a^2}$, either is also left constrained or is
constrained mildly, with 1-$\sigma$ constraints of 10\% (preferred fit), 15\%, or 20\%.  
The total number of parameters in this fit is 48, of which 20 (16 LO and NLO parameters and their $a$-dependence,
plus the usual 4 scale parameters) are tightly constrained.
We get reasonable
fits with all these versions.  The advantage of this approach
is that we can interpolate to the physical strange quark mass and extrapolate to the physical
light quark mass in the same fit.  We therefore use the preferred fit to subset {\it III}\/ for our central values for
quark masses and decay constants.  
The {\it ad hoc} nature of the higher order terms in this fit is not a problem because
we have already satisfied the chiral constraints at low quark mass. 
All that is required in the
region of $m_s$ is a fit that interpolates from 
our partially quenched valence masses and
nominal sea quark mass to the physical $m_s$ value.   
In other words, we do not need to 
rely in detail on chiral perturbation theory for the $m_s$ dependence, since we
can reach the physical value of $m_s$ in the simulation.
Effectively, we are depending only on two-flavor \chpt. Note however that we include
the more conventional chiral fits on subsets {\it I}\/ and {\it II}\/ among the alternatives
when estimating the systematic errors.

\subsection{Fit results}
\label{sec:fit-results}

\Figrefs{fpi-vs-m-II}{mpisq-over-m-vs-m-II} show our preferred NNLO fit to data subset {\it II}\/. We call
this fit  ``Fit B''; the corresponding fit on data subset {\it I}\/ is called ``Fit A.''
Fit B is a single fit to the data in both \figrefs{fpi-vs-m-II}{mpisq-over-m-vs-m-II},
as well as many more data points not shown.
The fit has a chi-square of 192 with 200 degrees of freedom,
giving  ${\rm CL} = 0.65$.  
This is a standard CL, with $\chi^2$ summed over all data points,
and number of degrees of freedom (d.o.f.) given by number of data points minus the number of
parameters.  If we include the Bayesian priors as effective ``data points,'' then
Fit B has a chi-square of 235 with 230 degrees of freedom, CL=0.39. The fact that this is also
an acceptable CL indicates that \schpt\ and
our assumptions about the $a$-dependence of fit parameters are
reasonably well behaved in this mass range. Fit A gives very similar results for decay constants and
quark masses, but
includes many fewer points (94 \vs 240) and has a lower confidence level (0.23). As discussed in
\secref{chiral-converge}, however, Fits A and B produce rather different 
values for the low energy constant $2L_6-L_4$,
indicating a large systematic uncertainty in that parameter. 
For our central values of the $L_i$, we average the results of
Fit A and B, and include the difference in the error.

\Figrefthree{fpi-vs-m-III}{fK-vs-m-III}{mpisq-over-m-vs-m-III} show $\pi$ and $K$ decay constants
and $\pi$ masses from the corresponding preferred NNNLO fit to data subset {\it III}\/ (``Fit C'').
 The LO and NLO parameters here are fixed, up to their statistical errors, by Fit B. 
Fit C has a chi-square of 383 for 368 degrees of freedom (CL=0.28); including the 
priors gives a chi-square of 418 for 402 degrees of freedom (CL=0.28). 
Central values of 
$f_K$, $f_\pi$, $f_K/f_\pi$, $m_s$, $\hat m$, $m_s/\hat m$, and $m_u/m_d$ are taken from this fit; while Fits A and B
are included as alternatives in estimating systematics.

Using the volume-dependence from
NLO \schpt, \eqs{chiral-log1}{chiral-log2}, the leading  finite volume 
effects can be removed from our data.
Such effects are rather small
to begin with ($<0.9\%$ on $M^2_{P^+_5}$ and $<1.4\%$ on $f_{P^+_5}$, based on fit B), and
this calculated volume-dependence 
is consistent with simulation results in the one case where two different 
volumes are available \cite{MILC_SPECTRUM2}.
One-loop finite volume effects have been removed from
the points and lines shown in 
\figrefto{fpi-vs-m-II}{mpisq-over-m-vs-m-III}. Possible residual errors from higher
order finite volume effects are discussed in \secref{FINITE-VOLUME}.

To extract continuum results for masses or decay constants from Fits A, B, or C, we first set the taste splitting
and the taste violating parameters to zero.  We then extrapolate the remaining, physical parameters to the continuum
linearly in $\alpha_S a^2$. For central values, we assume that the 
ratio of this quantity between fine and coarse lattices is 0.427
(see \eq{alpha-a2}).  
For the LO parameters we take the range of the ratio
to be 0.398 to 0.441 in estimating the systematic error, as in the discussion of \eq{alpha-a2}.  
But for all other parameters, we expand the range to
0.30 to 0.441 in recognition of the fact that the fits do not distinguish 
generic discretization errors, $\cO(\alpha_S a^2)$,
from taste-violating errors, $\cO(\alpha_S^2 a^2)$. We thus 
must include the range discussed following \eq{alpha2-a2}.

\tabref{fit-params} shows the central values 
of the continuum extrapolated parameters for Fits A, B, and C.
Note that the statistical errors on most of the parameters are
quite large.  This seems to be a consequence of the ``flat directions''
in the fitting function, as described in \secref{params}:  small fluctuations
in the data can produce large variations in the parameters.  However,
because of the correlations among the parameters, the statistical errors
of interpolated or extrapolated decay constants and masses are small,
comparable to those of the raw data.

\begin{table}[t]
\begin{center}
\setlength{\tabcolsep}{1.5mm}
\begin{tabular}{|c|c|c|c|}
\hline
&                 Fit A       & Fit B   &      Fit C   \\
\hline
$r_1\; \mu$&$\phantom{+}                        	 5.579(515) $&$\phantom{+}	 4.549(387) $&$\phantom{+}	 4.462(227)   $\\
$r_1\; f$  &$\phantom{+}                        	 0.186(14) $&$\phantom{+}	 0.185(21) $&$\phantom{+}	 0.185(15)   $\\
$(2L_6-L_4)\times10^3$&$\phantom{+}      	 0.244(156) $&$\phantom{+}	 0.705(157) $&$\phantom{+}	 0.763(89)   $\\
$(2L_8-L_5)\times10^3$&$       	 -0.038(96) $&$	 -0.330(113) $&$	 -0.392(76)   $\\
$L_4\times10^3$&$\phantom{+}               	 0.178(231) $&$\phantom{+}	 0.200(340) $&$\phantom{+}	 0.186(239)   $\\
$L_5\times10^3$&$\phantom{+}               	 1.834(247) $&$\phantom{+}	 1.949(263) $&$\phantom{+}	 2.054(179)   $\\
$\beta^{(m)}_1$&$              	 -0.566(80) $&$	 -0.279(43) $&$	 -0.260(79)   $\\
$\beta^{(m)}_2$&$              	 -0.314(195) $&$	 -0.994(193) $&$	 -1.050(124)   $\\
$\beta^{(m)}_3$&$\phantom{+}              	 0.208(36) $&$\phantom{+}	 0.149(33) $&$\phantom{+}	 0.145(19)   $\\
$\beta^{(m)}_4$&$              	 -0.281(22) $&$	 -0.150(13) $&$	 -0.081(9)   $\\
$\beta^{(m)}_5$&$\phantom{+}              	 0.554(336) $&$\phantom{+}	 1.658(367) $&$\phantom{+}	 1.861(310)   $\\
$\beta^{(f)}_1$&$\phantom{+}              	 0.237(62) $&$\phantom{+}	 0.188(36) $&$\phantom{+}	 0.257(85)   $\\
$\beta^{(f)}_2$&$\phantom{+}              	 0.135(142) $&$\phantom{+}	 0.131(209) $&$\phantom{+}	 0.128(141)   $\\
$\beta^{(f)}_3$&$\phantom{+}              	 0.189(36) $&$\phantom{+}	 0.182(33) $&$\phantom{+}	 0.150(54)   $\\
$\beta^{(f)}_4$&$              	 -0.059(36) $&$	 -0.058(24) $&$	 -0.062(15)   $\\
$\beta^{(f)}_5$&$              	 -0.098(286) $&$	 -0.115(415) $&$	 -0.109(389)   $\\
$\sigma^{(m)}$&             	 ---   &	 ---   &$	 -0.130(44)   $\\
$\rho^{(m)}$&               	 ---   &	 ---   &$\phantom{+}	 0.049(52) $\\
$\sigma^{(f)}$&             	 ---   &	 ---   &$	 -0.063(165)   $\\
$\rho^{(f)}$&               	 ---   &	 ---   &$	 -0.132(69)   $\\
\hline
\end{tabular}
\caption{Continuum extrapolated fit parameters 
for Fits A, B, and C, which are on mass subsets 
{\it I}\/, {\it II}\/, and {\it III}\/,
respectively.  No extrapolation errors are included; we show statistical errors only.
See \protect{\secref{schpt}} for definitions of the parameters.
\label{tab:fit-params}
}
\end{center}
\end{table}

Once the continuum chiral parameters are obtained, we set valence and sea quark masses
equal and obtain ``full QCD'' formulas for $m^2_\pi/(2\hat m')$,\ \ 
$m^2_K/(\hat m'+m'_s)$,\ \  $f_\pi$,\ \  and $f_K$ in terms of arbitrary quark masses $\hat m'$ and $m'_s$.  
The \tmpcyan lines in
\figrefto{fpi-vs-m-II}{mpisq-over-m-vs-m-III} 
show these as a function of $\hat m'$, with $m'_s$  held fixed at the value of the
simulation sea quark mass on the fine lattices.
As a consistency check,  we also show
in each case the result from extrapolation of full QCD points to the continuum at fixed
quark mass (\tmpcyan fancy squares). To generate these points, we use the chiral fits only to interpolate the coarse
data so that it corresponds to the same physical quark masses as the fine data.
There are just two such points in each plot because we have just two runs with different
sea quark masses on the fine lattice.  
Since discretization errors come both from taste violations and generic errors,
there is an ambiguity in the extrapolation used
to find these points.  We have assumed that
taste violations dominate and have extrapolated linearly in $\alpha^2 a^2$, \ie with a ratio of 0.35
in $\alpha^2 a^2$ between coarse and fine (see discussion following \eq{alpha2-a2}).
This agrees both with our order of magnitude estimates (taste violations $\approx\! 6\%$;
generic errors $\approx\!2\%$) and with a detailed 
analysis below (\secref{continuum-extrap}).

To proceed further we need to know the physical values of the quark masses.  These can be
obtained from \figref{mpisq-over-m-vs-m-II} or \figref{mpisq-over-m-vs-m-III}  by finding those
values of $\hat m$ and $m_s$ that give the $\pi$ and $K$ their physical QCD masses in the isospin
limit, $m_{\hat \pi}$ and $m_{\hat K}$ (defined in \eq{masses-EM}).  An iterative procedure is required
because both meson masses depend on both quark masses, although the dependence of $m_{\hat \pi}$ on
$m_s$ is mild, since $s$ only appears as a sea quark.  The nature of this extrapolation/interpolation
is most clearly seen in \figref{msq-vs-m-III},\footnote{An almost identical plot, 
but without the extrapolation to find $m_u/m_d$ appeared in Ref.~\protect{\cite{strange-mass}}.}
which is again Fit C, but now shown for squared meson masses
as a function of light quark mass.
For clarity, we plot data 
with only one choice of sea quark masses for the coarse and fine sets; the variation with light sea 
quark mass is quite small on this scale.
The \tmpred dashed lines show the fit after extrapolation to the continuum, going to
full QCD, and iteratively
adjusting the strange quark mass to its physical value, so that the pion and kaon reach their 
physical QCD values at the same value of $\hat m$.  

Note that nonlinearities in the data are quite small on the scale of \figref{msq-vs-m-III}. 
Linear fits to $m^2_{P_5^+}$  as a function of $m_x+m_y$
would change the physical quark mass values by only $2\%$ to $7\%$, 
depending on the range of quark masses included and whether or not we fit 
the decay constants simultaneously. 
(The correlation between masses and decay constants implies that the fits are correlated even
in this case, where they have no free parameters in common.)
However, the tiny statistical errors in our data imply that even small nonlinearities 
must be accurately represented in order to obtain good fits.
Indeed, linear fits have $\chi^2/{\rm d.o.f.}\! \sim\! 20$. For an example of the accuracy of 
a linear fit, see the lowest fit line (for Goldstone pions) in \figref{splittings}.

Once the physical $s$ quark mass $m_s$ is in hand, we can adjust the \tmpcyan lines in
\figrefto{fpi-vs-m-II}{mpisq-over-m-vs-m-III} to put the
strange mass at its physical value.  
This gives the dotted \tmpred lines.   Following the dotted \tmpred lines to the
physical light quark mass $\hat m$ gives our extrapolated results for $f_\pi$,
plotted as \tmpred fancy pluses in \figrefs{fpi-vs-m-II}{fpi-vs-m-III}.
The errors in the \tmpred fancy pluses are statistical only; the systematic
errors are shown separately. We also plot the ``experimental'' values of the
decay constants \cite{PDG}, where we put experimental in quotation marks to emphasize that the
decay constants are extracted from experiment using theoretical input and values of
CKM matrix elements, which themselves have uncertainties.\footnote{See \protect{\secref{results}} for additional 
discussion about the experimental value of $f_K$.}
 
\Figref{msq-vs-m-III}, also shows how $m_u$ is extracted:  
With $m_s$ and $\hat m$ determined, we can continue
the full QCD kaon line as a function of the light valence 
quark mass (holding the light sea quark mass fixed at $\hat m$)
until it reaches the value of $(m^2_{K^+})_{\rm QCD}$, \eq{mK+-EM}.  The continuation
is shown as a \tmpgreen dashed line.
For clarity, a magnified version of the relevant region is shown in \figref{msq-vs-m-III-blowup}.
There is a slight change in the slope of the dashed line at $m^2_{\hat K}$ because, below this mass (\tmpgreen
section),
the light sea quark mass is no longer changing.
Above this point, light valence and sea masses change together.
The ratio of the $x$ coordinates of the points where the kaon 
line intersects the physical $(m^2_{K^+})_{\rm QCD}$
and $m^2_{\hat K}$ values is $m_u/\hat m= 0.60$.

Given $m_u$, we can extract $f_K$, which is really $f_{K^+}$, not
the decay constant ``$f_{\hat K}$'' of an isospin-averaged kaon.
After extrapolating the chiral parameters to the continuum, we
set valence and sea strange quark masses equal to the physical $m_s$.
We now make a two-step extrapolation in the light quark mass,
as shown in \figref{fK-vs-m-III}. 
We first set the light sea mass equal to the light valence mass $m_x$ and extrapolate in 
$m_x$ down to $\hat m$ (\tmpred dotted line).  We then fix the sea quark mass
at $\hat m$ and continue the extrapolation in valence mass $m_x$ to $m_u$ (short \tmpgreen
dotted line). It is clear from the size of the systematic errors that this final short
extrapolation does not at present produce a significant change in $f_K$.  However, 
this distinction between $f_{K^+}$ and $f_{\hat K}$
will become more important as lattice computations
improve.

\subsection{Discussion}\label{sec:discussion}

Good fits are not possible without the
taste-violating terms in \schpt.  
\Figref{fpi-vs-m-no-taste-viols} corresponds exactly to \figref{fpi-vs-m-II} except that
in \figref{fpi-vs-m-no-taste-viols} the taste splittings in meson masses have been
set to zero and the taste-violating chiral parameters 
($\delta'_V$, $\delta'_A$, $L'$, and $L''$) have been eliminated.
Thus the fit in \figref{fpi-vs-m-no-taste-viols} is to
the ``continuum'' NNLO \chpt\ form,\footnote{As in the \schpt\ case, NNLO chiral logarithms are not
included.} which has only four fewer parameters  than the \schpt\ fit in \figref{fpi-vs-m-II} (\ie a total of 36).
We put ``continuum'' in quotation marks here because generic variations in physical parameters between coarse and
fine lattices are still allowed.  Further, even if we allow these generic variations to
be arbitrarily large, instead of the $\approx 2\%$ variation permitted in the standard fits,
we cannot obtain good fits without including the taste violations.
The fact that continuum \chpt\ fits are so poor reassures 
us that the good fit in \figref{fpi-vs-m-II} is not
simply a trivial consequence of having a lot of fit parameters --- one has to get the physics right.
Other test fits described below, such as fitting without the nonanalytic terms in the fit function
(\secref{chiral-logs}), give additional
reassurance, since they have equal or comparable numbers of parameters to Fit B but cannot describe the lattice data.

\subsubsection{Continuum extrapolation}
\label{sec:continuum-extrap}

For the decay constants, our preferred method of continuum extrapolation is to
extrapolate the chiral fit parameters, as described in \secref{fit-results}.  
An alternative method is to determine decay
constants in physical units at fixed lattice spacing, and then attempt to extrapolate these quantities.
There are two ways to find fixed lattice-spacing values: (1) simply use the complete
chiral fits to extract the decay constants on the coarse or fine sets, or 
(2) first set the taste-violating parameters ($\delta'_V$, $\delta'_A$, $L'$, $L''$) and splittings in the fit to zero,
and then extract the decay constants for each set.  The advantage of method (2) is that,
once taste-violations have been set to zero, remaining discretization errors should be dominantly of
the generic type, so we may extrapolate to the continuum linearly in $\alpha_S a^2$. In method (1), the decay
constants at fixed $a$ have both generic and taste-violating discretization errors, so there is an
ambiguity in extrapolating them to the continuum.  

\Figref{f-vs-a2} shows the decay constants at fixed $a$ using both methods, and the
various extrapolations to the continuum.
Once taste violations are removed (method (2)), the remaining discretization errors are quite small,
giving a $1\%$--$2\%$ change between coarse and fine, as expected.  Further, because this change is small,
it would not make a significant difference in the extrapolated answer if we were to replace the 
$\alpha_S a^2$ extrapolation with an $\alpha^2_S a^2$ one.  With method (1), the change from coarse to
fine is $5\%$--$6\%$, roughly the same size as the difference in the raw data between
these lattices: See \figrefto{fpi-vs-m-II}{mpisq-over-m-vs-m-III}.\footnote{The difference between
coarse and fine raw data appears to be significantly greater than 6\%  in \protect{\figref{fK-vs-m-III}},
but this is because the raw points also have different strange valence quark masses.}
There is also a noticeable difference ($\approx\!1.2\%$) in the extrapolated results with method (1), 
depending on whether $\cO(\alpha_S a^2)$ or $\cO(\alpha^2_S a^2)$ errors are assumed. 
This ambiguity would grow to $\approx\!2.2\%$ if we used the full allowed range of values for the
$\cO(\alpha_S a^2)$ or $\cO(\alpha^2_S a^2)$.  For $f_\pi$
and $f_K$, we therefore do not consider method (1)
results among the possible alternatives in assessing the systematic error.  We
use the parameter-extrapolated version (\tmpred diamond and fancy plus in \figref{f-vs-a2}) for central values,
and include the results of method (2) when estimating systematic errors.  Note that we
are rejecting method (1) because of its inherent ambiguity, not because it disagrees with the
other methods of extrapolation.  

On the other hand, for the ratio $f_K/f_\pi$ and for quark masses ($m_u/m_d$ here, and $m_s$, $\hat m$ and $m_s/\hat m$
in Ref.~\cite{strange-mass}), the results change little with lattice spacing, so the ambiguity in method (1) is tiny
(much less than other systematic errors).  Therefore, in those cases we
include all three methods of extrapolation in our systematic error estimates.

The nice consistency of \figref{f-vs-a2} with our understanding of the sources of discretization errors is comforting.
However, we caution the reader that some aspects of this picture are dependent on the assumptions that went into
our fits.  In particular, we have inserted Bayesian priors to enforce a (1 standard deviation) 
constraint that LO chiral parameters
change by at most $2\%$ from coarse to fine lattices.  When we 
relax this constraint to $4\%$, the difference of method (2)
results from coarse to fine increases to $3\%$--$4\%$; while method (1) differences remain at about $6\%$.
The relaxed version of the chiral fit is included in the systematic error analysis.  With just
two lattice spacings, however, we cannot remove this constraint entirely without the fit becoming
unstable, as has been  emphasized in \secref{params}.  
Instead, the key point here is that we {\it can}\/ obtain good fits of our 
entire data set based on the theory of taste violations, plus some smaller
generic errors.  This is to be contrasted  with our failed attempts to fit the data without including taste-violations,
even when generic errors are allowed to be arbitrarily  large.

\subsubsection{Evidence for chiral logarithms}
\label{sec:chiral-logs}
From \secref{NNLO}, we know that the coefficient of the chiral logarithm terms, the chiral
coupling $1/(16\pi^2f^2)$, is not
tightly constrained by the fits.  If the chiral coupling is allowed to be a free parameter,
fits to higher valence masses prefer values of $f$ near
$f_K$ in the coupling; while fits to lower masses prefer $f$ closer to $f_\pi$. On the other hand,
acceptable fits can be obtained for all our mass ranges with the chiral coupling fixed
anywhere between its value for $f_\pi$ and that for $f_K$.  
Given this freedom, can we claim that chiral logarithms are observed
at all? To answer this question, we consider a variety of alternative fits without chiral logarithms.

First of all, since \figref{msq-vs-m-III} appears so linear to the eye, one can ask whether a simple
linear fit would work.  The answer is no: linear fits of $m^2_{P_5^+}$ \vs $m_x+m_y$ 
have chi-square per degree of freedom $\sim\! 20$. 
The point is that the statistical errors in the data are so small, and the correlations are well enough
determined, that the small departures from linearity must be accurately represented by the fits.
These deviations are seen more clearly in \figrefs{mpisq-over-m-vs-m-II}{mpisq-over-m-vs-m-III}, where the
valence quark masses are divided out.  Similarly, even though the apparent curvature in the
decay constant data is not large (see \figrefs{fpi-vs-m-II}{fpi-vs-m-III}),  linear fits of  $f_{P^+_5}$
\vs $m_x+m_y$ are also terrible, with chi-square per degree of freedom $\sim\! 25$.

We next check whether the data can be fit by including all the higher order (non-linear) analytic terms,
but omitting the chiral logarithms.  
With the chiral logarithm functions $\ell$, \eq{chiral-log1}, and $\tilde \ell$, \eq{chiral-log2}, set to zero,
we attempt a fit directly comparable to our NNLO Fit B on mass subset {\it II}\/. This fit has 38 free parameters,
which is 2 less than Fit B because the taste-violating hairpin parameters $\delta'_A$ and $\delta'_V$ decouple
when $\ell=\tilde \ell = 0$.  
Despite the large number of parameters, the fit is very bad,
with $\chi^2/{\rm d.o.f.}=7.38$ for 202 degrees of freedom; ${\rm CL}<10^{-194}$.

Be\'cirevi\'c and Villadoro \cite{Becirevic:2003wk}, 
have pointed out that, for some current simulations on small volumes, the finite
size effects are much more important than the actual chiral logarithms.  We would not expect that
to be the case here since the lattice volumes are relatively large ($L\ge2.5\;{\rm fm}$) and
finite volume effects here are small (at most 1.4\% --- see \secref{fit-results}).  To check this
expectation, we first removed the finite volume effects from the data using Fit B, and then
fit again to the 38 parameter form with $\ell=\tilde \ell = 0$.
This fit is improved over the previous one, but still quite bad:
$\chi^2/{\rm d.o.f.}=3.08$ for 202 degrees of freedom; ${\rm CL}<10^{-43}$.
Our conclusion is that the effect of the staggered chiral logarithms is in fact observed in our data.

One can ask whether the finite volume effects are also directly observed.  The answer seems
to be yes:  A 40-parameter fit leaving out these effects (setting $\delta_1$ and $\delta_3$ to zero
in \eqs{chiral-log1}{chiral-log2}) but otherwise identical to Fit B has 
$\chi^2/{\rm d.o.f.}=1.95$ for 200 degrees of freedom; ${\rm CL}\approx 2\times 10^{-14}$.

\subsubsection{Are the lattice masses light enough for \schpt\ to be applicable?}
\label{sec:applicability}
To discuss this question, we first have to say what we mean by the 
physical quark masses at fixed $a$. For current purposes
we define the physical values of the lattice masses, $am_s$ and $a\hat m$, by method
(2), \ie as the quark
masses that give the pion and kaon their physical masses 
when all taste splittings and taste-violating chiral parameters are set to zero. 
This gives $am_s\approx 0.0457$, $a\hat m \approx 0.00166$ on
the coarse lattices and $am_s\approx 0.0289$, $a\hat m \approx 0.00105$
on the fine. We have not made a detailed study of the errors in these numbers,
but systematic errors are $\sim\! 6\%$, and statistical errors are 1\% or less.
We could alternatively define the physical $m_s$ at a given lattice spacing by method (1),
\ie as
the masses that give the physical mesons masses directly, including all effects of
taste violations in the chiral loops. The latter values of quark masses were quoted in
\cite{strange-mass,MILC_SPECTRUM2} and are about   
$\sim\!15\%$ smaller on the
coarse lattices and $\sim6\%$ smaller on the fine lattices than the ones quoted above. We choose the
method (2) definition here because we are going to be adding on the 
taste-splittings to meson masses explicitly.  We note that the difference between the methods
becomes becomes small, $1\%$--$2\%$, once we extrapolate to the continuum, and is included in
the systematic errors quoted in Ref.~\cite{strange-mass}.

We now consider meson masses on the coarse lattice for subset {\it II}\/.  This is the worst
case because subset {\it II}\/ contains
the highest valence quark masses for which we have applied a fully chiral
description, and the coarse lattices have the largest additional contributions to meson 
masses from taste violations. The valence masses here obey $m_x+m_y\le 0.7 m'_s \approx 0.77\, m_s$.
Since the smallest valence mass is $0.1\, m'_s$, the largest is $0.6\,m'_s \approx 0.66\, m_s$;
while the largest sea quark mass in the simulation is  $m'_s \approx 1.09\,m_s$.
For the following estimates, we assume linear dependence of squared meson masses
on quark masses, and take $486\,\MeV$ as the mass a ``kaon'' would have 
in the absence of electromagnetism and if
the light quark were massless.  
In other words, we use $\mu m_s=(486\,\MeV)^2$.   
This comes from \eq{masses-EM} and the ratio $m_s/\hat m=27.4$  \cite{strange-mass}.
Then the largest valence-valence Goldstone
meson mass in coarse subset {\it II} is $425\,\MeV$. Adding on the largest splitting
(the taste-singlet case) gives $623\,\MeV$; while an ``average'' taste splitting
(see discussion after \eq{chia2def}) gives $551\,\MeV$.  We do not think it unreasonable
to expect \schpt\ to work in this mass range, although it is not surprising ---
considering the small statistical errors of the data --- that NNLO
terms are needed.  Further, the comparison
with subset {\it I}\/ results is a good check, because the corresponding masses there are significantly lower:
$321\,\MeV$, $557\,\MeV$, and $475\,\MeV$, respectively.

Mesons with one or two sea quarks also appear in chiral loops. These are generally comparable
to the masses just discussed. But they are significantly larger
when the sea quark is an $s$, which is exacerbated by the fact that
the simulation value, $m'_s$, is larger than the physical mass.  On subset {\it II}\/,
this largest valence-sea (Goldstone) mass is $642\,\MeV$. 
Adding on the biggest splitting
gives $787\,\MeV$. This taste-singlet meson enters with a factor $1/16$ in the sum over
tastes; the ``average'' taste version is $731\,\MeV$.  On subset {\it I}\/, the masses become
$578\,\MeV$, $736\,\MeV$, and $676\,\MeV$, respectively.
Further, there are sea-sea contributions, which are independent of
the valence mass subset.  The most relevant here is
the taste-singlet $\eta$. Its mass depends (mildly) on the light quark
sea masses, but is $\sim\!765\; \MeV$ including splitting.  In addition there are $\eta'$-like
particles in the taste axial and vector channels whose masses are comparable to the taste-singlet $\eta$
but have smaller ($\cO(a^2)$) couplings.

The meson masses involving the $s$ sea quark are admittedly quite high 
to expect that even NNLO \chpt\ will be accurate. For example,
the largest mass mentioned above, $765\; \MeV$, corresponds to a $\chi_q$ value, \eq{chiqdef}, of
0.43. This suggests an error from neglected terms of order $(0.43)^3=8\%$.
But, just as for the valence subset {\it III}, the issue here for decay constants and quark masses is
only to get a good interpolation to the physical $s$ quark mass.  Indeed,
if the $s$ sea quark in the simulation had been chosen at the {\it a posteriori} determined physical mass, 
we would not have needed to use \chpt\ for the $s$ at all, but could use a $SU_L(2)\times SU_R(2)$ \chpt\ for the
light quark extrapolation.  The systematic error on the coarse lattice from adjusting $m'_s=1.09 m_s$
to $m_s$ may be crudely estimated as $(0.43)^3\times[(1.09)^3-1]\approx 2\%$.  Since $m'_s$ is closer
to $m_s$ on the fine lattice, some of this error will be extrapolated away when we go to the continuum limit.
On the other hand, chiral coefficients (at NLO and NNLO) that involve the sea quarks 
are not fit accurately because the ``lever arm'' is small: the sum of the sea quark masses changes by
less than a factor of 2 over our entire range of coarse lattices and only by 30\% for the fine
lattices.  A more reliable estimate of the error in adjusting the $s$ quark mass comes from
considering the range in results over the full list of alternative mass subsets, chiral fits, and continuum extrapolations.
It can be as large as half the total chiral error in our results for 
decay constants and quark masses (see line a1 in \tabrefs{decay-syst}{muomd-syst});
the remaining error comes from extrapolating in the light quark mass.

For the $L_i$, the situation is somewhat different.
Missing higher order terms in the $SU_L(3)\times SU_R(3)$ expansion mean that there is spurious analytic
dependence on the light quark masses that increases as meson masses get larger.
Here we are missing the NNLO chiral logarithms, so those terms determine the size
of the errors.  Letting $M$ be a generic meson mass, the absent terms are of order
$M^4\log(M^2)/(8 \pi^2 f_\pi^2)^2$.  Putting in the largest meson masses discussed above,
results in an estimate of the absolute error in the $L_i$ of a few times $10^{-4}$.
This is indeed the size of the errors we observe when we consider all the alternative
chiral fits discussed above and/or restrict the valence masses to subset {\it I}\/ instead of
subset {\it II}\/ (\secref{results}).  The low energy constant $2L_6-L_4$ may be an
exception: Since the errors in it are large on this scale, $\sim\!4\times 10^{-4}$, 
they may also not be very reliable. Difficulty in extracting  $2L_6-L_4$ is
again related to the small lever arm for the sea quark dependence. This can be seen in \figref{mpisq-over-m-vs-m-II-blowup},
an enlargement of a small region of \figref{mpisq-over-m-vs-m-II}.  As the sea quark mass is
changed, the differences are small --- comparable to statistical errors --- and not monotonic. Contrast this
with the monotonic sea quark dependence seen for $f_\pi$ in \figref{fpi-vs-m-II}.
A coarse simulation now in progress, with all three sea quark masses at
about $0.66m_s$, should help to reduce significantly the error in the sea quark mass dependence.

\subsubsection{Convergence of \chpt}
\label{sec:chiral-converge}
\Figref{chpt-converge-fpi} shows the convergence of
$SU(3)_L\times SU(3)_R$ \chpt\ for $f_\pi$ and $f_K$. All chiral parameters in this plot have been extrapolated to the
continuum. The NLO terms contribute $\!20\%$.  This is true
even for $f_\pi$, because $m_s$ does not vanish in the chiral limit.  The
convergence of $SU(2)_L\times SU(2)_R$ \chpt\ for $f_\pi$ can also be extracted from this
plot by starting with the ``chiral limit'' line instead of the ``LO''
line lowest order contribution to $f_\pi$. Note that the 
\tmpmagenta \tmpand \tmpcyan fancy squares, which are included as a consistency check,
are the only full QCD points that we can extrapolate
to the continuum at fixed mass. We have
lighter valence quarks on the fine lattices, and lighter valence and sea
quarks on the coarse lattices.  All such partially quenched points are
included in the fits that produce the lines in the plot.  \Figref{chpt-converge-fpi} comes
from the fit to data subset {\it II}\/ (Fit B).  If instead we restrict the fit to data subset {\it I}\/ (Fit A),
the picture is virtually unchanged.

\Figrefs{chpt-converge-msq-II}{chpt-converge-msq-I} are the corresponding plots for meson masses 
($m^2_{P^+}/(m_x+m_y)$).
\Figref{chpt-converge-msq-II} is generated from the fit to data subset {\it II}\/ (Fit B); 
while \figref{chpt-converge-msq-I}
uses data subset  {\it I}\/ (Fit A).  Here there is a significant difference between the 
two plots, with the latter showing
much smaller higher order corrections in $SU(3)_L\times SU(3)_R$ 
\chpt\ than the former. The difference illustrates the poor control
over the low energy constant $2L_6-L_4$ (see \secref{applicability}): We get
$2L_6-L_4= 0.70(17)\times10^{-3}$ with Fit B, and
$2L_6-L_4 = 0.24(16)\times10^{-3}$ with 
Fit A.\footnote{The statistical errors here are slightly larger than those
in \protect{\tabref{fit-params}} because the statistical errors associated
with the continuum extrapolation are included.}
Because this parameter multiplies $2\hat m +m_s$, its effect
does not vanish in the chiral limit of the light quark mass $\hat m$.  Its variation is largely canceled by
differences  in the LO parameter $\mu$ and the NNLO parameters $\beta_2^{(m)}$ and $\beta_5^{(5)}$, \eq{m-NNLO},
so that the full NNLO fit line and the extrapolated $\pi$ and $K$ values are quite close in the two fits.
This means that the ambiguity in $2L_6-L_4$ and the LO term is largely irrelevant 
to the extraction of quark masses (and, indirectly, decay constants);
the variation between the fits is of course included in systematic errors estimates of these quantities.

The qualitative expectation from \chpt\ is that coefficients in the expansion should be
$\cO(1)$ when we use the dimensionless expansion parameters $\chi_q$, \eq{chiqdef}.
Both Fit A and Fit B pass this test (the largest coefficient in either is 
$\beta^{(m)}_2\approx 1.66$ in Fit B --- see \tabref{fit-params}),
so we must accept the large systematic effect on $2L_6-L_4$ as inherent in the current data set.
Indeed, the size of the difference in the LO term, $\mu$,  between the two fits
is reasonable, given that the fits prefer a chiral coupling $1/(16\pi^2f^2)$ with $f$ moving from
$\approx\! f_\pi$ to $\approx\! f_K$ as the quark masses rise toward $m_s$ 
(see comments about the parameter $\omega$ in \secref{NNLO}).  Both fits here fix $f=f_\pi$, but Fit
B is effectively able to reduce the effect of the chiral coupling by reducing the LO parameter $\mu$
and compensating by increasing the NLO parameter $2L_6-L_4$.  Since the difference between $f^2_\pi$ and $f_K^2$
is more than $40\%$, the $\sim\!20\%$ difference between the $\mu$ from the two fits is not unexpected.
New simulations with lighter strange sea quark masses will allow us to take all quark masses deeper into
the chiral regime, as well as greatly increase the lever arm on the $m_s$ dependence, and 
should help to resolve this issue.

\subsubsection{Mass dependence of renormalization scheme}
\label{sec:mass-dependence}

Strictly speaking, it is incorrect to
use a scale determined by a quantity like $r_1$ or the $\Upsilon^\prime$--$\Upsilon$ mass difference
in chiral fits to lattice data,
since such quantities themselves have some (small) sea quark mass dependence
not included in chiral perturbation theory \cite{Sommer:2003ne}. We have investigated
this effect by changing to a mass-independent renormalization scheme:   Instead of fixing
the (relative) lattice scale for particular sea quark mass values from $r_1/a$ at those values,
we can use, on each lattice, the value of $r_1/a$ after extrapolation to the physical 
mass values\footnote{Here we consider both the quark mass 
values determined by method (1) and those determined by method 
(2) --- see \protect{\secref{applicability}}.} using the fit \eq{R1_FIT_EQ}. This produces the same value
of $a$ on all the coarse (or, separately, fine) lattices, independent of the sea quark masses.
We include the difference between the mass independent scheme and our standard approach in our assessment of
systematic errors.  

For decay constants and quark masses, the change in scheme is {\it a priori}\/ unlikely to
make much difference because the physical
point is unaffected --- all that may change is the extrapolation to it.  
Further, any low-order analytic dependence on sea quark mass introduced though $r_1$ would  automatically be
compensated by changes in the analytic chiral fit parameters. So the only problem would be due to
nonanalytic quark mass dependence, which is probably quite small because $r_1$ is a short-distance
quantity, at or near the perturbative region.  We thus consider the variation in scheme
simply as another alternative version of the chiral fits.  This means that it would affect the final
systematic error only if it produced the largest difference from the central value over all
the alternatives. In fact it is fairly small, as expected (see line a3 in
 \tabrefs{decay-syst}{muomd-syst}).

The situation is logically quite different for the low energy constants.  Here, analytic
sea quark mass dependence in $r_1$ would directly change the output values of 
$L_4$ and $2L_6-L_4$, which multiply sea quark masses.
We therefore consider the scheme dependence as a systematic error in its own right, and
add any error found in quadrature with other systematic errors.  In practice, however,
this effect is still smaller than other errors (see \tabref{Li-syst}).

\subsubsection{Residual finite volume effects}\label{sec:FINITE-VOLUME}

At the precision we are working (especially for $f_K/f_\pi$), it is important
to consider whether finite volume effects coming from terms beyond one-loop in \schpt\
could be non-negligible.  Indeed, 
Colangelo and Haefeli \cite{Colangelo:2004xr} have recently investigated 
such effects in full continuum QCD. For volumes and masses comparable
to those used here, they find large higher order 
corrections to the volume dependence, roughly 30\% to 50\% of the one-loop results. 

In asymptotically large volumes,
the finite-volume effects in \schpt\ are suppressed 
relative to those in continuum \chpt\ for the same (Goldstone) masses because most of the 
pions entering chiral loops have larger masses.
This can be easily seen in \eq{fpi-NF2}: the lightest (Goldstone) pion appears with
a weight 1/16 as large as in the continuum.  
However, at our current volumes, masses, and lattice spacings, the relation between \schpt\ and
\chpt\ finite volume effects is complicated, with the former just as likely to be
larger than the latter as smaller.
\Eqs{fpi-NF2}{eigenvalues-NF2}, together with our numerical
results for splittings (\tabref{splittings}) and the taste-violating hairpin parameters
(\eq{hairpin-results}, below), show how the asymptotic rule can be violated.  Since $\Delta_A$ is
the smallest splitting, and $\delta'_A$ is non-negligible and negative, 
$m_{\eta'_A}$ may not be much larger than the Goldstone pion mass. 
Then, due to the factors of 4 in \eq{fpi-NF2},
finite volume corrections coming from the $\eta'_A$ can be as large or
larger than the continuum corrections.
As $a\to0$, the term
$\ell(m^2_{\eta'_A})$ would be canceled by $\ell(m^2_{\pi_A})$, but this cancellation may not
be effective for finite volume effects at a given value of $a$, because the volume
effects are sensitive to small mass differences.  
Since one-loop finite volume effects on our lattices are comparable to those in the continuum,
we have no {\it a priori} reason to expect that our results are
protected from the higher order effects \cite{Colangelo:2004xr}
seen in the continuum.

Assuming that the higher order finite volume effects in \schpt\ are roughly the same size
as those in full continuum QCD when the volumes and the masses of the mesons 
in the loops are the same, we can estimate the resulting systematic
errors. For our data, the biggest one-loop finite volume effects appear when both the
valence masses and $\hat m'$ are small (giving a light $\eta'_A$ in the two-flavor case, 
\eq{eigenvalues-NF2}, or a light $\eta_A$ in the three flavor case, \eq{eigenvalues}).
The worst case occurs in the coarse run with $a\hat m'=0.007$;
the run with $\hat m'=0.005$ has smaller finite volume effects because 
$L\approx 3.0\;$fm there, instead of $L\approx 2.5\;$fm for other runs.
From the calculations in Ref.~\cite{Colangelo:2004xr},\footnote{We are
indebted to Gilberto Colangelo for providing us with the results for higher order effects in
$f_\pi$ and $f_K$ at the values of volume and meson mass relevant to our computations}
we estimate that the residual higher order finite volume effect is at most
$0.47\%$ in $f_\pi$, $0.24\%$ in $f_K$, and $0.23\%$ in $f_K/f_\pi$. 
More stringent bounds on the errors can be obtained by removing from the data set 
those points that have the largest finite volume corrections. Eliminating
8 of 240 points from mass subset {\it II}\/ 
(5 from coarse run $0.007/0.05$, 2 from coarse run $0.01/0.05$,
and 1 from fine run $0.0062/0.031$), we lower the largest one-loop finite volume
effect on $f_{P^+_5}$ or $M^2_{P^+_5}$ from $1.35\%$ to $0.81\%$.  Not surprisingly,
since the reduced data set retains most of the lowest valence mass points and all of the
lowest sea mass points, it
produces nearly identical results (within $0.05\%$) as the original set.  But a repeat
of the analysis using Ref.~\cite{Colangelo:2004xr} now bounds the residual finite volume
error by $0.29\%$ in $f_\pi$, $0.15\%$ in $f_K$, and $0.14\%$ in $f_K/f_\pi$.   
These are negligible compared to our other systematic
errors. In the future, however, as quark masses 
in staggered simulations decrease further, it will be necessary either to have a 
better handle on these higher order effects in \schpt\
or to go to significantly larger volumes.

\subsubsection{Fourth root of the determinant}\label{sec:fourth-root}
In order to eliminate the quark doubling that is still present in the
staggered action, the simulations here take the fourth root of the quark
determinant for each flavor in order to reduce the quark tastes from
4 to 1 per flavor.  There is apparently no ultra-local lattice action
that would correspond to the effective action that results from taking this
fourth root.  The possibility thus exists that physical non-localities
will remain in the continuum limit, potentially spoiling the description of QCD
by the staggered action.  The good agreement of the staggered results 
with experiment and with continuum chiral behavior plus understood
discretization effects (both in current and
previous work \cite{BIG_PRL,MILC_SPECTRUM2,MASON_ALPHA}) lead us to
believe that this is not a problem, but the question is not settled.  

The comparison of simulation data with \schpt\ forms allows us to make
a crude but somewhat more direct test of the fourth root trick.
\Eqs{m-NLO}{f-NLO}, as written, take into account the fourth root by dividing
each sea quark loop contribution by 4, to leave 1 taste per flavor.   
It is a simple exercise to generalize \eqs{m-NLO}{f-NLO} to make
the number of tastes remaining a free parameter.  We can then ask what number
of tastes per flavor is preferred by the simulation data. With a fit otherwise identical to
our standard NNLO fit, 
we find $1.44(15)$ for the preferred number of tastes per flavor on data 
subset {\it II}\/, and $1.28(12)$ on the lighter masses in subset {\it I},
where the errors are statistical only.  If we allow the chiral coupling 
to vary also (choice (3) fits, \secref{NNLO}), we get $1.35(18)$ 
and $1.22(14)$ on subsets {\it II}\/ and {\it I}\/, respectively. In the latter case, the coefficient $\omega$
corresponds to an $f$ in the chiral coupling that is about halfway between
$f_\pi$ and $f_K$ ($\omega\approx0.82(11)$), which is reasonable. Given that the fits in any case
do not tightly constrain the chiral logarithm terms
(see \secref{NNLO}), we consider these results satisfactory.

We note that there has been recent numerical \cite{NUMERICAL} and analytic \cite{ADAMS}
work indicating even more
directly that the fourth root trick should work as expected.
On the other hand,
there have been two other recent papers
that purport to show problems with
locality \cite{NONLOCAL}.  We do not believe the latter work is
worrisome because it does not 
take into account the taste structure of staggered quarks.  Instead of trying to 
project onto a single taste to find the fourth root of the determinant, those
papers look only at the fourth root of the Dirac operator itself.  That procedure,
in our opinion, is almost guaranteed to find a nonlocal result, just as it would in 
trying to reduce eight Wilson fermions to two, which certainly 
has an alternative, local solution.

\subsubsection{Taste violating hairpins}\label{sec:hairpin-results}
Before turning to our physical results, we quote the values of the
two taste-violating hairpin parameters coming from the fits. Together with the
splittings, \tabref{splittings}, these parameters appear in 
\schpt\ calculations for other physical quantities, such as heavy-light decay
constants \cite{CA_CB3}. Averaging values from Fits A and B, we find, on the coarse
lattices:
\begin{eqnarray}\label{eq:hairpin-results}
r_1^2 a^2\delta'_A &=& -0.28(3)(5) \nonumber \\
r_1^2 a^2\delta'_V &=& -0.11(8)({}^{+21}_{-4}) \ ,
\end{eqnarray}
where the errors are statistical and systematic, respectively. The latter error comes
from the variation over all acceptable chiral fits on 
mass subsets {\it I}\/ and {\it II}\/.  The parameter $a^2\delta'_A$
is comparable in size to the taste-violating splittings (\tabref{splittings});
while $a^2\delta'_V$ is consistent with zero but poorly determined.
The values of $a^2\delta'_A$ and $a^2\delta'_V$ on the fine lattices
are not fit separately but are constrained to be 0.35 times as large for
central-value fits. (See discussion following \eq{alpha2-a2}.)

\section{Final Results and Conclusions}
\label{sec:results}

The central values and error estimates for $f_\pi$, $f_K$, and $f_K/f_\pi$ are collected
in \tabref{decay-syst}.  Central values come from Fit C (mass subset {\it III}\/). The scale
errors are found by repeating the analysis after moving our value $r_1=0.317$ fm by plus or
minus one standard deviation ($\pm 0.007$ fm); see \secref{simulations}. The change in decay
constants under this variation in the scale  is slightly less than the nominal $2.2\%= 0.007/0.317$.
This is due to a cancellation coming from the corresponding readjustment of the quark masses
needed to give the mesons their physical masses.

\begin{table}[t]
\begin{center}
\setlength{\tabcolsep}{1.5mm}
\def\arraystretch{.8}
\begin{tabular}{|c|c|c|c|}
\noalign{\vspace{0.5cm}}
\hline
&$f_\pi$  & $f_K$ & $f_K/f_\pi$  \\
\hline
  central value &  $129.46$       & $156.63$  & $1.2099$ \\
\hline
\multicolumn{4}{|c|}{errors}\\
\hline
 statistics & $\phantom{+}0.87$ & $\phantom{+}0.98$ & $\phantom{+}0.0042$  \\
\hline
 scale  &     $+2.35$ & $+2.58$ & $+0.0027$ \\
        &     $-2.36$ & $-2.51$ & $-0.0020$ \\
\hline
 (indirect) EM effects &  $\phantom{+}0.01$ & $\phantom{+}0.10$  & $\phantom{+}0.0009$ \\
\hline
 chiral/continuum & $+2.37$ & $+2.19$  & $+0.0125$ \\
extrapolation     & $-2.58$ & $-2.59$  & $-0.0112$ \\
\hline
\hline
\multicolumn{4}{|c|}{chiral/continuum error ``slices''}\\
\hline
 a &                $+0.79$ & $+0.60$ & $+0.0093$ \\
   &                $-2.50$ & $-1.84$ & $-0.0075$ \\
\hline
 a1 &               $+0.35$ & $+0.27$ & $+0.0075$ \\
    &               $-1.05$ & $-0.42$ & $-0.0037$ \\
\hline
 a2 &               $+0.63$ & $+0.02$  & ----      \\
    &               $-0.56$ & $-1.20$  & $-0.0057$ \\
\hline
 a3 &                ---    &  ---     & $+0.0024$ \\
    &               $-1.11$ & $-1.05$  &   ---     \\
\hline
\hline
 b &                $+2.19$ & $+2.02$  & $+0.0089$ \\
   &                $-1.33$ & $-1.89$  & $-0.0095$ \\
\hline
 b1 &               $+0.12$ & $+0.09$  & $+0.0017$ \\
    &               $-0.52$ & $-0.62$  & $-0.0019$ \\
\hline
 b2 &               $+0.69$ & $+2.00$  & $+0.0089$ \\
    &               $-0.63$ & $-1.60$  & $-0.0065$ \\
\hline
 b3 &               $+0.39$ & $+0.89$  & $+0.0032$ \\
    &               ---    &  ---  &  --- \\
\hline
\hline
 c &                $+0.27$ & $+0.26$  & $+0.0015$ \\
   &                $-2.52$ & $-2.79$  & $-0.0007$ \\
\hline
\end{tabular}
\caption{Central values and error estimates for $f_\pi$, $f_K$, and $f_K/f_\pi$.
All errors are absolute amounts, not percentages.  Decay constants and their errors are
in $\MeV$. Unsigned errors are taken as symmetric. 
The chiral/continuum error ``slices'' show variation under reduced sets of
possible alternative fits/extrapolations; see text. 
\label{tab:decay-syst}
}
\end{center}
\end{table}

The (indirect) EM errors just come from changes in our results for quark masses due to the assumed range
of $\Delta_E$ (see \secref{EM}); clearly this effect is very small.  
Direct EM effects that pertain to the comparison of decay constants with experiment
are much larger, of order several percent --- see Ref.~\cite{PDG}.  However the direct
effects are not relevant here because $f_\pi$ and $f_K$ are defined in the absence of
electromagnetism. 

The chiral/continuum errors are found by taking the maximum deviation from the central value
over all versions of the chiral fits described in \secref{params} and \secref{mass-dependence}, and
all versions of the continuum extrapolations described in \secref{fit-results} and
\secref{continuum-extrap} (including ranges in assumptions about how $\alpha_Sa^2$ and
$\alpha^2_s a^2$ change from coarse to fine lattices --- see discussions
\eqs{alpha-a2}{alpha2-a2}),
as well as variation in the perturbative parameter 
$R_m$ described in \secref{perturbation-theory}.
Because the continuum and chiral extrapolations are connected within \schpt, it is not
meaningful to quote separate errors for each.  However, since a large number of alternatives
are considered here, we believe it  will  be helpful to the reader to report the variations in 
physical results as one moves along various ``slices'' through the alternatives. 
The slices shown in \tabref{decay-syst} are defined as follows:
\begin{itemize}

\item[]{a.}\ \ All alternative chiral fits on all mass subsets, but only with the preferred method of
continuum extrapolation (extrapolation of chiral parameters), and only with preferred values of the
ratios of $\alpha_S a^2$ and of $\alpha_S^2 a^2$ (\eqs{alpha-a2}{alpha2-a2}).

\item[]{a1.}\ \ Same as a, but restricted to mass subset {\it III}. This is mainly an estimate of the
errors involved in interpolating around $m_s$. 

\item[]{a2.}\ \ Same as a, but restricted to chiral fits where the chiral coupling $\omega/(16\pi^2 f_\pi^2)$
is allowed to vary. with $\omega=1.0\pm0.1$  (see \secref{NNLO}).

\item[]{a3.}\ \ Same as a, but restricted to fits where the scale is chosen in a mass-independent manner
(see \secref{mass-dependence}).

\item[]{b.}\ \ Alternative values of the ratios of $\alpha_S a^2$ and of $\alpha_S^2 a^2$ used in continuum
extrapolation and/or alternative method  of extrapolation (method (2) -- \secref{continuum-extrap})
and/or alternative value of $R_m$ (\secref{perturbation-theory}).  
The preferred chiral fit is kept (Fit C).

\item[]{b1.}\ \ Same as b, but restricted to the preferred value (0.35) 
of ratio of $\alpha_S^2 a^2$ for all taste-violating quantities.

\item[]{b2.}\ \ Same as b, but only the ratio of $\alpha_S^2 a^2$ is varied (in the range
$0.3$--$0.4$) and only for taste-violating quantities 
that are not directly measured ($\delta'_A$, $\delta'_V$, $L'$, and $L''$ --- see 
\secref{NLO}).  The preferred continuum extrapolation (extrapolation of chiral fit parameters)
is used.

\item[]{b3.}\ \ Same as b, but only $R_m$ is varied, and the 
preferred continuum extrapolation  is used.

\item[]{c.}\ \  Alternative method  of extrapolation (1) is used and ratio of $a^2$ 
varies over union of ranges of  $\alpha_S a^2$ and of $\alpha_S^2 a^2$ --- \secref{continuum-extrap}.

\end{itemize}
As discussed in \secref{continuum-extrap}, method (1) continuum extrapolation (slice c) is not included among
our systematic alternatives because of the large ambiguity in how to perform the
extrapolation. \tabref{decay-syst} shows, however, that it produces deviations comparable to the full
chiral/continuum extrapolation error.

We add in quadrature the signed errors from the chiral/continuum extrapolation, the scale determination, and from direct EM
effects,  giving a total positive and a total negative systematic error. We then take the larger of the two
as a final symmetric error.
Note that chiral extrapolation errors and scale errors contribute almost equally
to the systematic error on $f_\pi$ and $f_K$; 
while scale errors are unimportant for the ratio.
The final results for decay constants are:
\begin{eqnarray}\label{eq:f_results}
f_\pi & = &  129.5 \pm 0.9\pm 3.5 \; \MeV \nonumber\\
f_K & = &  156.6 \pm 1.0\pm 3.6 \; \MeV \nonumber\\
f_K/f_\pi  & = & 1.210(4)(13)\ ,
\end{eqnarray}
where the first error is statistical and the second is systematic.

In \secref{NNLO} we argued that fits that allowed the chiral coupling to vary by more than 10\% (``choice (3)''
fits with arbitrary $\omega$) should be excluded from the analysis. 
If we were to include all choice (3) fits in the systematic error analysis, the error on $f_\pi$ would
increase from $3.5\,\MeV$ to $4.3\,\MeV$; that on $f_K/f_\pi$ would increase from $0.013$ to $0.022$; while
that on  $f_K$ would be unchanged.

Our results are in good agreement with the experimental numbers
\cite{PDG}:  $ f_\pi = 130.7 \pm 0.4\, \MeV$,
$f_K = 159.8 \pm 1.5\, \MeV$,
 $f_K/f_\pi= 1.223(12)$.
Note that the experimental determination of
$f_K$ has a rather large error.  That is because it depends
not only on the precisely measured leptonic decay width of the kaon,
but also on $V_{us}$, which has a significant uncertainty.
The errors on our result for $f_K/f_\pi$
are small enough that one may turn the comparison around, and use our answer
together with the measured leptonic decay widths to constrain  $V_{us}$ \cite{Marciano:2004uf}.
With Eq.~(16) 
in Ref.~\cite{Marciano:2004uf}, $|V_{ud}|=0.9740(5)$, and the current result for $f_K/f_\pi$,
we obtain 
$$|V_{us}|=0.2219(26)\ .  $$
The error is completely dominated by current lattice errors,  which we have
added in quadrature. Neglecting $|V_{ub}|^2$, the unitarity relation is then
\begin{equation}\label{eq:unitarity}
|V_{ud}|^2+ |V_{us}|^2 =0.9979(15)
\end{equation}
The 2$\sigma$ violation
that comes from using the PDG value $|V_{us}|=0.2196(26)$ \cite{PDG} becomes  a
$1.4\sigma$ effect here.  We note also that our result is compatible with the
very recent KTeV determination \cite{Alexopoulos:2004sw}: 
$|V_{us}|=0.2252(8)(21)$.

The values for $f_\pi$ and $f_K$ in \eq{f_results} should be considered as updates of those 
presented 
in Ref.~\cite{BIG_PRL}. The current results are based on an expanded data set.  In addition, the
analysis in Ref.~\cite{BIG_PRL} was performed differently: The data was first extrapolated
to the continuum at fixed quark mass and then fit to continuum \chpt\ forms. \schpt\ was used only
in estimating the systematic error of the extrapolation procedure.  A correction for
finite volume effects could not be made with the older approach; instead a finite volume
error had to be included.  The present results and those in Ref.~\cite{BIG_PRL} 
agree within their respective systematic errors.

\begin{table}[t]
\begin{center}
\setlength{\tabcolsep}{1.5mm}
\def\arraystretch{.8}
\begin{tabular}{|c|c|}
\noalign{\vspace{0.5cm}}
\hline
&$m_u/m_d$  \\
\hline
  central value &  $\phantom{+}0.429$ \\
\hline
\multicolumn{2}{|c|}{errors}\\
\hline
 statistics & $\phantom{+}0.004$ \\
\hline
 scale  &     $\phantom{+}0.002$ \\
\hline
 EM effects & $+0.084$ \\
            & $-0.076$ \\
\hline
 chiral/continuum & $+0.012$ \\
extrapolation     & $-0.006$ \\
\hline
\hline
\multicolumn{2}{|c|}{chiral/continuum error slices}\\
\hline
 a &                $+0.012$ \\
   &                $-0.005$ \\
\hline
 a1 &               $+0.002$ \\
    &               $-0.005$ \\
\hline
 a2 &               $+0.012$ \\
    &               $-0.004$ \\
\hline
 a3 &               $+0.003$ \\
    &                ---     \\
\hline
\hline
 b &                $+0.004$ \\
   &                $-0.002$ \\
\hline
 b1 &               $+0.000$ \\
    &               $-0.002$ \\
\hline
 b2 &               $+0.000$ \\
    &               $-0.002$ \\
\hline
 b3 &               $+0.002$ \\
    &               ---      \\
\hline
\hline
 c &                  ---    \\
   &                $-0.002$ \\
\hline
\end{tabular}
\caption{Same as \protect{\tabref{decay-syst}}, but for $m_u/m_d$.
\label{tab:muomd-syst}
}
\end{center}
\end{table}

Errors for our direct determination of $m_u/m_d$ are shown in \tabref{muomd-syst}.
Adding the scale and chiral/continuum extrapolation errors in quadrature, and symmetrizing
as for the decay constants, we get the total
simulation error.
Our final result is
\begin{equation}\label{eq:muomd}
m_u/m_d = 0.43(0)(1)(8) \ ,
\end{equation}
where the errors are from statistics, simulation systematics, 
and direct EM effects, respectively.  
We have allowed for EM effects in a wide range
$0 \le \Delta_E=\delta_E\le 2$ (see \eqs{masses-EM}{mK+-EM}).
If instead we were to assume the result of Ref.~\cite{Bijnens:1996kk} ($\Delta_E=0.84\pm0.25$),  
we would obtain $m_u/m_d = 0.44(0)(1)(2)$.  
Including all choice (3) fits in the systematic error analysis would increase the simulation
systematic error from $0.01$ to $0.02$.

Even with the generous range of possible EM effects, \eq{muomd} clearly bounds $m_u$ away from zero.
An alternative way of expressing this is to determine the value of $\Delta_E$ that would
be required
in order to allow for $m_u=0$.  We find that it would take an absurdly large violation
of Dashen's theorem, $\Delta_E\approx 8.4$.

Values for quark masses at scale $2\,\GeV$, as well as the
ratio $m_s/\hat m$, were reported in Ref.~\cite{strange-mass}.
Since that work used the same lattice data, chiral fits, and error analysis as that
described above, we repeat the results here for completeness:
\begin{eqnarray}\label{eq:quark-mass-results}
m_s^\msbar &=&  76(0)(3)(7)(0)\;\MeV\ ,\nonumber \\
\hat m^\msbar  &=&   2.8(0)(1)(3)(0)\; \MeV\ , \nonumber \\
m_s/\hat m  &=&  27.4(1)(4)(0)(1)
\end{eqnarray}
where the errors are from statistics, simulation, perturbation theory,
and electromagnetic effects, respectively.  

Combining the current result for
$m_u/m_d$ with the perturbative mass renormalization calculated in \cite{strange-mass}
(or, equivalently, with $\hat m^\msbar$  in \eq{quark-mass-results}), we obtain:
\begin{eqnarray}\label{eq:mu-and-md}
m_u^\msbar =  1.7(0)(1)(2)(2)\;\MeV \nonumber \\
m_d^\msbar =  3.9(0)(1)(4)(2)\;\MeV  \ ,
\end{eqnarray}
where the errors have the same meaning as in \eq{quark-mass-results}, and the scale
is again $2\,\GeV$.  The separate EM errors in $m_u$ and $m_d$ are highly, and negatively,
correlated, and therefore consistent with the large EM error in $m_u/m_d$.

The results for $m_u/m_d$ and $m_s/\hat m$ in \eqs{muomd}{quark-mass-results}
appear inconsistent with the relation between $m_s/m_d$ and $m_u/m_d$ shown in Fig.~1 of Ref.~\cite{Kaplan:1986ru}.
However, that appears to be due to NNLO effects not included in \cite{Kaplan:1986ru}. 
Indeed, Amoros \et~\cite{Amoros:2001cp} obtain $m_u/m_d=0.46(9)$ with a  NNLO phenomenological
analysis.  Further, our results for the two ratios are consistent with the 
NNLO relation shown in Fig.~3 of Ref.~\cite{Amoros:2001cp}.  

Since $m_u$ is bounded well away from 0, the issue of the 
physicality of $m_u=0$ \cite{CREUTZ}
does not arise directly here.  However, should the
existence of non-perturbative, additive shifts in masses
proposed in Ref.~\cite{CREUTZ} be confirmed, there could be some 
lattice scheme dependence in the quark masses and
ratios in \eqs{quark-mass-results}{mu-and-md}.  
We would expect that such non-perturbative effects at the scale 
of the cutoff would be small at the mass values found here, 
but there is no proof of this. Comparison with three-flavor 
results with other lattice regularizations will be important in
resolving this question. 

\tabref{Li-syst} shows the systematic errors for the Gasser-Leutwyler low energy constants, $L_i$.
Central values are obtained from averaging the results of Fit A and Fit B (on mass sets {\it I}\/
and {\it II}\/ respectively); those results are repeated here for
convenience from \tabref{fit-params}.  The difference between these fit results and the central value is 
the largest contribution to the chiral/continuum extrapolation error for $2L_8-L_5$ and $2L_6-L_4$.
As discussed in  \secrefs{NNLO}{mass-dependence}, we include two additional systematic
errors here, to be added in quadrature with the scale and chiral/continuum extrapolation
errors: the NNLO error caused by taking $\mu\to\mu_{\rm tree}$ in the NLO terms,
and the effect of using a slightly mass-dependent renormalization scheme.

\begin{table}[t]
\begin{center}
\setlength{\tabcolsep}{1.5mm}
\def\arraystretch{.8}
\begin{tabular}{|c|c|c|c|c|}
\noalign{\vspace{0.5cm}}
\hline
&$L_5$  & $L_4$ & $2L_8-L_5$ & $2L_6-L_4$  \\
\hline
  central value &  $\phantom{+}1.89$       & $\phantom{+}0.19$  & $-0.18$ & $ \phantom{+}0.47$ \\
\hline
\multicolumn{5}{|c|}{errors}\\
\hline
 statistics & $\phantom{+}0.28$ & $\phantom{+}0.29$ & $\phantom{+}0.11$    & $\phantom{+}0.16$  \\
\hline
 scale  &     $+0.01$ & $+0.06$ & $+0.03$ & $+0.03$ \\
        &     $-0.00$ & $-0.05$ & $-0.03$ & $-0.03$ \\
\hline
$\mu_{\rm tree}$ & $\phantom{+}0.13$ & $\phantom{+}0.01$ & $\phantom{+}0.01$ & $\phantom{+}0.03$ \\
\hline
 mass dependent  &     $+0.06$ & $+0.14$ & $+0.03$ & --- \\
 scheme          &   ---       & ---     &   ---   & $-0.01$ \\
\hline
 chiral/continuum & $+0.24$ & $+0.21$  & $+0.15$ & $+0.38$ \\
extrapolation     & $-0.15$ & $-0.19$  & $-0.20$ & $-0.31$ \\
\hline
\hline
\multicolumn{5}{|c|}{chiral/continuum error slices}\\
\hline
 Fit A&               $\phantom{+}1.83$ & $\phantom{+}0.18$ & $-0.04$ & $\phantom{+}0.24$ \\
 Fit B&               $\phantom{+}1.95$ & $\phantom{+}0.20$ & $-0.33$ & $\phantom{+}0.70$ \\
\hline
 a &                $+0.21$ & $+0.19$ & $+0.15$ & $+0.33$ \\
   &                $-0.14$ & $-0.17$ & $-0.18$ & $-0.27$ \\
\hline
 a2 &               $+0.03$ & $+0.11$  & $+0.02$ & $+0.07$ \\
    &               $-0.07$ & $-0.06$  & $-0.02$ & $-0.04$ \\
\hline
\hline
 b &                $+0.06$ & $+0.10$  & $+0.03$ & $+0.01$ \\
   &                $-0.04$ & $-0.09$  & $-0.02$ & $-0.04$ \\
\hline
 b1 &               $+0.00$ & $+0.00$  & $+0.00$ & $+0.00$ \\
    &               $-0.00$ & $-0.00$  & $-0.01$ & $-0.00$ \\
\hline
 b2 &               $+0.06$ & $+0.07$  &  ---    & $+0.01$ \\
    &               $-0.03$ & $-0.09$  & $-0.02$ & $-0.02$ \\
\hline
 b3 &               $+0.00$ & $+0.10$  &  ---    & $+0.00$ \\
    &               ---     &  ---     & $-0.02$ & ---     \\
\hline
\end{tabular}
\caption{Central values and error estimates for $L_i$ (multiplied by $10^3$) at chiral scale $\Lambda_\chi=m_\eta$.
We show differences from the central values everywhere
except for the lines marked Fit A and Fit B,
where we give the results from those fits.
See text for explanations of the various ``error slices.''
\label{tab:Li-syst}
}
\end{center}
\end{table}

The chiral/continuum ``error slices'' in \tabref{Li-syst} have the same meaning as
for the decay constants, except that a1 and c no longer apply. (Slice a1 shows differences
with mass set {\it III}\/, which is not included in this part of the analysis,
and slice c is not relevant since these quantities are themselves fit parameters.)
Further, slice a3, the effects of the mass-dependent scheme, has now been promoted to
a separate error.

After adding the systematic errors in quadrature and symmetrizing as before, we obtain:
\begin{eqnarray}\label{eq:Li_results}
L_5 &=& 1.9(3)(3) \; \times 10^{-3} \nonumber \\
L_4 &=& 0.2(3)(3) \; \times 10^{-3} \nonumber\\
2L_8 - L_5 &=& -0.2(1)(2) \; \times 10^{-3} \nonumber\\
2L_6 - L_4 &=& 0.5(2)(4) \; \times 10^{-3} \ ,
\end{eqnarray}
Systematic errors here are dominated by differences over
acceptable fits.  The chiral scale is taken as $\Lambda_\chi=m_\eta$ throughout.
Including all choice (3) fits in the systematic error analysis would not change the errors.

Reference~\cite{DONOGHUE} makes the following continuum estimates:
$L_5 = 2.3(2)\!\times\!10^{-3}$, $L_4 \approx L_6 \approx 0$;
while Ref.~\cite{COHEN} gives $L_5 = 2.2(5)\!\times\!10^{-3}$,
$L_4=0.0(5)\!\times\!10^{-3}$ and $L_6=0.0(3)\!\times\!10^{-3}$ (which they
call ``conventional estimates.'')
Here we have converted all the $L_i$ to $\Lambda_\chi=m_\eta$ scale
using \eqs{Li-scale}{Ci}.

The result for $2 L_8 -L_5$ is well outside the range that would allow for
$m_u=0$ \cite{Kaplan:1986ru,COHEN,Nelson:tb} 
in the context of \chpt:
\begin{equation}\label{eq:L85range}
-3.4\times 10^{-3}\; \ltwid \; 2L_8\!-\!L_5\; \ltwid\, -1.8\times 10^{-3} \ .
\end{equation}
We note, however, that the constraint on $m_u$ coming from $2 L_8 -L_5$ is not
independent from the direct determination above.  Knowing $2 L_8 -L_5$ would
fix $m_u$ in NLO up to EM effects.  The range in \eq{L85range} comes
from unknown NNLO (and EM) terms.  Since our fits give us some control
over NNLO effects, the direct determination seems preferable, and can become
quite precise if one uses more information on EM effects. This information
may come from phenomenology, \eg Ref.~\cite{Bijnens:1996kk}, or from lattice
simulations, perhaps along the lines of Refs.~\cite{Duncan:1996xy} or \cite{Duncan:2004ys}.

Our approach to computing low energy constants has much in common with earlier
work by Nelson, Fleming, and Kilcup \cite{Nelson:tb}, who also performed a partially quenched
analysis using 3 flavors of dynamical staggered quarks.   The main advances in the current analysis are: (1)
use of the improved dynamical staggered action and finer lattice spacings, putting us closer
to continuum physics, and (2) use of \schpt\ to control lattice artifacts, which are
still quite large, despite (1).  Our result for $2L_8-L_5$ is marginally consistent with that by 
Nelson \et\ \cite{Nelson:tb};
converting their result to chiral scale $m_\eta$, we get $2L_8-L_5 = -0.57(1)(14)\times10^{-3}$.

The current work will be improved by additional simulations now in progress, including coarse lattices
at lower strange quark mass ($am'_s=0.03$) and fine lattices at lower light quark mass ($a\hat m' = 0.1 am'_s = .0031$).
These simulations should enhance our control of the chiral extrapolation, the interpolation around the $s$ quark mass, 
and the extraction of low energy constants. In addition, we are beginning a parallel analysis
on a large quenched data set.  If the corresponding \schpt\ forms can describe that data well, it will increase
our confidence that the interaction of discretization and chiral effects is understood. 
Beyond that, planned simulations at still finer lattice
spacings will provide a better handle on both generic and taste-violating discretization errors,
thereby significantly reducing the final systematic errors.

\section*{ACKNOWLEDGMENTS}
We thank Christine Davies, Peter Lepage, Junko Shigemitsu and Matt Wingate for essential
discussions and correspondence.  We are grateful to
Oliver B\"ar, Gilberto Colangelo, George Fleming, Maarten Golterman, Rajan Gupta, Shoji Hashimoto, and 
Laurent Lellouch for discussion and constructive criticism. 
CB thanks the Aspen Center for Physics,  and DT thanks the
Institute for Nuclear Theory at the University of Washington
for hospitality during part of the course of
this work.
Computations for this work were performed at
the San Diego Supercomputer Center (SDSC),
the Pittsburgh Supercomputer Center (PSC),
Oak Ridge National Laboratory (ORNL),
the National Center for Supercomputing Applications (NCSA),
the National Energy Resources Supercomputer Center (NERSC),
Fermilab,
and Indiana University.
This work was supported by the U.S. Department of Energy under grants
DOE -- DE-FG02-91ER-40628,      
DOE -- DE-FG02-91ER-40661,      
DOE -- DE-FG02-97ER-41022       
and
DOE -- DE-FG03-95ER-40906       
and National Science Foundation grants
NSF -- PHY01--39939              
and
NSF -- PHY00--98395.            

\vfill\eject

\begin{figure}[tbh]
\resizebox{5.0in}{!}{\includegraphics{masses_b676_nofix.ps}}
\caption{
Pion masses with random-wall and Coulomb-wall sources and 
point and Coulomb-wall sinks from the coarse set with sea quark
lattice masses 0.01,0.05 (see Table~\protect{\ref{RUNTABLE}}).
The \tmpred crosses are random-wall source and Coulomb-wall sink,
and the \tmpgreen octagons are Coulomb-wall source and point sink (summed
over spatial sites to project out the zero momentum states).
The \tmpblue bursts are from a random-wall source and point sink,
and the \tmpmagenta squares have a Coulomb-wall source and sink.
The lower set of ``WW'' points include an excited state in the fit.
The symbol size is proportional to the confidence level of the fit,
with the symbol size in the labels corresponding to 50\%.
\label{masses_fig}}
\end{figure}
\vfill\eject

\begin{figure}[tbh]
\resizebox{5.0in}{!}{\includegraphics{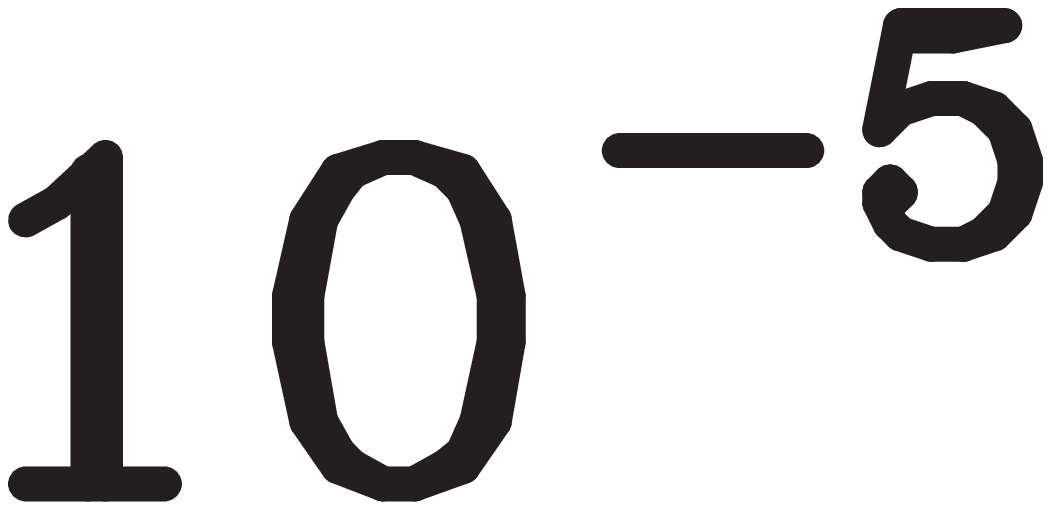}}
\caption{
Same as Fig.~\protect{\ref{masses_fig}} but for
pion propagator amplitudes.
The lower set of ``WW'' points again include an excited state in the fit.
The ``PW'' symbols have been displaced slightly to the right to separate
them from the ``WP'' points.
\label{amps_fig}}
\end{figure}
\vfill\eject

\begin{figure}[tbh]
\resizebox{5.0in}{!}{\includegraphics{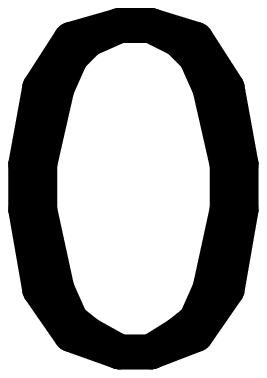}}
\caption{ Ratio of pion propagators.
Here $P_{WP}$ is the Coulomb wall source and point sink pion
propagator, {\it etc}\/.  The point source was implemented with a random
wall as discussed in the text.
\label{prop_ratio_fig}}
\end{figure}
\vfill\eject

\begin{figure}[tbh]
\resizebox{5.0in}{!}{\includegraphics{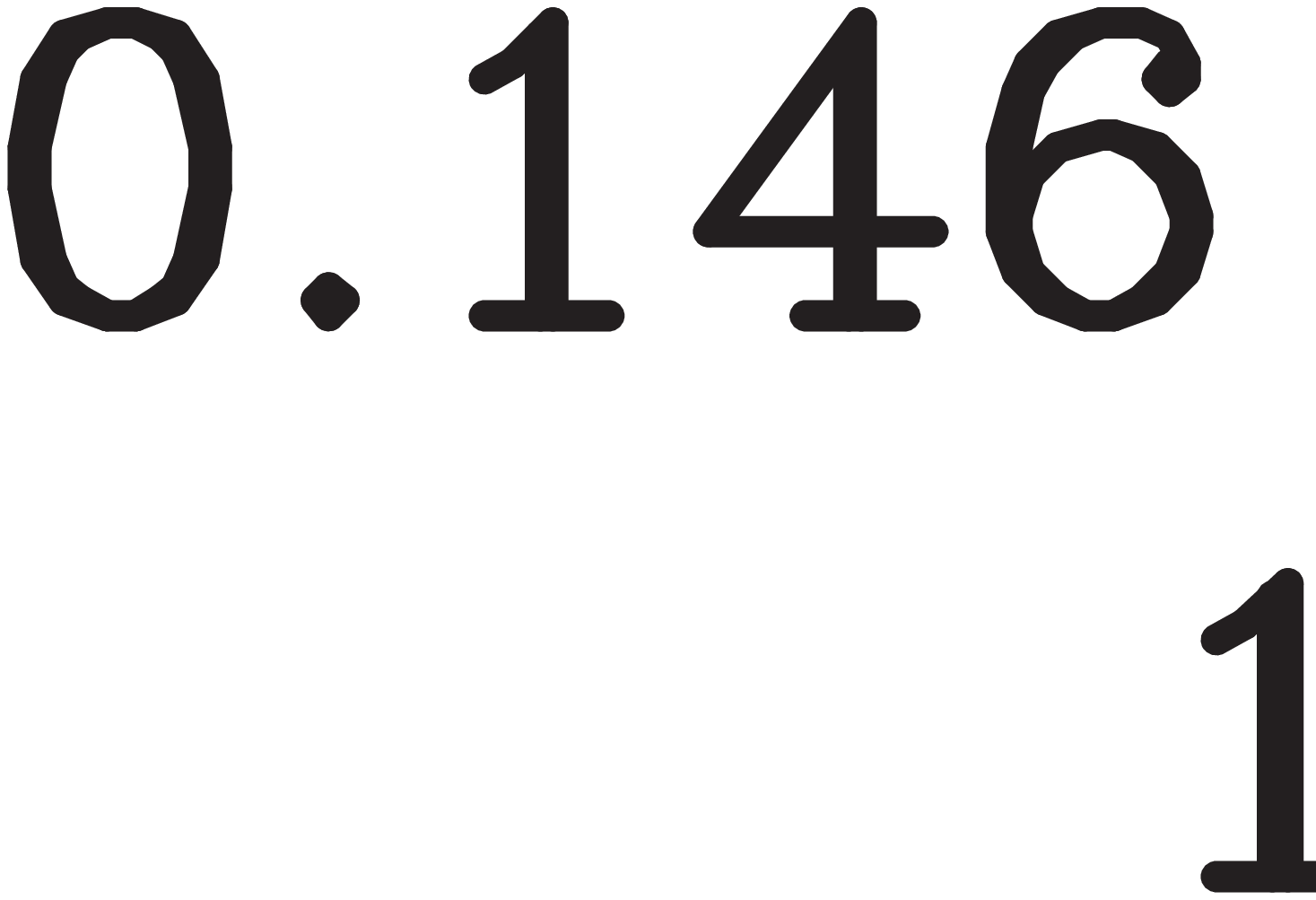}}
\caption{
Pion masses (\tmpred octagons) and amplitudes (\tmpblue crosses) as a function of the minimum time 
distance in the fit, 
from the fine set with sea quark
lattice masses 0.0062,0.031 (see Table~\protect{\ref{RUNTABLE}}).
The amplitudes have been multiplied by 175.}
\label{dmin-fine}
\end{figure}
\vfill\eject

\begin{figure}[tbh]
\resizebox{5.0in}{!}{\includegraphics{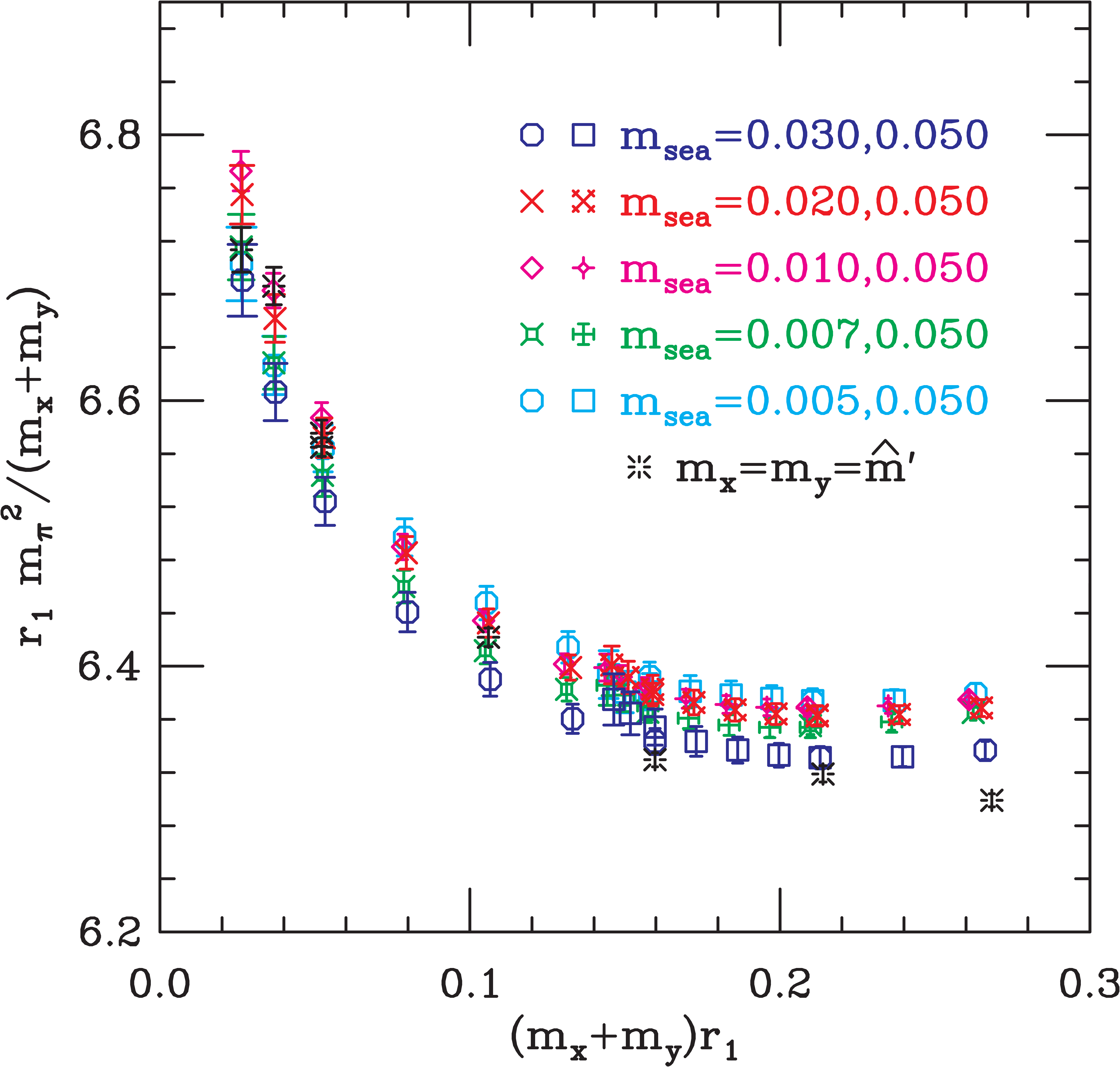}}
\caption{Pseudoscalar masses with $a \approx 0.125$ fm.
The horizontal axis is the sum of the valence quark mass
(in units of $r_1$).
For each set of values of $m_{\rm sea}$, 
the first symbol shows ``pion'' points with $m_x=m_y$; while
the second shows ``kaon'' points with $m_y=m'_s$. Bursts are pion points
 with valence masses equal to sea quark masses.  
\label{mpisq_coarse_fig}}
\end{figure}
\vfill\eject

\begin{figure}[tbh]
\resizebox{5.0in}{!}{\includegraphics{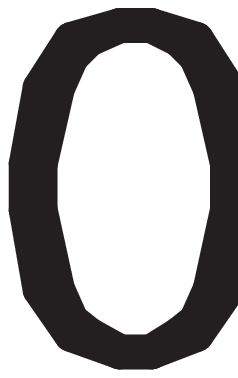}}
\caption{Decay constants in units of $r_1$ with $a \approx 0.125 $ fm.
The abscissa and symbols are the same as in Fig.~\protect{\ref{mpisq_coarse_fig}}.
\label{fpi_coarse_fig}}
\end{figure}
\vfill\eject

\begin{figure}[tbh]
\resizebox{5.0in}{!}{\includegraphics{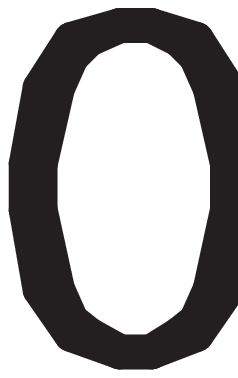}}
\caption{``Pion'' masses ($m_x=m_y$) and ``kaon''
masses ($m_y=m'_s$) with sea quark masses
$\hat m' = 0.4 m'_s$ and
$\hat m' = 0.2 m'_s$ at $a \approx 0.125 $ fm and $a \approx 0.09$ fm.   A ``point by point'' extrapolation to
$a=0$ (fancy magenta squares: $\hat m' = 0.4 m'_s$, and cyan squares: $\hat m' = 0.2 m'_s$) is also included.
\label{mpisq_extrap_fig}}
\end{figure}
\vfill\eject

\begin{figure}[tbh]
\resizebox{5.0in}{!}{\includegraphics{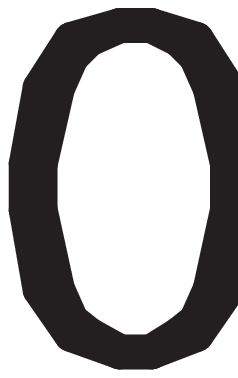}}
\caption{Same as Fig.~\protect{\ref{mpisq_extrap_fig}}
but for decay constants.
\label{fpi_extrap_fig}}
\end{figure}
\vfill\eject

%

\begin{figure}[tbh]
\resizebox{5.0in}{!}{\includegraphics{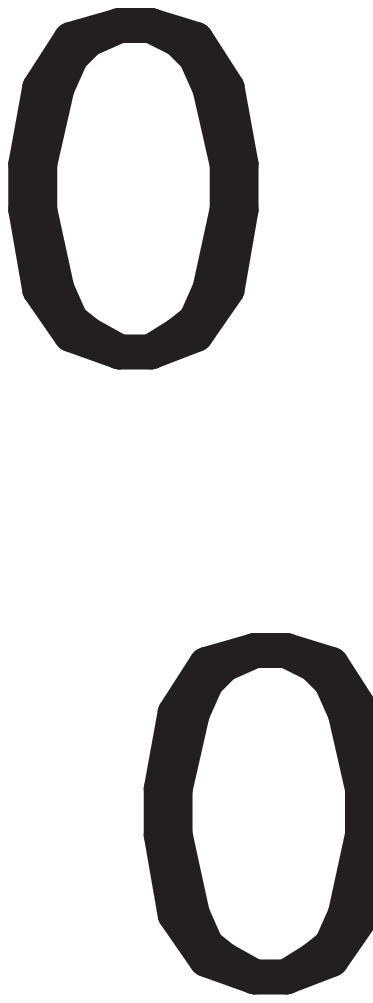}}
\caption{Squared masses of charged pions for various tastes on the coarse lattices.
We use $r_1$ to set the scale.
Tastes that are degenerate by $SO(4)$ symmetry are fit together.}
\label{fig:splittings}
\end{figure}
\vfill\eject

\begin{figure}[tbh]
\resizebox{5.0in}{!}{\includegraphics{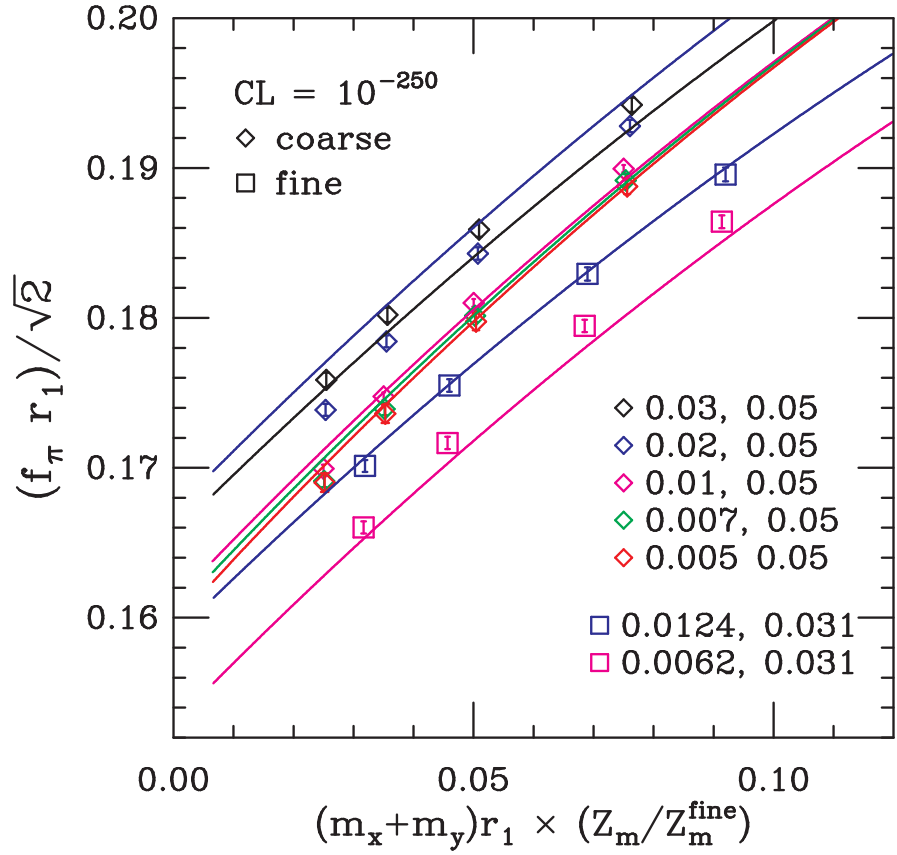}}
\caption{Fit of partially quenched data to continuum form.  Data for both 
$f_\pi$ and $m_\pi^2/(m_x+m_y)$
with various $m_x$ and $m_y$ values are included in the fit, but 
only $f_\pi$ points with $m_x=m_y$ are shown.}
\label{fig:fpi-vs-m-no-taste-viols}
\end{figure}
\vfill\eject

\begin{figure}[tbh]
\resizebox{5.0in}{!}{\includegraphics{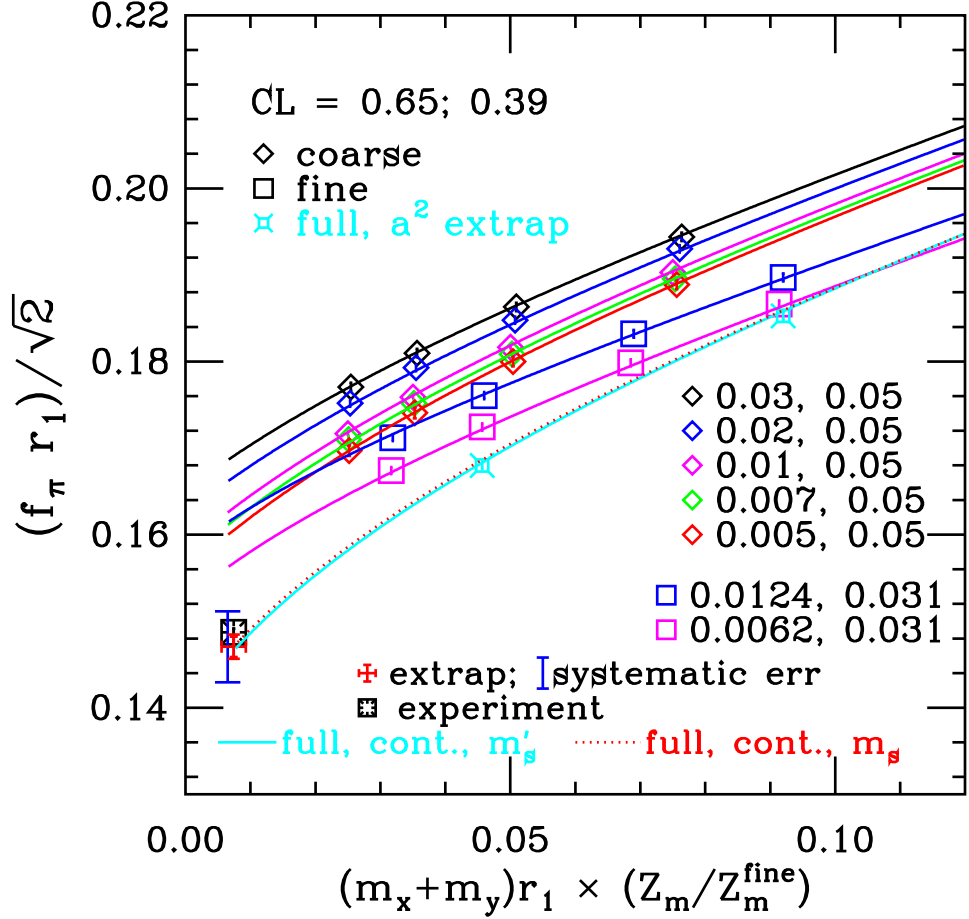}}
\caption{
``Pion'' decay constants with $m_x=m_y$ \vs quark mass, in units of $r_1$.  The relative
mass renormalization of coarse and fine lattices has been included so that data from both
may be presented on the same plot. Lines come from ``Fit B,'' a {\it single}\/ NNLO fit, \protect{\eqs{m-NNLO}{f-NNLO}},
to entire data subset {\it II}\/ (decay constants and masses). 
The \tmpcyan solid line and \tmpred dotted line represent the fit function in ``full QCD'' (valence and sea
masses set equal) after extrapolation of
parameters to the continuum.  The \tmpcyan solid line keeps the $s$ quark mass equal to $m'_s$ on the fine
lattices; while the \tmpred dotted line (just barely visible above the  \tmpcyan solid line)
replaces  $m'_s$ with the physical mass $m_s$.  
The \tmpcyan fancy squares result from extrapolation of full QCD points to the continuum at fixed
quark mass; their agreement with the \tmpcyan solid line is a consistency check. 
Points and fit lines have been
corrected for finite volume effects.
}
\label{fig:fpi-vs-m-II}
\end{figure}
\vfill\eject

\begin{figure}[tbh]
\resizebox{5.0in}{!}{\includegraphics{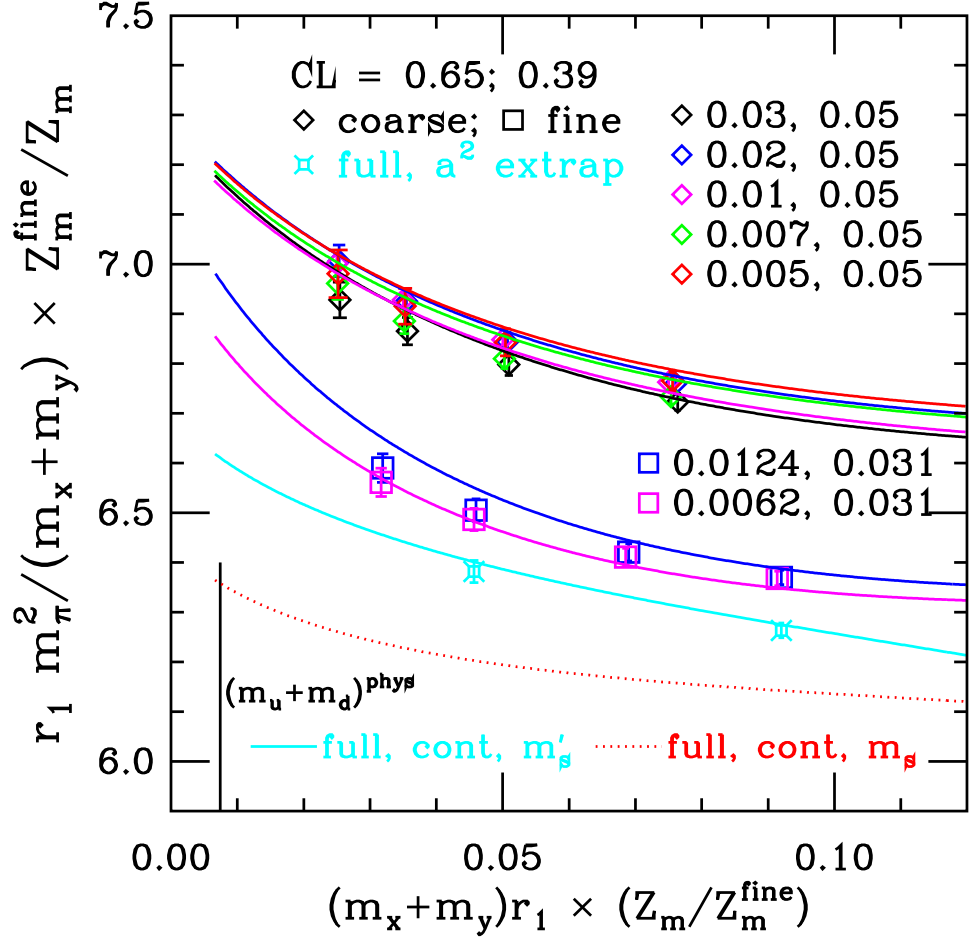}}
\caption{
Same as \protect{\figref{fpi-vs-m-II}} (Fit B), but squared ``pion'' masses divided by quark mass are shown.  
Because taste splittings are smaller for the fine lattices,
the average meson mass changes 
more rapidly with quark mass, and there is greater curvature at small quark mass.
}
\label{fig:mpisq-over-m-vs-m-II}
\end{figure}
\vfill\eject

\begin{figure}[tbh]
\resizebox{5.0in}{!}{\includegraphics{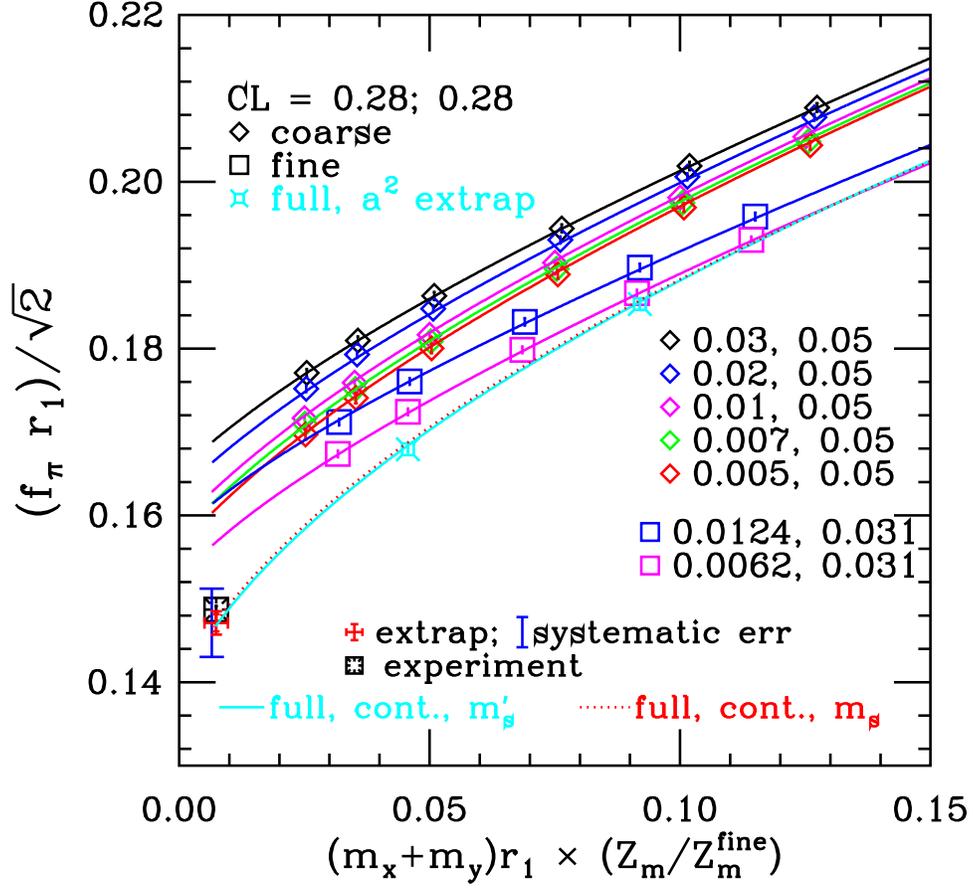}}
\caption{
``Pion'' decay constants with $m_x=m_y$ in data subset {\it III}\/. Lines come from Fit C, a {\it single} NNNLO fit
to masses and decay constants.
The \tmpcyan solid line, \tmpred dotted line, and 
\tmpcyan fancy squares have the same meaning as in 
\protect{\figref{fpi-vs-m-II}}.
 Points (and fit lines) have been
corrected for finite volume effects. 
}
\label{fig:fpi-vs-m-III}
\end{figure}
\vfill\eject

\begin{figure}[tbh]
\resizebox{5.0in}{!}{\includegraphics{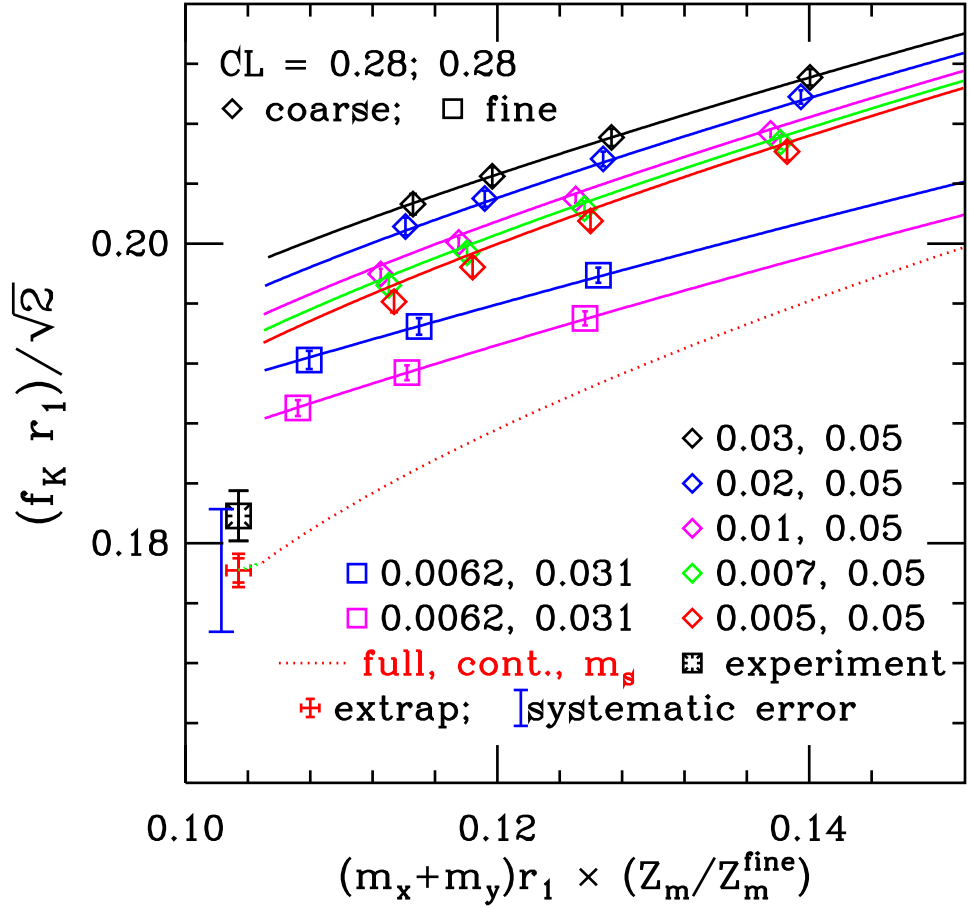}}
\caption{
Similar to \protect{\figref{fpi-vs-m-III}} (Fit C), but for
``kaon'' decay constants, with $m_y$ fixed as closely as possible on each set 
to $m_s^{\rm phys}$. The \tmpred dotted line has parameters extrapolated to the continuum,
the strange valence and sea masses set to the physical value, and light sea quark
mass equal to light valence mass $m_x$. The \tmpgreen dotted extension has the light sea quark
mass fixed at $\hat m$, and the valence mass $m_x$ continuing down to $m_u$ (thus giving
$f_{K^+}$).
}
\label{fig:fK-vs-m-III}
\end{figure}
\vfill\eject

\begin{figure}[tbh]
\resizebox{5.0in}{!}{\includegraphics{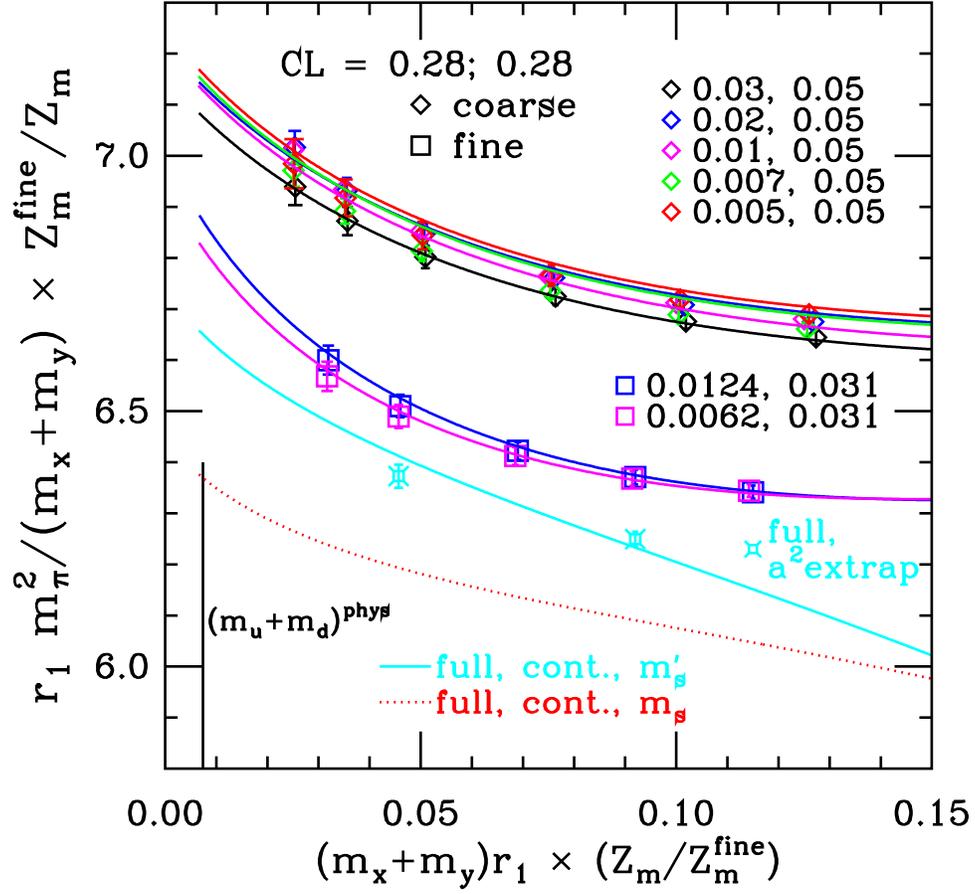}}
\caption{
Same as \protect{\figref{fpi-vs-m-III}} (Fit C), but for squared ``pion'' masses divided by quark mass.  
}
\label{fig:mpisq-over-m-vs-m-III}
\end{figure}
\vfill\eject

\begin{figure}[t]%
\begin{center}%
\resizebox{5.0in}{!}{\includegraphics[scale=0.35,angle=0,clip]{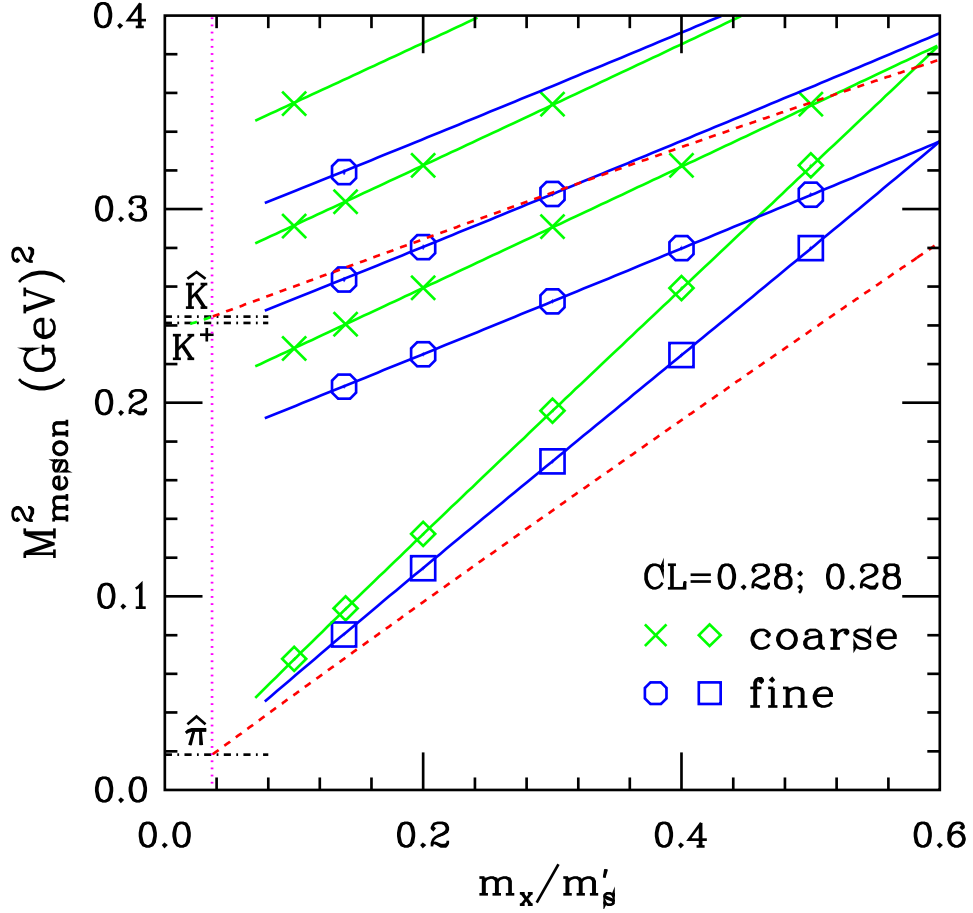}}
\end{center}%
\caption{Squared meson masses in subset {\it III}\/, 
as a function of $m_x/m'_s$. The lines are from Fit C.
 We show results from two lattices: a coarse
lattice with sea quark masses $a\hat m'=0.01$, $am'_s=0.05$,
and a fine lattice with $a\hat m'=0.0062$,
$am'_s=0.031$.  Three sets of ``kaon'' points
with $m_y=m'_s, 0.8 m'_s,  0.6 m'_s$, are plotted for
each lattice.
``Pion'' points have $m_x=m_y$.
The statistical errors in the points
are not visible on this scale.
The \tmpred dashed lines give the continuum-extrapolated fit
(now as a function of $m_x/m_s$), and the \tmpmagenta vertical dotted line shows
the physical $\hat m/m_s$ obtained. The extension (shown in \tmpgreen$\!\!$)
of the \tmpred dashed kaon line until it intersects the
QCD $K^+$ value then gives $m_u/m_s$, from which we find $m_u/\hat m$ or $m_u/m_d$.}
\label{fig:msq-vs-m-III}%
\end{figure}%
\vfill\eject

\begin{figure}[t]%
\begin{center}%
\resizebox{5.0in}{!}{\includegraphics[scale=0.35,angle=0,clip]{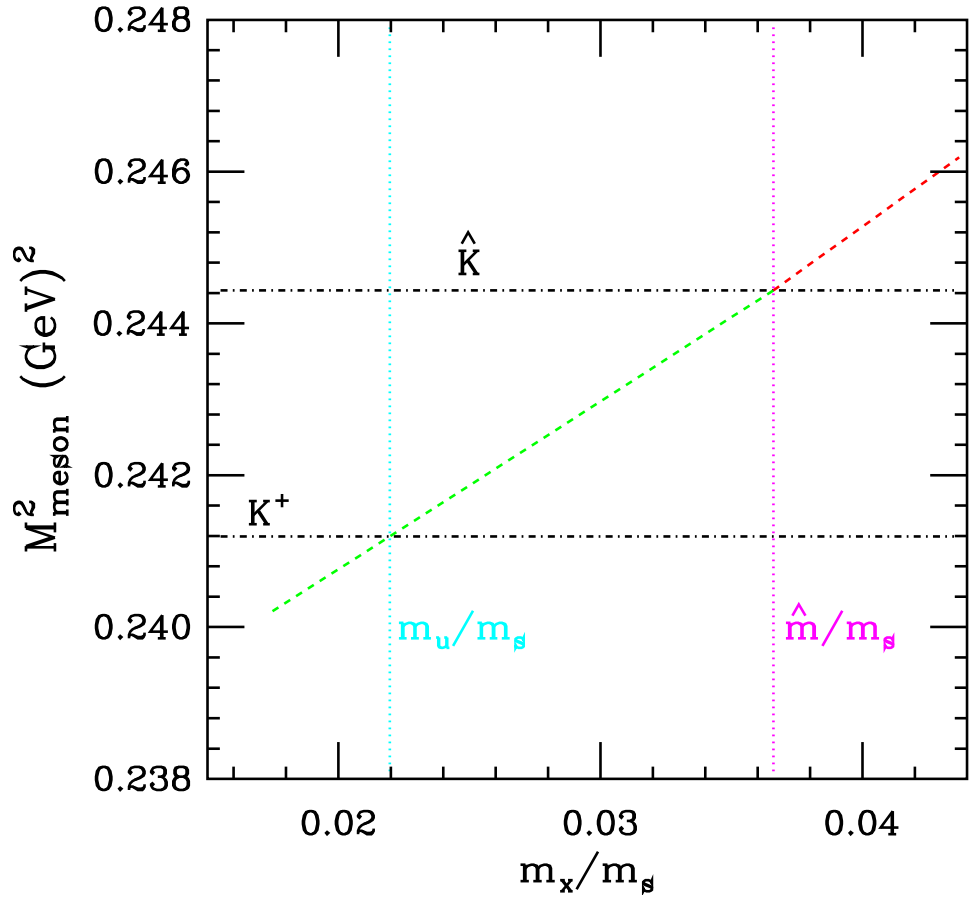}}
\end{center}%
\caption{A magnification of the region around the $\hat K$ and $K^+_{\rm QCD}$ masses in \protect{\figref{msq-vs-m-III}}.
The dotted vertical lines give  $m_u/m_s$ and $\hat m/m_s$.}
\label{fig:msq-vs-m-III-blowup}%
\end{figure}%
\vfill\eject

\begin{figure}[tbh]
\resizebox{5.0in}{!}{\includegraphics{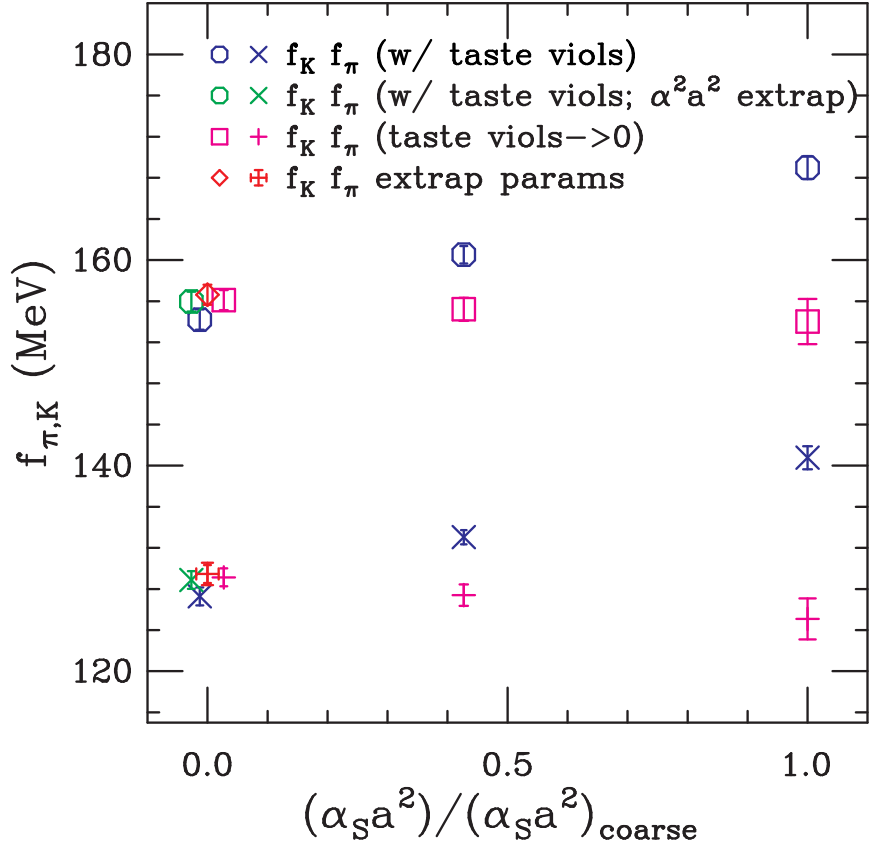}}
\caption{
Dependence of chirally-extrapolated $f_\pi$ and $f_K$ on lattice spacing.  
The \tmpblue octagons and crosses show values at fixed lattice spacing (using ``method (1)'')
and those extrapolated
to the continuum linearly in $\alpha a^2$.  For the \tmpgreen octagons and crosses, the extrapolation
is linear in $\alpha^2 a^2$.
The \tmpmagenta squares and pluses have the taste-violating effects at fixed lattice spacing
removed with \schpt\ (``method (2)''); the points are then extrapolated 
to the continuum linearly in $\alpha a^2$. The \tmpred diamond and fancy plus are the results of
extrapolating the chiral fit parameters to the continuum.  Extrapolated points at $a=0$ have been moved slightly
horizontally for clarity.  Fit C is used everywhere. }
\label{fig:f-vs-a2}
\end{figure}
\vfill\eject
      
\begin{figure}[tbh]
\resizebox{5.0in}{!}{\includegraphics{mpisq-over-m_vs_m_extrap_fixD_z.958_.036.025_blowup.ps}}
\caption{
Enlargement of a small region of \protect{\figref{mpisq-over-m-vs-m-II}}.
}
\label{fig:mpisq-over-m-vs-m-II-blowup}
\end{figure}
\vfill\eject

\begin{figure}[tbh]
\resizebox{5.0in}{!}{\includegraphics{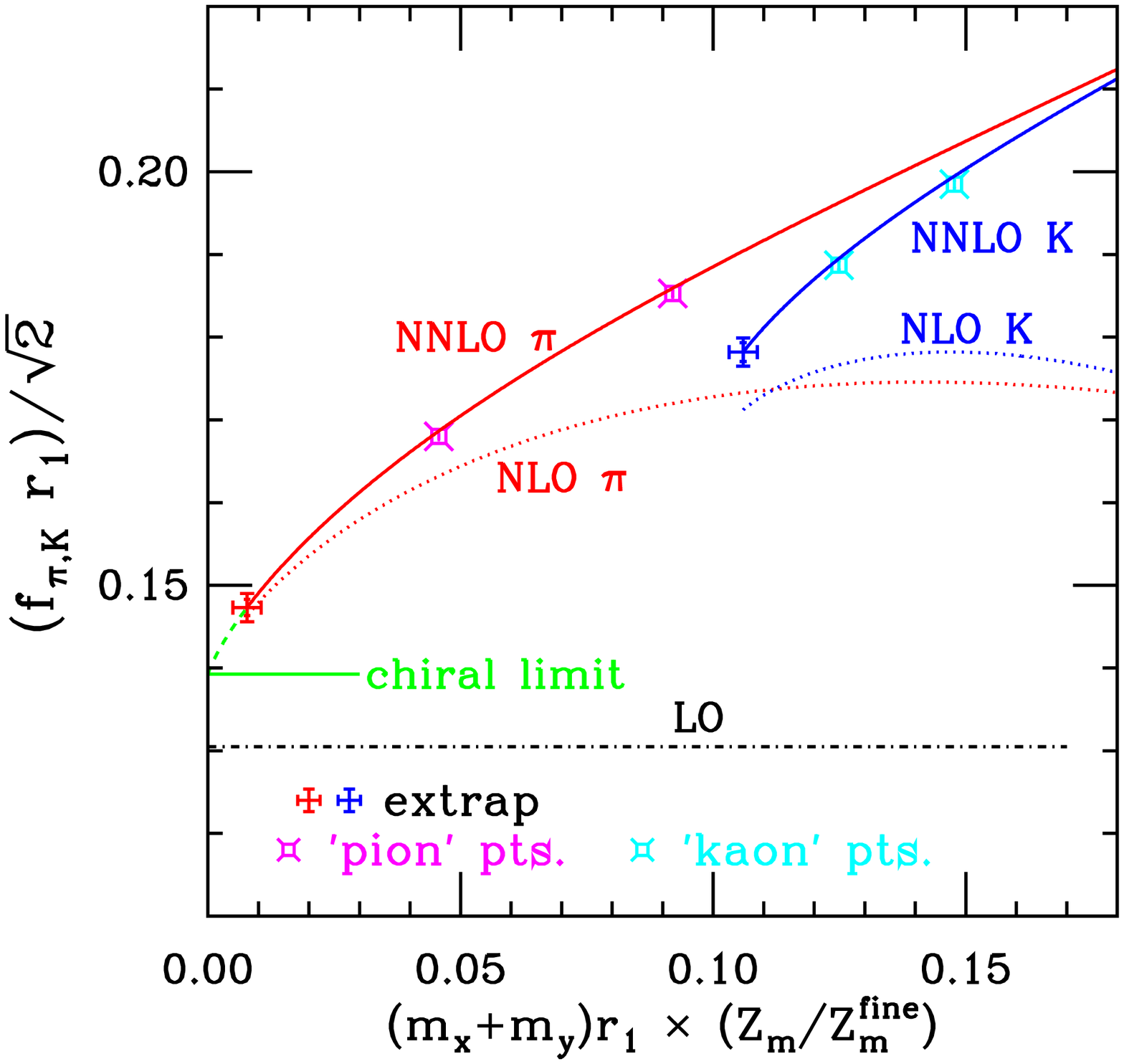}}
\caption{
Convergence of $SU(3)_L\times SU(3)_R$ \chpt\ for decay constants.  Our results on mass set {\it II}\/ for $f_\pi$ and $f_K$
at LO (dash-dotted \tmpblack line), NLO (dotted \tmpred \tmpand \tmpblue curves), 
and NNLO (solid \tmpred \tmpand \tmpblue curves) are shown.  The chiral parameters have been
extrapolated to the continuum.  Convergence of $SU(2)_L\times SU(2)_R$ \chpt\  can be seen by
comparing the $f_\pi$ results to the chiral limit, obtained (\tmpgreen dashed curve) by extrapolating
the NNLO $f_\pi$ result to $\hat m=0$. The \tmpmagenta \tmpand \tmpcyan fancy squares are found by extrapolating
our full QCD data to the continuum limit at fixed quark mass.   All results come from Fit B.
}
\label{fig:chpt-converge-fpi}
\end{figure}
\vfill\eject

\begin{figure}[tbh]
\resizebox{5.0in}{!}{\includegraphics{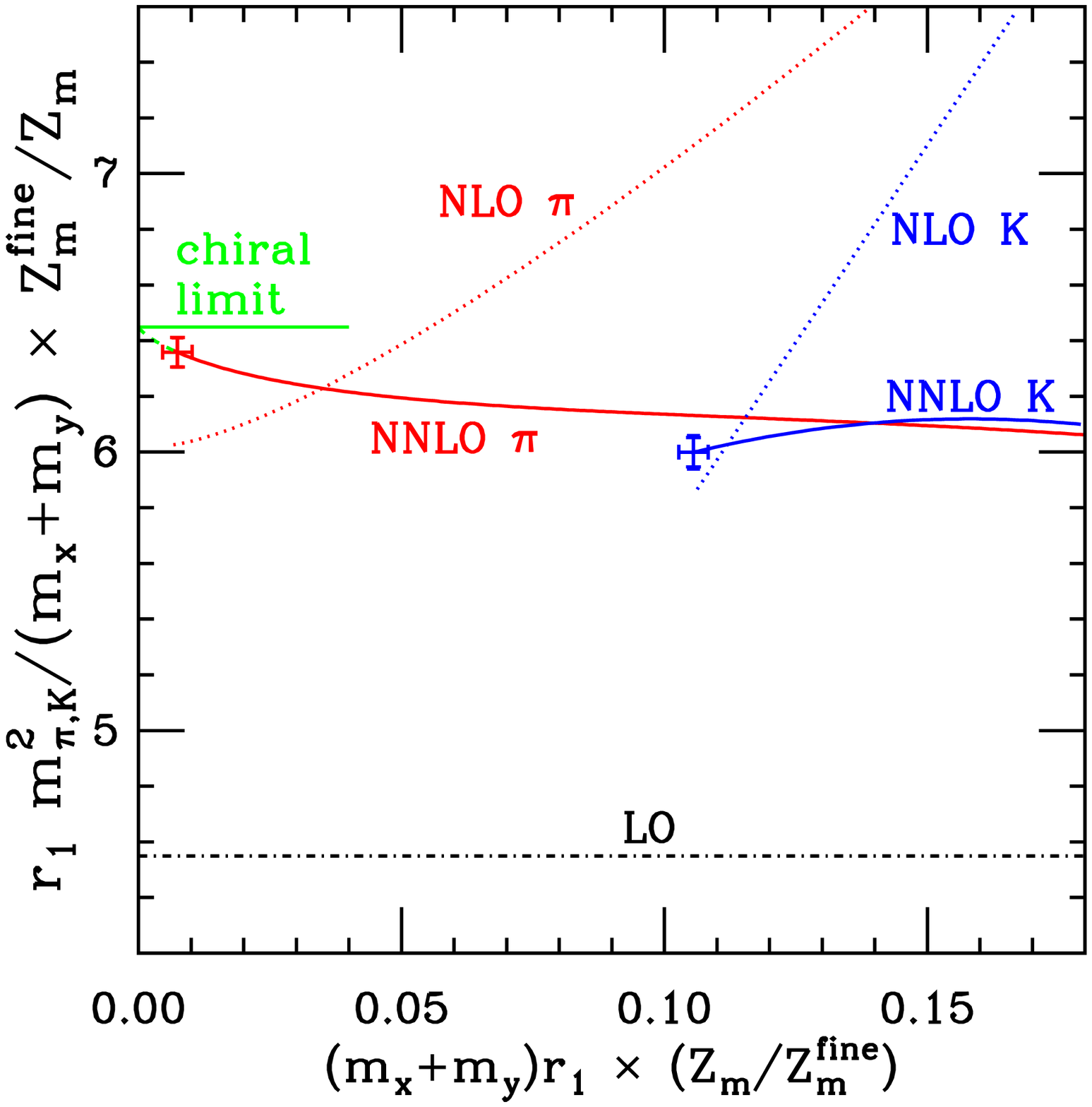}}
\caption{
Same as \protect{\figref{chpt-converge-fpi}}, but for $m_{P+}^2/(m_x+m_y)$. This uses Fit B (data subset {\it II}\/).
}
\label{fig:chpt-converge-msq-II}
\end{figure}
\vfill\eject
      
\begin{figure}[tbh]
\resizebox{5.0in}{!}{\includegraphics{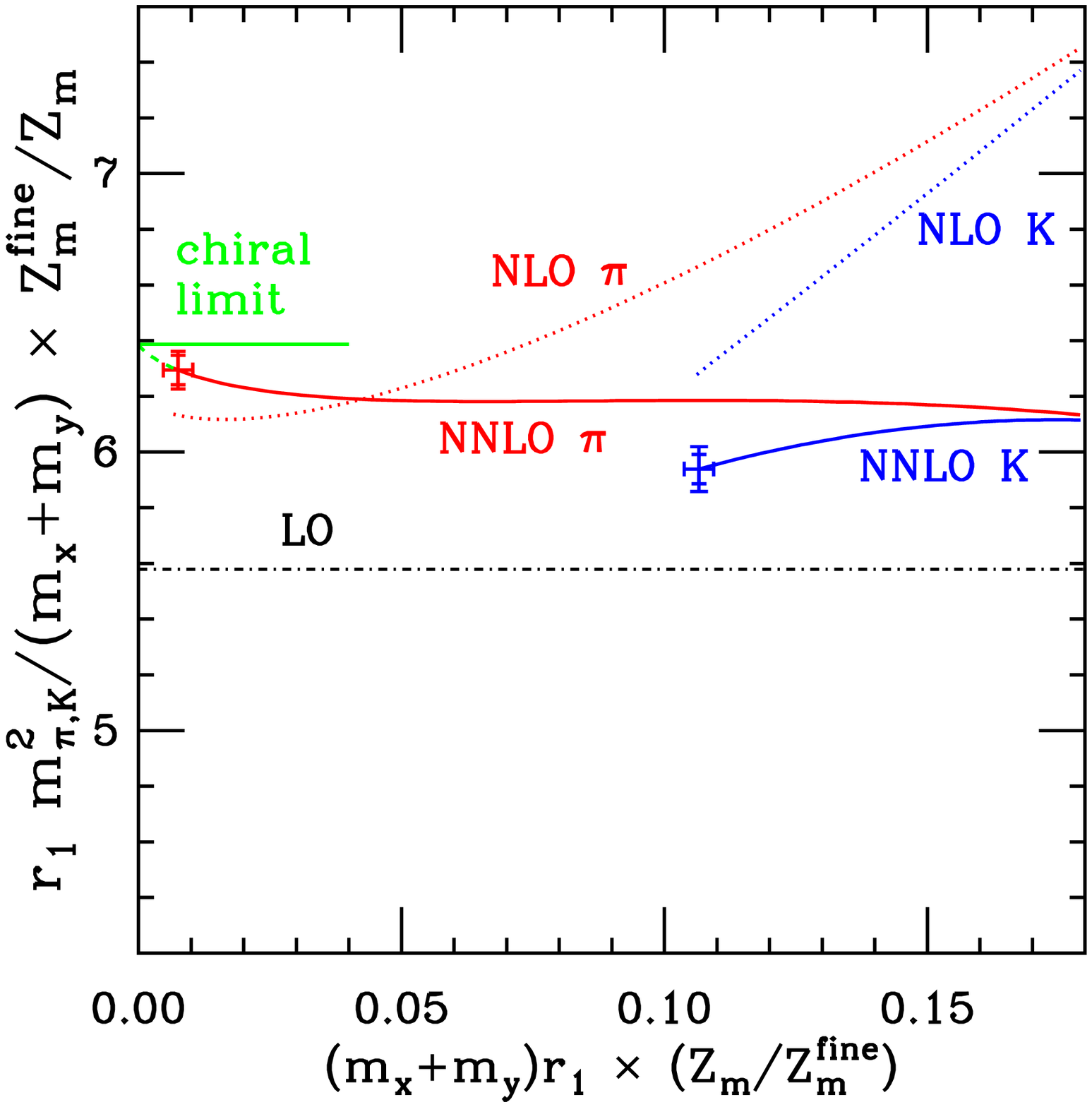}}
\caption{
Same as \protect{\figref{chpt-converge-msq-II}}, but for Fit A (data subset {\it I}\/).
}
\label{fig:chpt-converge-msq-I}
\end{figure}
\vfill\eject
\end{document}